\documentclass[twocolumn]{aastex63}
\newcommand{\psj}{Planet. Sci. J.}

\submitjournal{AJ}

\shortauthors{Vokrouhlick{\' y} et~al.}

\graphicspath{{./}{figures/}}

\usepackage{tikz}
\usetikzlibrary{arrows,shapes,positioning,shadows,trees}

\tikzset{
  basic/.style  = {draw, text width=4cm, drop shadow, font=\sffamily, rectangle},
  root/.style   = {basic, rounded corners=2pt, thin, align=center, fill=blue!10},
  level 2/.style = {basic, rounded corners=6pt, thin,align=center, fill=pink!40, text width=11em},
  level 3/.style = {basic, thin, align=center, fill=white!60, text width=8em}
}

\usepackage[utf8]{inputenc}
\usepackage{dirtytalk}

\usepackage{natbib}

\begin{document}

\title{Orbital and absolute magnitude distribution of Jupiter Trojans}

\correspondingauthor{David Vokrouhlick\'y}
\email{vokrouhl@cesnet.cz}

\author[0000-0002-6034-5452]{David Vokrouhlick\'y}
\affiliation{Astronomical Institute, Charles University, V Hole\v{s}ovi\v{c}k\'ach 2,
             CZ 18000, Prague 8, Czech Republic}
\author[0000-0002-4547-4301]{David Nesvorn{\' y}}
\affiliation{Department of Space Studies, Southwest Research Institute, 1050 Walnut St., Suite 300,
             Boulder, CO 80302, United States}
\author[0000-0003-2763-1411]{Miroslav Bro{\v z}}
\affiliation{Astronomical Institute, Charles University, V Hole\v{s}ovi\v{c}k\'ach 2,
             CZ 18000, Prague 8, Czech Republic}
\author[0000-0002-1804-7814]{William F. Bottke}
\affiliation{Department of Space Studies, Southwest Research Institute, 1050 Walnut St., Suite 300,
             Boulder, CO 80302, United States}
\author[0000-0001-6730-7857]{Rogerio Deienno}
\affiliation{Department of Space Studies, Southwest Research Institute, 1050 Walnut St., Suite 300,
             Boulder, CO 80302, United States}
\author{Carson D. Fuls}
\affiliation{Lunar and Planetary Laboratory, The University of Arizona, 1629 E. University
             Boulevard, Tucson, AZ 85721-0092, USA}             
\author{Frank C. Shelly}
\affiliation{Lunar and Planetary Laboratory, The University of Arizona, 1629 E. University
             Boulevard, Tucson, AZ 85721-0092, USA}

\begin{abstract}
 Jupiter Trojans (JTs) librate about the Lagrangian stationary centers L4 and L5 associated with
 this planet on a typically small-eccentricity and moderate-inclination heliocentric orbits.
 The physical and orbital properties of JTs provide important clues about the dynamical evolution
 of the giant planets in the early Solar System, as well as populations of planetesimals in their
 source regions. Here we use decade long
 observations from the Catalina Sky Survey (station G96) to determine the bias-corrected orbital and
 magnitude distributions of JTs. We distinguish the background JT population, filling smoothly the
 long-term stable orbital zone about L4 and L5 points, and collisional families. We find that the
 cumulative magnitude distribution of JTs (the background population in our case) has a steep
 slope for $H\leq 9$, followed with a moderately shallow slope till $H\simeq 14.5$, beyond which
 the distribution becomes even shallower. At $H=15$ we find a local power-law exponent $0.38\pm 0.01$. 
 We confirm the asymmetry between the magnitude limited background populations in L4 and
 L5 clouds characterized by a ratio $1.45\pm 0.05$ for $H<15$. Our analysis suggests an asymmetry in
 the inclination distribution of JTs, with the L4 population being tighter and the L5 population being
 broader. We also provide a new catalog of the synthetic proper elements for JTs with an updated
 identification of statistically robust families (9 at L4, and 4 at L5). The previously known
 Ennomos family is found to consist of two, overlapping Deiphobus and Ennomos families.
\end{abstract}


\section{Introduction} \label{intro}
Jupiter Trojans, objects librating about the stationary Lagrangian centers L4 
and L5 accompanying this planet, have a special status among the populations of small
bodies in the Solar System. Their fundamental role is twofold. First,
they represent an outpost of the distant heliocentric populations that are difficult 
to be reached by ground-based observations. Understanding the Jupiter Trojans has
therefore broader implications about the small-body populations far beyond Jupiter. 
Second, because of their strong gravitational coupling to Jupiter, the properties
of the Trojan populations provide unique information about the orbital history of
this planet, and in fact the early fate of the giant planets in general. Additionally, a
special circumstance of the Jupiter Trojans is their duality: they are organized
in two swarms, one about the leading center L4 and the other about the trailing center
L5. Once the objects became part of their respective Trojan clan, they became
disconnected from each other (both dynamically and collisionally). However, before
that moment, they were in all likelihood part of the same population of planetesimals
born in the trans-Neptunian region.%

In this paper, we adopt the presently favored capture hypothesis about the
Trojan origin associated with Jupiter's chaotic orbital evolution during the giant
planets reconfiguration \citep[e.g.,][]{nice2005,nvm2013,nes2018}. Alternative
models, such as in situ Trojan formation, Jupiter's large-scale inward migration leading to 
gas-drag assisted capture,
or a capture related to the terminal runaway mass increase of Jupiter due to a collapse
of its gaseous envelope, all have more difficulty matching JT constraints than the aforementioned
orbitally-triggered capture mechanism. \citep[see also reviews by][]{astiii,sb2014,astiv}.

We argue that substantial differences in 
the parameters of the L4 and L5 swarms may provide 
important clues about (i) the capture mechanism, and/or (ii) the post-capture evolution of
one (or both) of the swarms. Aspects that have grown enigmatic over the past decades are
(i) the putative population asymmetry between the two swarms, and (ii) differences in their
inclination distribution (and a closely related issue of the Trojan families).

Denoting the magnitude-limited populations of the L4 and L5 Trojan swarm by $N_4$ and
$N_5$, we define their ratio $f_{45}=N_4/N_5$ to help us investigate a potential asymmetry.
As an aside, we note that \citet{grav2011} and \citet{grav2012} do not report any statistically
significant difference in albedo-size dependence of the L4 and L5 populations, yet they note small
differences in albedo distribution and taxonomy. This means that one needs to be cautious about interpreting
the magnitude-limited asymmetry as an equal measure of the size-limited asymmetry.  It is possible that
a small charge in albedo for each population may influence the link (see Sec.~\ref{concl}).
 
Deviation of $f_{45}$ from unity was considered as soon as the known Trojan population
has grown sizable. Still, studies up to late 1990s had to face small samples of
detected Trojans in both swarms and, in absence of a solid debiasing efforts, asymmetric
observational incompleteness \citep[implying the
intrinsic $f_{45}$ value was different from what plain observations provided, e.g.,][]{shoe1989}.
The situation started to change in the 2000s from a flood of new data from powerful and well characterized sky surveys.
For example, \citet{szabo2007} analyzed a sample of about 900 Jupiter Trojans (and candidate
Trojans) in nearly five years of the Sloan Digital Sky Survey (SDSS) observations. These
authors found $f_{45}=1.6\pm 0.1$ down to the inferred $H\simeq 13.8$ magnitude completeness.

In a second example, \citet{ny2008} used observations of the 8.2~m Subaru telescope in a campaign conducted during the
2001 season to characterize the L4 and L5 populations to small sizes. They adopted an empirical
distribution function of the ecliptic longitude and latitude to characterize sky-plane density
of Trojans near their respective libration center. Using this approach, these authors found
$f_{45}=1.85\pm 0.42$ for $D>2$~km (assuming a fixed albedo of $0.04$). 

In a third example, \citet{grav2011}
analyzed space-borne observations of more than 2000 known and candidate Jupiter Trojans taken
by the Wide-field Infrared Survey Explorer (WISE) during its half-year cryogenic phase and
about the same timespan of the post-cryogenic extension. The strength of the
WISE multi-band observations in infrared is their ability to determine the size of the
targets. Albedos can then be inferred, provided the absolute magnitudes are the targets are 
well characterized. In what these authors called a preliminary debiasing effort, \citet{grav2011} found
$f_{45}=1.4\pm 0.2$ for $D>10$~km Jupiter Trojans.

As a preamble of our work, we present a highly simplified guess of the asymmetry parameter
$f_{45}$ from today's data based on the following reasoning (we shall largely substantiate this
result in Sec.~\ref{res}). We find it plausible that the photometric completion of the Trojan
clouds as for today is close to $H_{\rm c}\simeq 13.8$ magnitude (this is near the $10$~km limit
considered by \cite{grav2011} assuming a mean geometric albedo of $p_V\simeq 0.075$). This value is
consistent with the modestly lower completion limit of Catalina Sky Survey (CSS) observations
between 2013 and 2022; we will analyze these data below (Sec.~\ref{res}). It also matches
the results from \citet{hm2020}. The Minor Planet Center
(MPC) orbital database contains $N_4\simeq 2040$ and $N_5\simeq 1410$ Trojans with $H\leq H_{\rm c}$
after subtracting populations of the major families identified in Appendix~A (Table~\ref{mb_fams_2023}).
Adopting a naive $\simeq \sqrt{N}$ population uncertainty, which is probably an underestimate, we would
estimate that $f_{45}=1.45$ with an uncertainty $\simeq f_{45}\,\left[(1/N_4)+(1/N_5)\right]^{1/2}\simeq 0.05$,
thus $f_{45}=1.45\pm 0.05$ for $H\leq H_{\rm c}$. This result is close to the value reported
in \citet{grav2011}.

Another aspect of the L4 and L5 swarm asymmetry, often discussed by the previous studies, concerns
the orbital inclination distribution. For instance, \citet{jewitt2000} analyzed data from a week-long campaign
using the University of Hawaii 2.2~m telescope to observe a sector of the L4 swarm and detected
93 Trojans. Assuming a simple Gaussian sky-plane density distribution function, and approximate
relation of the sky-plane motion of the detected Trojan to the inclination of its heliocentric
orbit, these authors determined the biased-corrected inclination distribution of the L4 Trojans in the $0^\circ$
to $30^\circ$ range. They observed a bi-modality with maxima at $\simeq 9^\circ$ and $\simeq 19^\circ$
degrees. Earlier observations of the L5 Trojans did not show this structure \citep[see already][]{dvh1979}. Instead, the
L5 swarm inclinations were recognized to extend to larger values with a maximum near $\simeq 27^\circ$.
\citet{sb2014} sum up the inclination data, working with the observed biased populations,
and refer to the above-mentioned differences as to an unsolved problem. We note that this 
inclination difference was also discussed and analyzed in \citet{pirani2019a}. 

These reported
anomalies can be explained by the presence of the Trojan families, mainly Eurybates, Arkesilaos
and Hektor in L4 and the Ennomos/Deiphobus clan in L5 (see, e.g., Figs.~\ref{fig7} and \ref{fig_prop_mb},
or data in the Table~\ref{mb_fams_2023}). As discussed below, the families are strongly concentrated
in inclination and the principal clans may contain a substantial number of members. In this
sense, the inclination problem is also closely related to the population problem mentioned above. An objective
comparison of the L4 and L5 populations must subtract the families contribution before performing any 
analysis. Nevertheless, the issues with Trojan inclinations lead to more issues to consider,
namely the number and properties of the Trojan families in the L4 and L5 populations. There
seems to be differences between the two swarms that warrant an explanation.

Accordingly, understanding the differences in the
Jupiter Trojan L4 and L5 populations requires analysis of {\it both} the
orbital and magnitude (or size) frequency distributions. 
In Sec.~\ref{datag96} we review the presently known Jupiter Trojan populations and
discuss what has been detected by CSS. The latter will form the basis of our analysis.
The reason is that most CCS data were carefully characterized, which allows us to calculate a 
bias-corrected model. Some of the complex bias features are illustrated in
Sec.~\ref{datag96}, with further details provided in the Appendix~B. In Sec.~\ref{model} we introduce
our model for the
orbital architecture of the Jupiter Trojans, as well as the assumption about their magnitude
distribution. In both cases, we distinguish two components: (i) the background continuous population, and
(ii) Trojans in the discrete families. As to the latter, we consider eight prominent families
(five in the L4 swarm, and three in the L5 swarm), all newly identified in the Appendix~A. In
Sec.~\ref{complete} we describe the complete (bias-corrected)
Trojan populations from the bias-affected observations. This will include:
(i) the definition of the detection probability as well as how it maps onto our model parameters
(Sec.~\ref{secbias}), and (ii) numerical tools that can find the model that best-matches the available
observations (Sec.~\ref{opti}). Our results are summarized in Sec.~\ref{res}, with our conclusions 
discussed in Sec.~\ref{concl}. As a part of our project, we determine a new catalog
of Jupiter Trojan proper orbital elements and identify an updated set of Trojan families, both which are 
provided in Appendix~A.
  
\section{The observed Trojan population}\label{datag96}
In this section, we will discuss the currently known population of Jupiter Trojans.
Next, we will determine what fraction has been detected by CSS, and briefly discuss the most
important selection effects.
\smallskip

\noindent{\it Population of known Jupiter Trojans.-- }We determined the currently known
population of Jupiter Trojans using a two step scheme. First we downloaded the orbital
catalog {\tt MPCORB.DAT} from the Minor Planet Center website (the February 22, 2023 version).
Next we selected all inputs using a simple criteria in osculating orbital elements
which were available for a common epoch of MJD 60,000.  The initial database represented
about 1,245,000 orbits, but about 2\% were rejected because they reflected poorly-determined 
single-opposition orbits given at different, mostly earlier epochs. 
The orbital selection was: (i) semimajor axis $a$ in
between $4.95$~au and $5.45$~au, (ii) eccentricity smaller than $0.6$, (iii)
inclination smaller than $50^\circ$, and (iv) absolute value of the
resonant angle $\sigma$ larger than $20^\circ$ and smaller than $130^\circ$. This search resulted
in 7,929 orbits around the L4 libration center and 4,189 orbits around the L5 libration center.
A small number of these orbits, however, correspond to unstable Jupiter family 
comets masquerading as JTs. This led us to numerically propagate all selected orbits, together with all
planets and the Sun, for a 1~Myr timespan. We found that 26 and 12 candidate Trojans
near L4 and L5, respectively, escaped from the libration region about their Lagrange center, 
were discarded from the simulation by striking Jupiter, or
were ejected from the Solar system. The remaining
7,903 bodies near L4 and 4,177 bodies near L5 were found to reside in the resonant zone
of their respective center, with their $\sigma$ librating for the entire timespan of
the integration. Note that additional losses might occur if the integration was extended, 
but we believe the selected population is sufficient for our needs.
\begin{figure}[t!]
 \begin{center}
  \includegraphics[width=0.47\textwidth]{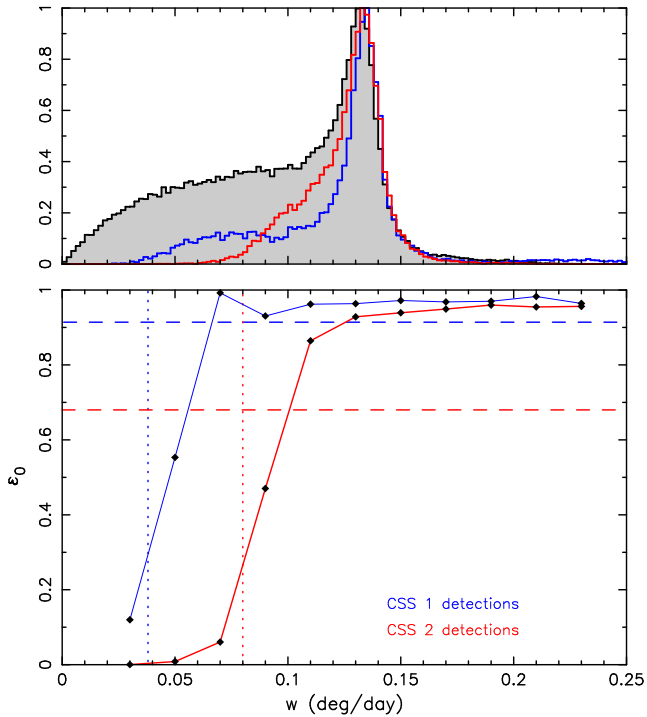}
 \end{center}  
 \caption{Upper panel: Intrinsic distribution of the apparent motion of Jupiter Trojans
  in the fields-of-view of CSS phase~II (determined from a modelled synthetic population
  of objects) is shown by the black histogram with gray shading. The color-coded histograms are
  the apparent motion distributions of the detected Trojans (red for L4 and blue for 
  L5 Trojans); all distributions normalized to maximum. Detected populations miss the
  tail of very slowly moving objects (see below).
  Lower panel: Mean value of the bright-end detection probability $\epsilon_0$ of the CSS
  observations during the phase~I (red) and phase~II (blue) as a function of the apparent
  motion $w$ on the CCD detector of G96 (abscissa in deg~day$^{-1}$). The horizontal dashed
  lines are global average values of $\epsilon_0$ over all $w$ values, the vertical dotted lines
  denote a critical $w$ value for which $\epsilon_0$ drops to $\simeq 0.25$. The drop in the detection
  efficiency below $\simeq 0.04$ deg~day$^{-1}$ (phase~I) and $\simeq 0.08$ deg~day$^{-1}$
  (phase~II) is related to a finite pixel size of the detector, preventing the identification
  of a clear tracklet for a slowly moving Trojan image.}
 \label{fig00}
\end{figure}
\begin{figure*}[t!]
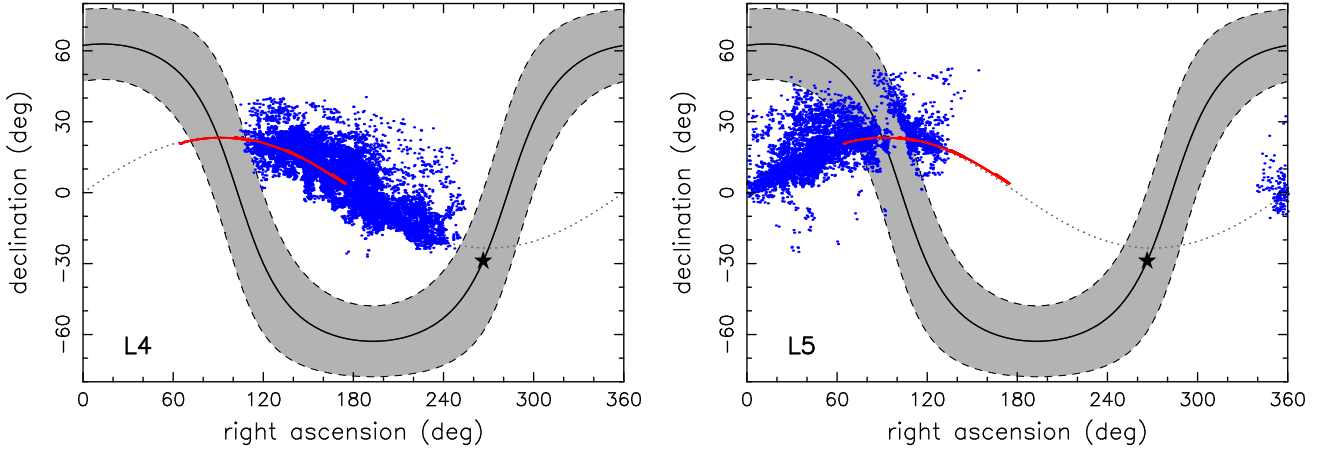

 \begin{center}
 \begin{tabular}{cc}
  \includegraphics[width=0.47\textwidth]{f2a.eps} &
  \includegraphics[width=0.47\textwidth]{f2b.eps} \\ 
 \end{tabular}
 \end{center}  
 \caption{Jupiter Trojan detections by the CSS operations in between Jan~2, 2013 and
  May~14, 2016 (phase~I). There are 14,303 detections of 4,551 individual bodies librating
  about the L4 center (left panel) and 7,230 detections of 2,460 individual bodies librating
  about the L5 center (right panel). The blue symbols are topocentric right ascension
  (abscissa) and declination (ordinate) of the observations. The red line is the Jupiter
  track in the same period of time near the ecliptic plane (dotted line). The solid black line
  is the projection of the galactic plane surrounded by the $\pm 15^\circ$ latitude strip
  (gray region). The black star is the direction to the galactic center.}
 \label{fig0}
\end{figure*}
\begin{figure*}[t!]
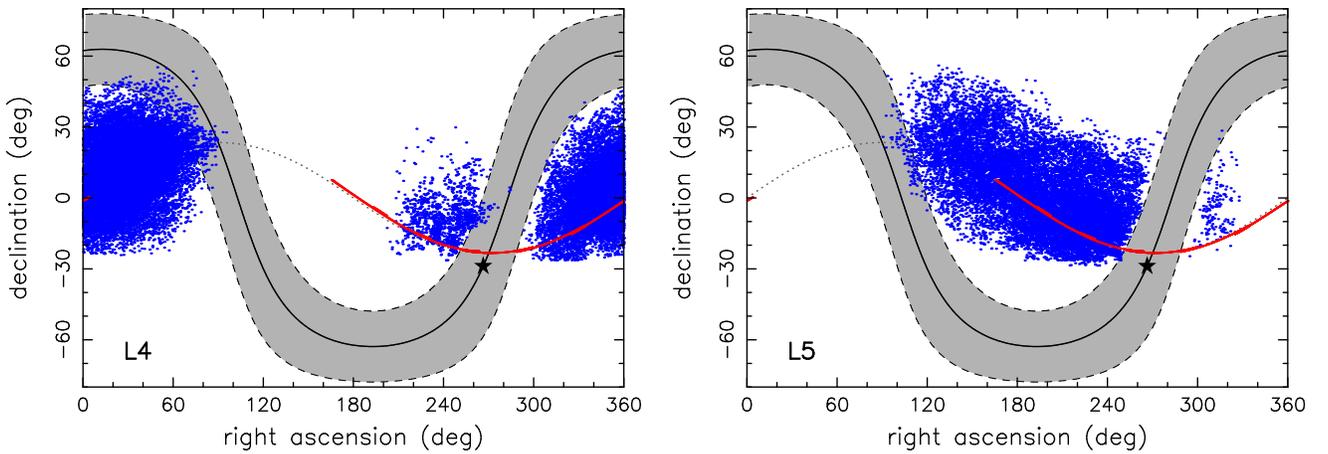

 \begin{center}
 \begin{tabular}{cc}
  \includegraphics[width=0.47\textwidth]{f3a.eps} &
  \includegraphics[width=0.47\textwidth]{f3b.eps} \\ 
 \end{tabular}
 \end{center}  
 \caption{Jupiter Trojan detections by the CSS operations in between May~31, 2016 and
  June~14, 2022 (phase~II). There are 40,832 detections of 6,307 individual bodies librating
  about the L4 center (left panel) and 18,293 detections of 3,041 individual bodies librating
  about the L5 center (right panel). The blue symbols are topocentric right ascension
  (abscissa) and declination (ordinate) of the observations. The red line is the Jupiter
  track in the same period of time near the ecliptic plane (dotted line). The solid black line
  is the projection of the galactic plane surrounded by the $\pm 15^\circ$ latitude strip
  (gray region). The black star is the direction to the galactic center. The observations primarily
  avoid pointing toward crowded stellar fields near the galactic center.}
 \label{fig1}
\end{figure*}
\begin{figure*}[t!]
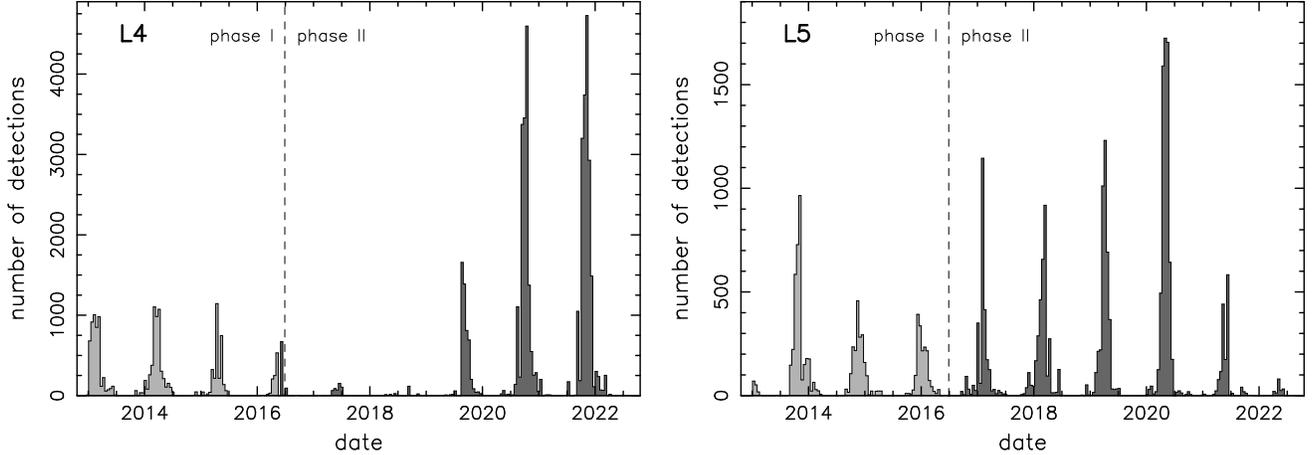

 \begin{center}
 \begin{tabular}{cc}
  \includegraphics[width=0.47\textwidth]{f4a.eps} &
  \includegraphics[width=0.47\textwidth]{f4b.eps} \\ 
 \end{tabular}
 \end{center} 
 \caption{Number of Trojan detections throughout the CSS operations in phases I and II: left for 
  L4 cloud, right for L5 cloud. Apart from the obvious annual variation, due to the Earth revolution
  about the Sun changing the viewing geometry of the Trojan clouds at night sky, there is also
  a longer term variation due to the Jupiter motion. This dictates when a particular cloud, L4 or L5,
  appears projected in the direction of the galactic plane. For instance the major minimum of L4
  objects detection around 2018 is when the cloud overlaps with the galactic center direction.
  Finally, data during the phase II operations of CSS are roughly four times more numerous than
  during the phase I operations, due to four times larger field of view.}
 \label{fig2}
\end{figure*}
\smallskip

\noindent{\it Catalina Sky Survey observations of Jupiter Trojans.-- }We now use the Trojans
identified in the MPC database to determine which of them were detected by Catalina Sky Survey.%
\footnote{\url{https://catalina.lpl.arizona.edu/}}
CSS, managed by Lunar and Planetary Laboratory of the University of Arizona, is one of the two most prolific
survey programs that have been in action over the past decade \citep[e.g.,][]{CSSEPSC2019}. Its principal
mission is
to discover and further track the near-Earth objects, aiming at characterization of a 
significant fraction of the population with sizes as small as $140$~m. However, CSS is an
invaluable source of data for other Solar System studies as well \citep[e.g.,][leaving aside numerous
applications of CSS data in studies of variable stars and optical transients]{yf2023}, and this includes also
the Jupiter Trojan population.

In this paper, we use observations of the CSS 1.5 m survey telescope located at Mt. Lemmon
(MPC observatory code G96). \citet{nes2023} carefully evaluated
the asteroid detection probability for the G96 operations in the period between January 2013
and June 2022. Additionally, this interval has been divided in two phases: (i) observations before
May~14, 2016 (phase~I), and (ii) observations after May~31, 2016 (phase~II). Not only is the second
phase longer, allowing the survey to collect more observations (the available database contains
$61,585$ well characterized frames --sequences of four, typically $30$~s exposure images-- during
the phase~I and $162,280$ well characterized frames during the phase~II), but the primary difference
stems from an important upgrade of the CCD camera in the second half of May~2016. The new camera
has four times larger field of view, and even slightly better photometric sensitivity, allowing thus
to cover a much larger latitude region about the ecliptic (its only drawback is the large pixel
size as discussed below). Investigating this database, we found the following.  In phase~I, there 
were (i) 14,303 detections of 4,551 individual bodies
librating about the L4 center and 7,230 detections of 2,460 individual bodies librating
about the L5 center. In phase~II, there were 40,832 detections of 6,307 individual
bodies librating about the L4 center and 18,293 detections of 3,041 individual bodies
librating about the L5 center. As a result, the CSS operations
during the 2013 - 2022 period were powerful enough to detect 75-80\%
of the known Trojans. The remaining 20-25\% cases that escaped CSS attention were discovered and/or
detected by other surveys (such as Pan-STARRS), or were detected by CSS only after
June~14, 2022 (the last date for which we have characterized the survey's performance).
\smallskip

\noindent{\it Selection effects that have an influence on CSS observations.-- }The
completeness of the population data in the CCS observations is affected by two broad categories
of effects: (i) the {\it geometric bias}, expressing completeness of the CSS 
fields of view coverage of the sky zone onto which Trojans project, and (ii) the
{\it photometric bias}, constraining our ability to detect objects
in the Trojan population due to telescope/detector limitations. The photometric bias
itself represents a set of complicated phenomena, whose origin is twofold: (a)
the telescope and detector ability to record signal up to a certain limit, beyond which
the observed Trojan is too faint and therefore undetected, and (b) the ability to identify
the Trojan as a moving object on a stellar background. We use a detailed and
properly calibrated formulation of the CSS detection efficiency developed by \citet{nes2023}
within a project to create a new model of the near-Earth object (NEOs) population.
While the faintness problem is the same, the critical issues of the identification problem for 
NEOs and Trojans
are just the opposite. Denoting the apparent motion of the observed object on the CCD camera 
by $w$, a delicate issue to face in NEO detection is the limit of its large values ($w\geq 5$
deg~day$^{-1}$, say). Yet, this case is important because the smallest NEOs may only be 
detected during their close encounters with the Earth when the apparent motion is high.
In Trojans, the opposite limit of very small $w$ values may be a source of a trouble.
This is because when $w$ is too small, $w\leq 0.04$ deg~day$^{-1}$ in the phase~I and
$w\leq 0.08$ deg~day$^{-1}$ in the phase~II (Fig.~\ref{fig00}), the record of the Trojan does not allow 
one to construct a recognizable tracklet. The identification may even fail for very bright
objects. We therefore extended the analysis of \citet{nes2023}
to allow reliable determination of the detection efficiency in the very small limit
of $w$. 

As an example of our analysis, we show in Fig.~\ref{fig00} how the bright-object 
limit of the detection probability $\epsilon_0$ depends on $w$ in both phases of the
CSS operations. Interestingly, phase~I may reach Trojans down to smaller $w$ values
than in phase~II, which is otherwise superior in many respects. This is because the larger 
pixel size of the new camera installed in May 2016 requires faster moving Trojans to spread their
trail on the detector over a sufficient number of different pixels, defining thus a characteristic
tracklet.

Next, we briefly describe the various geometric and additional photometric aspects of the bias 
in Trojan observations (overall, one may argue that observations of planetary Trojans, 
including those of Jupiter, could serve as a textbook example of a plethora of bias phenomena).
An insight into this component of the bias is provided by Figs.~\ref{fig0}
to \ref{fig2}. There are two basic periods modulating this part of the bias,
namely (i) the annual effect (in fact $\simeq 1.09$~yr effect of Jupiter's synodic
period with respect to the Earth), simply dictating when the respective cloud appears at
opposition on night sky, and (ii) a $\simeq 11.9$~yr cycle of Jupiter's heliocentric
revolution. This latter has an influence of position of libration centers L4 and L5
on the sky in absolute terms. Locations near the galactic plane are observationally
prohibited and could cause absence of the data (see, e.g., small number of L4 Trojan
detection from 2017 to 2019 on Fig.~\ref{fig2}). As a result,
a survey lasting less than $\simeq 11.9$~yr may suffer from highly
unbalanced observational conditions for the two clouds. The CSS data available to us
span $9.5$~yr. While not ideal, the geometric limitations are fortunately not
severe, and neither L4 nor L5 Trojans are under-represented in the observations.
\begin{figure*}[t!]
 \plottwo{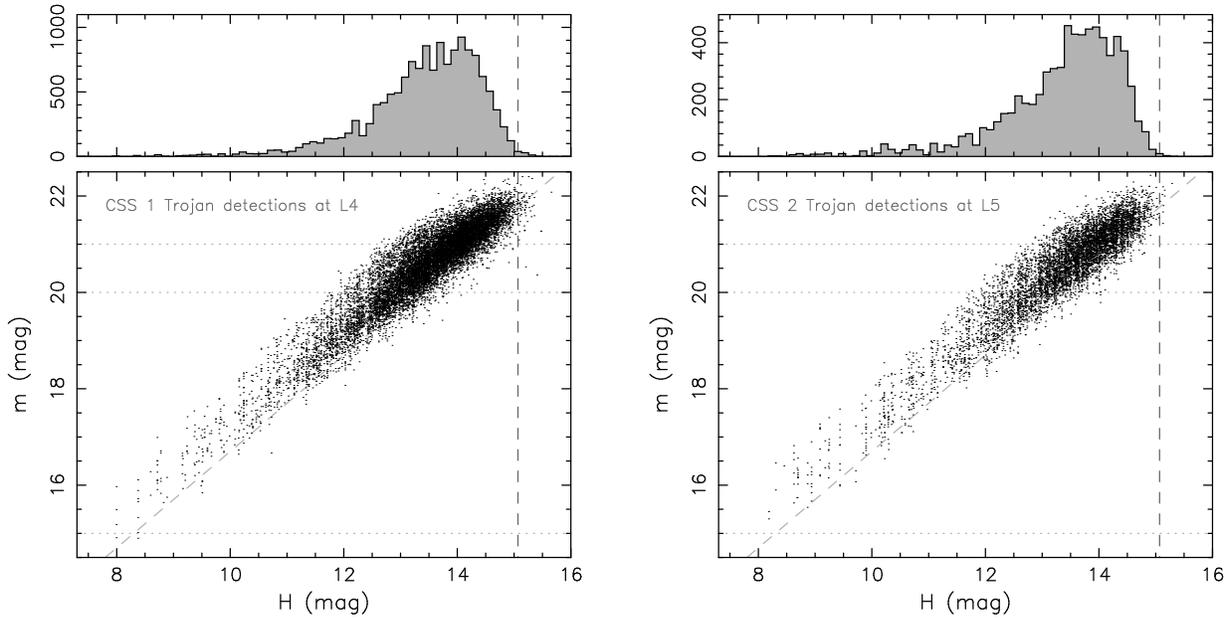}{f5b.eps}
 \caption{Upper panel: Histogram of the absolute magnitude $H$ values for detections of L4 (left)
  and L5 (right) Trojans during the CSS phase I operations (the ordinate is number of such detections
  withing the bin). The vertical dashed line shows magnitude at the center of a bin in which number
  of detections dropped below $5$\% of the bin with maximum number of detections. This value is
  about $\simeq 15.1$ for both L4 and L5 clouds.
  Bottom panel: Correlation between the absolute magnitude $H$ at the abscissa and visual magnitude $m$ at
  the ordinate for detections of L4 (left) and L5 (right) Trojans during the CSS phase I operations.
  At an ideal opposition, $m-H\simeq 6.7$ (neglecting the phase function correction) from simple distance
  arguments (the dashed line). The nearly two magnitude spread toward larger $m$ values are detections
  at different phase angles. The dotted lines in between visual magnitudes 20 and 21 indicate
  typical 50\% photometric efficiency of detections. Object brighter than $m\simeq 15$ (lower dotted line)
  may saturate exposures and be undetected due to their slow motion. Luckily, there are detection loses
  due to this effect for Trojan populations, because even the largest of them at opposition barely
  reach this limit. The only photometric limit, indicated by the vertical dashed line from the upper
  panel, thus occurs at the faint-end of the spectrum.}
 \label{fig3a}
\end{figure*}
\begin{figure*}[t!]
 \plottwo{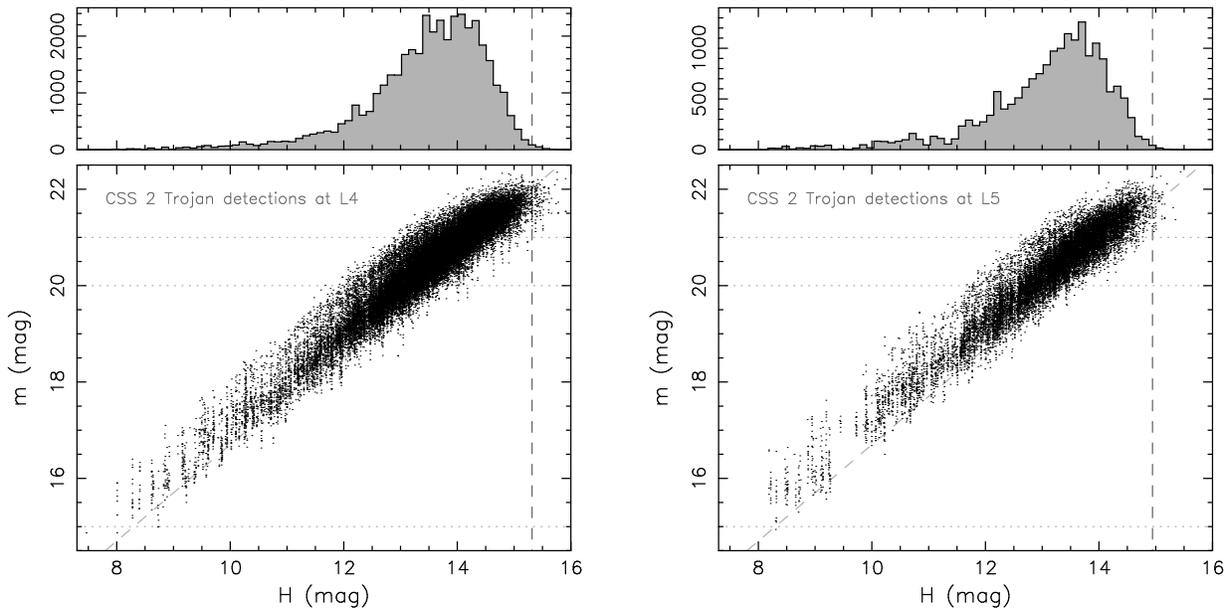}{f6b.eps}
 \caption{The same as is Fig.~\ref{fig3a}, but now for the CSS phase II operations containing roughly
  three times more detections. The vertical dashed lines are now different for the two Trojan clouds:
  $\simeq 15.35$ for L4 and $\simeq 14.95$ for L5.}
 \label{fig3b}
\end{figure*}
\begin{figure*}[t!]
 \plottwo{f7a.eps}{f7b.eps}
 \caption{Cartesian representation of $\zeta-\zeta_{\rm J}$ for 7,903 L4 Trojans (left) and
  4,177 L5 Trojans (right) from the MPC catalog as of February 2023 (using osculating elements
  at MJD 60,000). The red symbols show the major families in the respective Trojan clouds:
  Eurybates at L4, and Ennomos/Deiphobus at L5 using their identifications in the Appendix~A. At the zero
  level, the population distribution is independent of the polar angle $\varphi$ in these
  coordinates, depending only on the radial coordinate $A$ (see Eq.~\ref{properi}).}
 \label{fig4}
\end{figure*}
\begin{figure*}[t!]
 \plottwo{f8a.eps}{f8b.eps}
 \caption{Cartesian representation of $z-e_{\rm J}\exp(\pm\imath \pi/3)$ for 7,903 L4 Trojans (left) and
  4,177 L5 Trojans (right) from the MPC catalog as of February 2023 (using osculating elements
  at MJD 60,000). The red symbols show the major families in the respective Trojan clouds:
  Eurybates at L4, and Ennomos/Deiphobus at L5 using their identifications in the Appendix~A. At the zero
  level, the population distribution is independent of the polar angle $\psi$ in these coordinates,
  depending only on the radial coordinate $B$ (see Eq.~\ref{propere}).}
 \label{fig5}
\end{figure*}

The second bias component is photometric (see Figs.~\ref{fig3a} and \ref{fig3b}).
In fact, limitations of detectability may affect both bright and faint objects.
This is, for instance, the case for main belt observations by CSS, which did
not detect the brightest bodies (e.g. Ceres or Juno). Luckily, this problem does
not appear to affect the Trojan data, for which even the largest population members are faint
enough that saturation and confusion with stationary objects does not occur. Obviously,
a given telescope/detector configuration will always suffer from limitations at the
faint end. 

For CSS operations, the characteristic 50\% detection efficiency ranges
between 20 and 21 apparent visual magnitude (depending on the night conditions).
Neglecting the phase function correction, we may roughly estimate the corresponding absolute
magnitude limit using Pogson's relation $m-H\simeq 5\log(r\Delta)$, where $r\simeq 5.2$~au
and $\Delta \simeq 4.2$~au are heliocentric and geocentric distances at opposition.
This provides $m-H\simeq 6.7$ shown by the dashed line in Figs.~\ref{fig3a} and \ref{fig3b}.
Trojans having magnitudes up to $H\simeq 14$ should thus be reached by CSS with a still fairly
well characterized detection efficiency, but this may also apply to bodies that are magnitude
fainter. The rough estimate of Trojan population completeness near this $H$ value is also
supported by the fact that no new object with $H\leq 14.2-14.3$ was discovered within the past
five years \citep[see][who give $H\simeq 13.8-13.9$ as a completeness limit for Jupiter
Trojans]{hm2020}. 

A comparison of the two panels in Fig.~\ref{fig3b} indicates and interesting
(and important) difference between G96 observations of the L4 vs L5 Trojans: the L4 detections
sample the population about $\simeq 0.35$ magnitude deeper in terms of absolute magnitude
compared to the L5 detections (see the dashed vertical lines). This is a result of yet
another bias. During phase~II of the CSS operations, Jupiter moved on its elliptic
orbit between the mean anomaly $\simeq 160^\circ$ at the beginning to $\simeq 310^\circ$
at the end. On average, the L4 Trojans were thus at smaller heliocentric and geocentric
distances. Given the constant detection limit of G96 in terms of the apparent magnitude,
the survey was able to reach slightly smaller L4 Trojans compared to L5 Trojans.

Some more illustrations of the selection effects in the CSS observations are provided in
the Appendix~B, where we also conduct a test justifying our simplifying
assumptions about the detection probability function. The fundamental issue of how to map
the above-mentioned selection effects and the detection probability on the CSS images to 
the detection probability in the parameter space of the observed population of Jupiter Trojans
is postponed to Sec.~\ref{secbias}.

\section{Orbital architecture and magnitude distribution of the Trojan population}\label{model}

\subsection{Simple parametrization for model fitting}

Several previous studies have attempted to debias the Trojan observations, with the 
goal being to determine the complete population down to some size or absolute magnitude. 
The orbital distribution
properties of the Trojans were most often characterized with observer-related parameters,
namely by the distribution
of the heliocentric ecliptic longitude and latitude relative to Jupiter or the corresponding
libration center. Typically a simple double-Gaussian distributions were adopted
\citep[e.g.,][ and others]{szabo2007, wb2015}. Some studies, see \citet{jewitt2000},
attempted to go beyond this approach by introducing a very simple approximation
of the orbital inclination distribution.
\begin{figure*}[t!]
 \plottwo{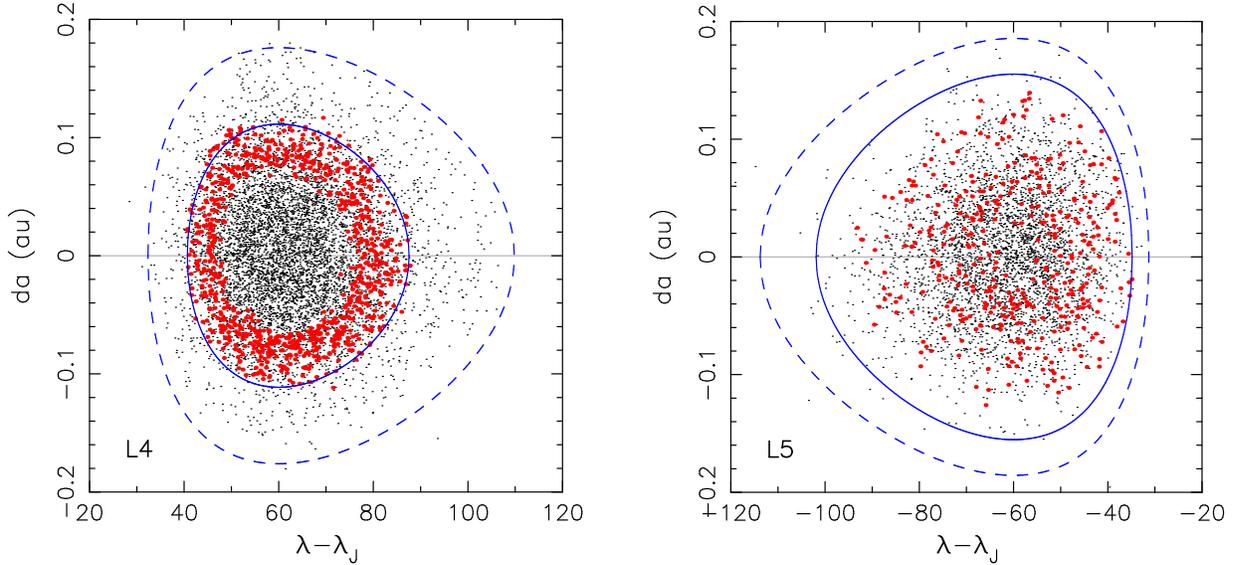}{f9b.eps}
 \caption{Projection of the MPC Trojan clouds as of February 2023 onto the $\lambda-\lambda_{\rm J}$
  (abscissa) and $da=a-a_{\rm J}$ (ordinate) plane (using osculating elements at MJD 60,000, and
  $a_{\rm J}=5.204$~au). The left panel for L4, the right panel for L5. Red symbols among L4 population
  highlight the Eurybates family members, those among the L5 population show the more dispersed
  Ennomos/Deiphobus family members. The blue lines are the isolines of the first integral of
  the Trojan motion Eq.~(\ref{propera}) for selected values of $C$ parameter: (i) $C=1.68$ (solid) and
  $C=1.95$ (dashed) in L4 panel, and (ii) $C=1.88$ (solid) and $C=2$ (dashed) in L5 panel. In spite of
  it huge simplicity, representation using Eq.~(\ref{propera}) helps to parameterize the population for
  our purposes.}
 \label{fig6}
\end{figure*}

Here we intend to describe orbital architecture of the Trojan population in greater detail
than previous attempts, namely with the most relevant orbital parameters. This description 
will be more complete, and we will also attempt to discern the
background population from the populations in major families (such as Eurybates
or Ennomos). This removal is needed to describe the putative asymmetry in the L4 and L5 populations.

\subsubsection{Orbital distribution}
Orbital elements of a different level of sophistication were developed by both analytical and
numerical methods. Here, we do not need to work with precise parameters whose
stability will last over very long timescales. Rather, we need parameters that are
basically osculating at an epoch close to the decade of CSS observations (the close link
to the osculating elements will make them easy to implement in the debiasing methods;
Sec.~\ref{secbias}), and which capture the basic features of Jupiter's perturbations. 
With that in mind, here is our choice (we note that \citet{vin2015} and \citet{vin2019}
used variables $e_{\rm p}$ and $I_{\rm p}$ similar to ours $A$ and $B$ to search for
principal Trojan families).
\smallskip

\noindent{\it Orbital inclination.-- }We link the heliocentric orbital inclination $I$ and longitude
of node $\Omega$ together into a complex variable $\zeta$, such that ($\imath=\sqrt{-1}$)
\begin{equation}
 \zeta=\sin I \exp(\imath\Omega)=\zeta_{\rm J}+A\exp(\imath \varphi)\; , \label{properi}
\end{equation}
where $\zeta_{\rm J}$ is the same variable for Jupiter orbit. The second term on the right hand side
of (\ref{properi}) describes the orbit free of the main Jupiter perturbing effect; namely, the
amplitude $A$ crudely represents what would be the proper sine of inclination.
\smallskip

\noindent{\it Orbital eccentricity.-- }Next, we link the heliocentric eccentricity $e$ and
longitude of perihelion $\varpi$ together into a complex variable $z$, such that
\begin{equation}
 z=e\exp(\imath(\varpi-\varpi_{\rm J}))=e_{\rm J}\exp(\pm\imath \pi/3)+ B\exp(\imath \psi)\; ,
  \label{propere}
\end{equation}
where $e_{\rm J}$ and $\varpi_{\rm J}$ are the respective variables for Jupiter orbit. As above, the
amplitude $B$ of the second term on the right hand side here crudely represents what would be
the proper eccentricity \citep[see, e.g.,][]{bs1987,li2021}.
\smallskip
\begin{figure*}[t!]
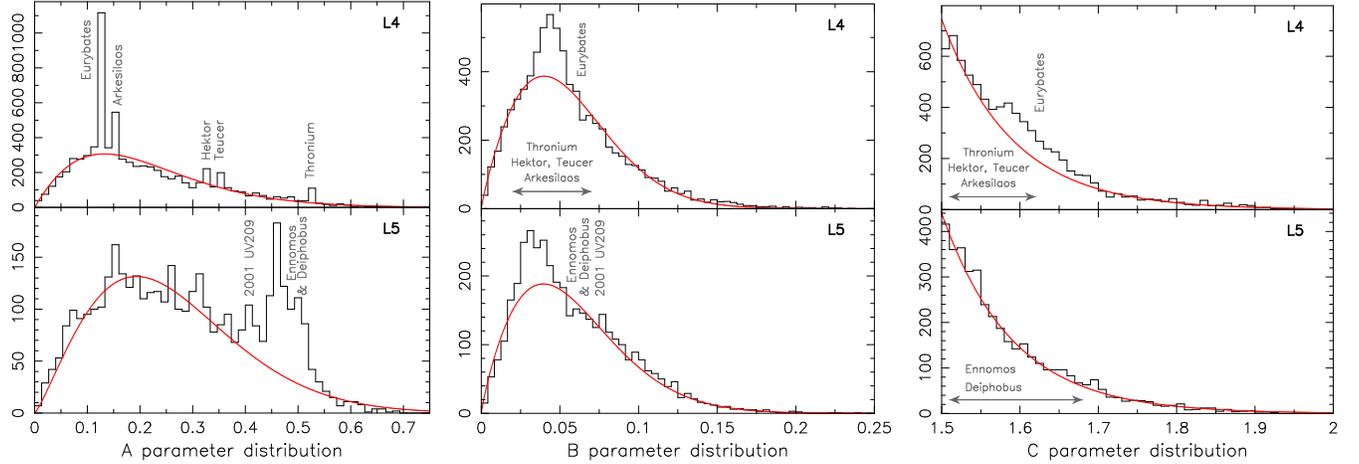

 \begin{center}
 \begin{tabular}{ccc}
  \includegraphics[width=0.31\textwidth]{f10a.eps} &
  \includegraphics[width=0.32\textwidth]{f10b.eps} &   
  \includegraphics[width=0.31\textwidth]{f10c.eps} \\
 \end{tabular}
 \end{center}
 \caption{The observed population of Jupiter Trojans in the MPC catalog as of February
  2023 represented using the distribution of the $A$, $B$ and $C$ orbital parameters: L4
  cloud on top, L5 cloud at bottom. As to their definition see Eqs.~(\ref{properi}), (\ref{propere}) and
  (\ref{propera}). The red lines are approximate fits of the background population
  using the simple few-parametric functions in
  Eqs.~(\ref{adist}), (\ref{bdist}) and (\ref{cdist}), avoiding the contribution of the major families.
  In particular, we used: $\alpha_1=1$, $\alpha_2=1.3$ and $s_{\rm A}=0.095$ for the inclination distribution,
  $\beta_1=0.87$, $\beta_2=1.75$ and $s_{\rm B}=0.04$ for the eccentricity distribution, and
  $\gamma=1$ and $s_{\rm C}=0.045$ in the L4 population, and
  $\alpha_1=1.2$, $\alpha_2=1.6$ and $s_{\rm A}=0.15$ for the inclination distribution,
  $\beta_1=0.85$, $\beta_2=1.6$ and $s_{\rm B}=0.038$ for the eccentricity distribution, and
  $\gamma=1$ and $s_{\rm C}=0.045$ in the L5 population. The most prominent Trojan families
  contribute to the total population in specific values of the proper elements. As usual,
  they are most noticeable in the inclination distribution (left panel; Thronium family was formerly
  called 1996~RJ). The population in Eurybates family accounts
  for more then $10$\% of the observed population in L4.  The Ennomos/Deiphobus clan (see the
  Appendix) may represent a slightly smaller share among the L5 Trojans.}
 \label{fig7}
\end{figure*}

\noindent{\it Orbital semimajor axis and longitude in orbit.-- }\citet{sd2003}, see
already \citet{y1983}, and \citet{m1999,m2001} developed the simplest,
but very useful, representation of the resonant motion of Jupiter's Trojans. Averaging
the perturbing Hamiltonian over periods of Jupiter's heliocentric revolution and shorter,
they further neglected orbital eccentricity and inclination terms and showed that Trojan
libration near L4 and L5 equilibria has an integral of motion
\begin{equation}
 \left(\frac{da}{a_{\rm J}}\right)^2=\frac{8\mu}{3} \left[C-f\left(\sigma\right)\right]\; ,
  \label{propera}
\end{equation}
where
\begin{equation}
 f\left(\sigma\right)=\frac{1+4|\sin \sigma/2|^3}{2|\sin \sigma/2|}\; ,
\end{equation}
with $\sigma=\lambda-\lambda_{\rm J}$, $da=a-a_{\rm J}$ ($a_{\rm J}\simeq 5.206$~au), and
$\mu=m_{\rm J}/(m_\odot + m_{\rm J})$. The value of the integration constant $C$ satisfies
$C\in (C_{\rm min},C_{\rm max})$, where for tadpole orbits $C_{\rm min}=1.5$ and $C_{\rm max}=2.5$.
The Eq.~(\ref{propera}) expresses, at the zero level, the resonant correlation
between the involved orbital elements, namely the semimajor axis $a$ and the longitude
in orbit $\lambda$ (the neglected terms are proportional to $e^2$ and $I^2$). As above,
the parameter $C$ crudely represents the proper semimajor axis $da_{\rm P}$, such that
$da_{\rm P}\simeq a_{\rm J}\sqrt{8\mu\,\Delta C/3}$, where $\Delta C=C-C_{\rm min}$ is the
excess of $C$ over its minimum value $C_{\rm min}=1.5$.
\smallskip

\noindent{\it Assumed background orbital distribution of the Jupiter Trojans.-- }The
currently known population of Jupiter Trojans in L4 and L5 clouds is shown, 
using the above described variables, in Figs.~\ref{fig4}, \ref{fig5} and \ref{fig6}.
Obviously, this is the observationally biased population. In order to characterize the
complete (biased-corrected) population, we need to describe it using a relatively simple functional
form. Its free parameters will be adjusted in the process of (i) turning the complete
synthetic population to the biased synthetic population (applying the CSS biases), and
(ii) comparison of the latter with the observations.
\begin{figure*}[t!]
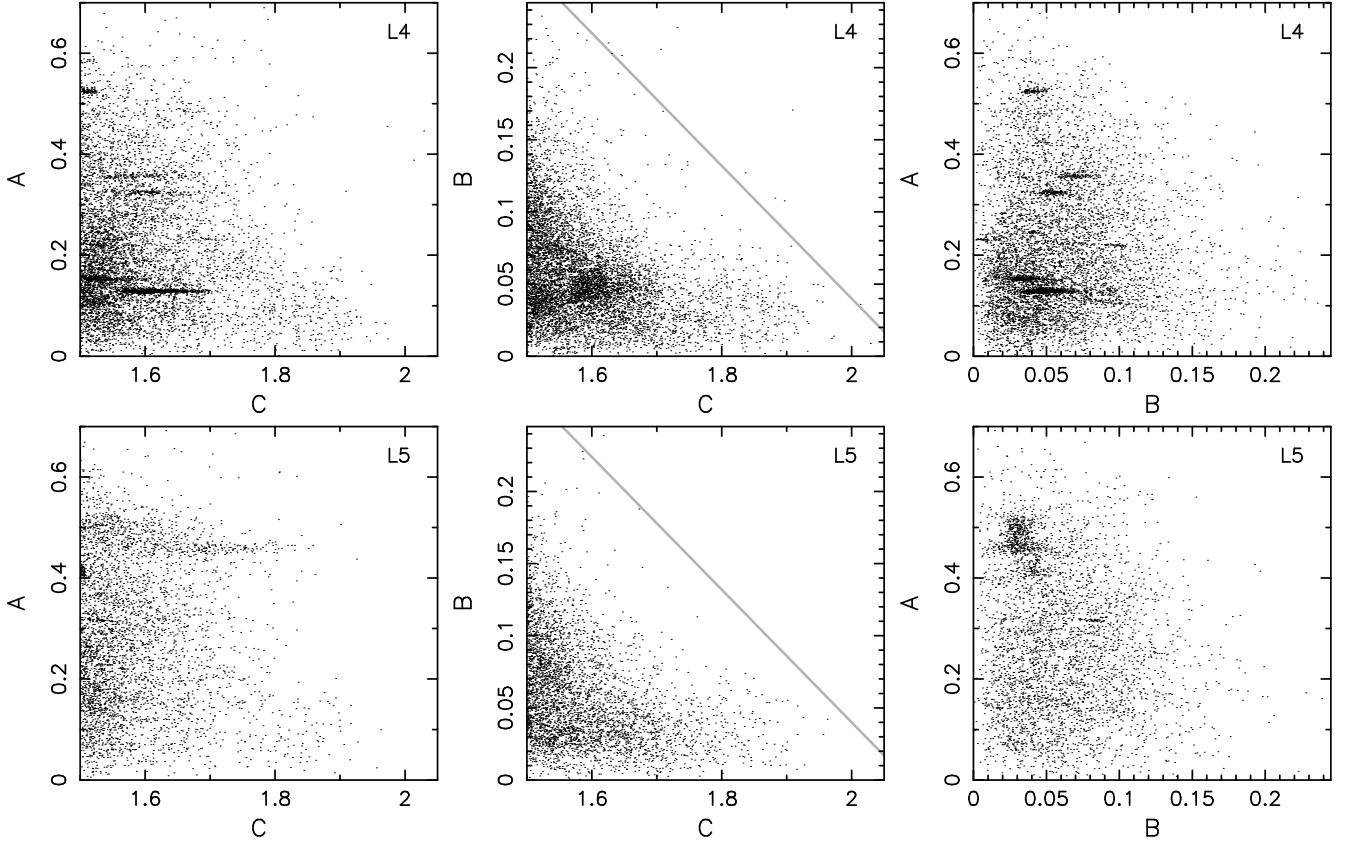

 \begin{center}
 \begin{tabular}{c}
  \includegraphics[width=0.98\textwidth]{f11a.eps} \\
  \includegraphics[width=0.98\textwidth]{f11b.eps} \\  
 \end{tabular}
 \end{center}
 \caption{The observed population of Jupiter Trojans in the MPC catalog as of February
  2023 represented by a 2D projections in the $A$, $B$ and $C$ orbital parameters: (i)
  $A$ vs $C$ (left), (ii) $B$ vs $C$ (middle), and (iii) $A$ vs $B$ (right). Top panels
  for the L4 cloud, bottom panels for the L5 cloud. Two distinct features are worth to
  be noted: (i) correlation between $C$ and $B$ in the middle panels (such that larger $B$
  require smaller $C$), which is the expression of orbital stability constraint, and (ii)
  orbital clusters, most distinctly seen in the $A$ vs $B$ plane in the right panels
  \citep[see also Fig.~9 in][]{vin2015,vin2019}. The gray line in the middle panels approximates
  the stability limit (we improve this guess by a direct numerical experiment;
  Figs.~\ref{fig_stab1a} and \ref{fig_stab1b}). The latter are the Trojan families, collections
  of fragments from the collisional disruption of parent bodies. Curiously,
  the L4 cloud shows a clear evidence of 9 families, starting with the best
  known Eurybates case \citep[e.g.,][]{br2011,retal2016,mar2022}. The families in L5, while
  also existing, seem to be blurred by either higher ejection velocities, dynamical diffusion
  at their particular location or some other effect.}
 \label{fig7bis}
\end{figure*}
\begin{figure*}[t!]
 \plotone{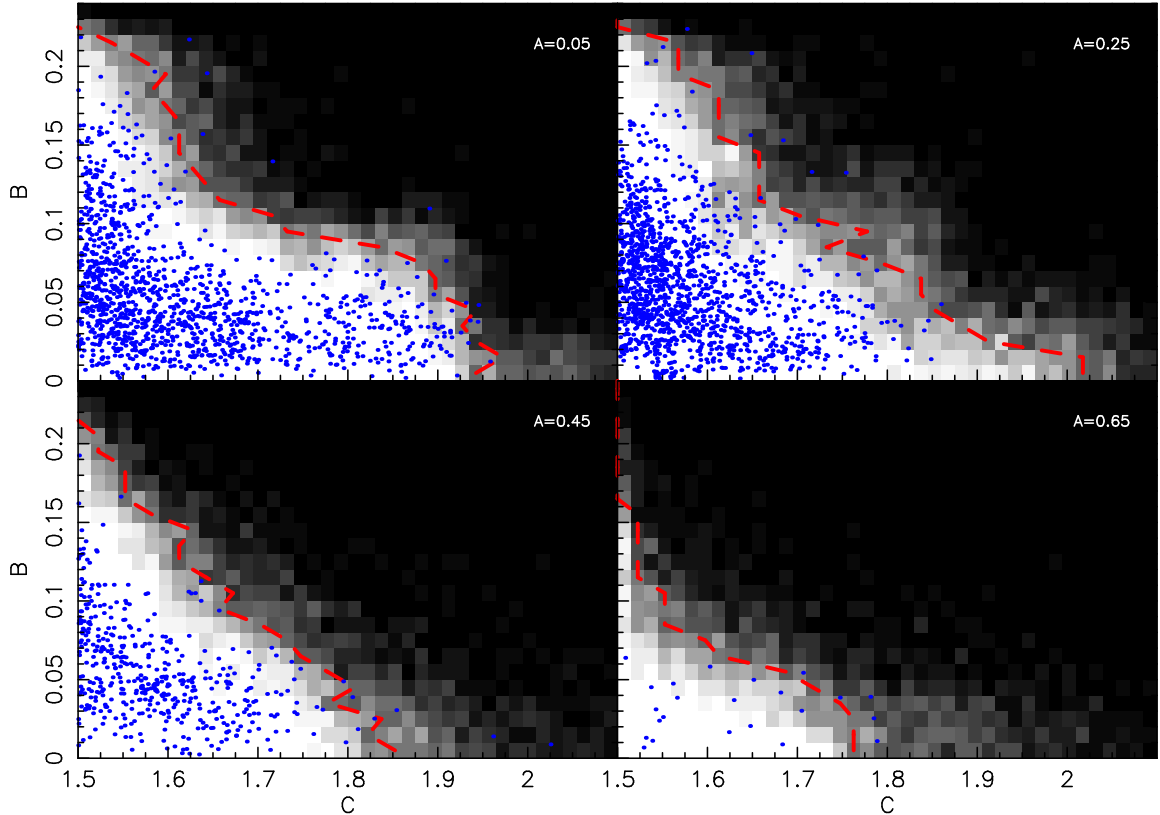} 
 \caption{Results of a numerical experiment probing stability of Trojan orbits in the L4 cloud.
  Orbits of 28 clone variants were started in different bins in the $C$ vs $B$ plane (i.e.,
  librations amplitude vs eccentricity), with four different fixed values of $A$ (i.e., inclination) given
  by the label. Residence in the libration zone about the L4 stationary point was monitored
  throughout the simulation reaching $100$~Myr period of time. All orbits in the white cells are stable,
  all orbits in the black cells are unstable (i.e., escaped from the Trojan region within the
  integration). The gray bins were found partially stable and the scale of grayness is linearly
  proportional to the number of clone orbits escaped. The stability limit is shown by the red dashed line
  connecting the first bins with at $50$\% partial instability for the given $B$ value. The blue
  symbols show the known Trojan population within the $\pm 0.05$ value of $A$ as referred to the
  nominal value.}
 \label{fig_stab1a}
\end{figure*}
\begin{figure*}[t!]
 \plotone{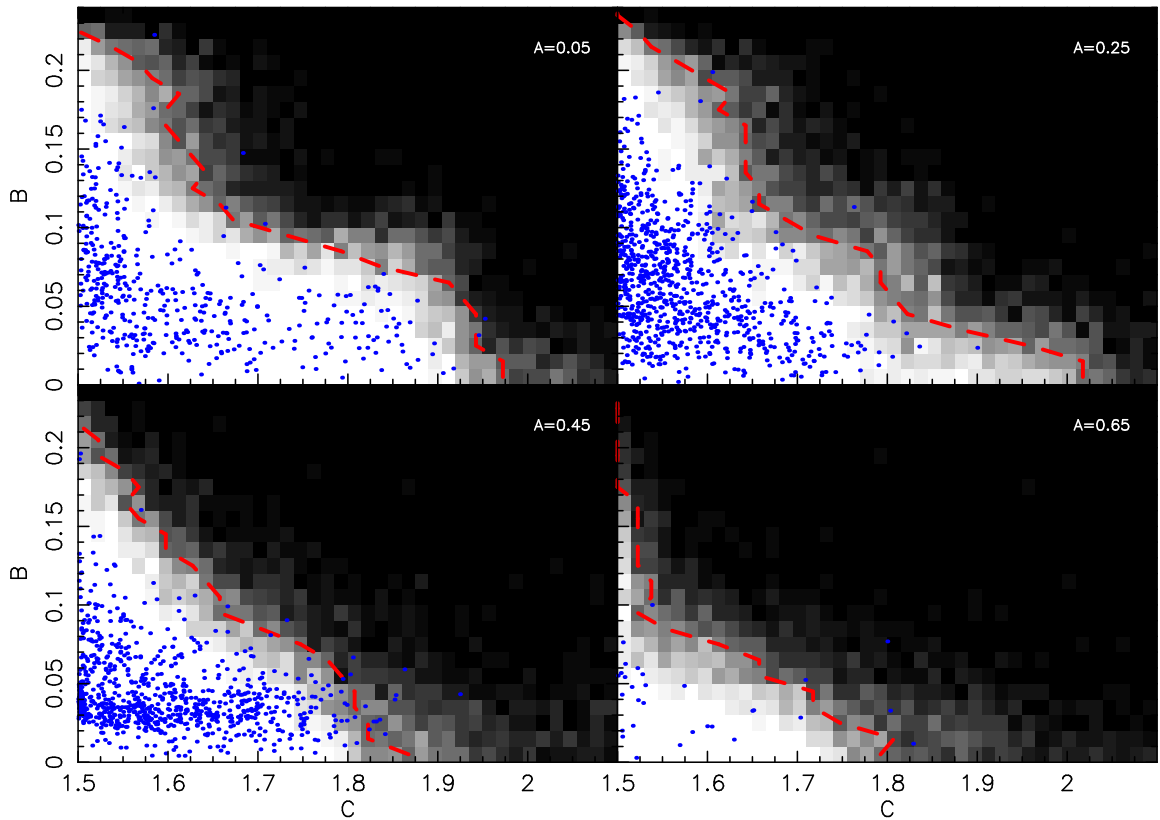}
 \caption{The same is in Fig.~\ref{fig_stab1a}, but now for the L5 Trojan population.}
 \label{fig_stab1b}
\end{figure*}

Let us start with the background population.
A quick look at Figs.~\ref{fig4} and \ref{fig5} reveals that the Trojan distribution density is
basically independent of the phase angles $\varphi$ and $\psi$ defined in Eqs.~(\ref{properi})
and (\ref{propere}). Rather, it only depends on the respective radial coordinates $A$ and $B$.
Similarly, the Trojan distribution density in the $(da,\sigma=\lambda-\lambda_{\rm J})$ plane
(Fig.~\ref{fig6}) basically depends on the $C$ parameter, being homogeneous within a given
$(C,C+dC)$ bin (this is because the principal orbital shear occurs within the bin, and possibly
across neighboring bins only). Therefore, at the zero approximation, we may represent the orbital
architecture of the Trojan clouds at L4 and L5 using distribution function depending on the $A$, $B$ and
$C$ parameters only. Additionally, to keep things simple, we neglect correlations between these parameters
such that the orbital distribution function $dN_{\rm back}(A,B,C)$, specifying number of Trojans in
the bin $(A,A+dA;B,B+dB;C,C+dC)$, may be split into independent contributions
\begin{equation}
 dN_{\rm back}(A,B,C) = dN_{\rm back}(A)\,dN_{\rm back}(B)\,dN_{\rm back}(C)\; .  \label{abcdist}
\end{equation}
The observed, albeit biased, population hints the possible functional form:
\begin{equation}
 \frac{dN_{\rm back}(A)}{dA} \propto A^{\alpha_1}\exp\left[-\frac{1}{2}\left(\frac{A}{s_{\rm A}}\right)^{\alpha_2}\right]
  \; ,  \label{adist}
\end{equation}
\begin{equation}
 \frac{dN_{\rm back}(B)}{dB} \propto B^{\beta_1} \exp\left[-\frac{1}{2}\left(\frac{B}{s_{\rm B}}\right)^{\beta_2}\right]
  \; , \label{bdist}
\end{equation}
and
\begin{equation}
 \frac{dN_{\rm back}(C)}{dC} \propto \exp\left[-\frac{1}{2}\left(\frac{C-C_{\rm min}}{s_{\rm C}}\right)^\gamma\right]
  \; , \label{cdist}
\end{equation}
may hold for the background population (Fig.~\ref{fig7}), over which the contribution of
significant families are linearly superimposed (their distribution is simpler, because we assume
the families uniformly occupy a very limited interval --possibly a single bin-- in $A$, $B$ and $C$
proper elements; see below). The fair match of the biased population justifies the use of
Eqs.~(\ref{adist}), (\ref{bdist}) and (\ref{cdist}) also for the complete (debiased) populations
near L4 and L5. Note, however, that this is a particular, operational choice of our model,
which could be reassessed or improved in the future studies.
Obviously, the parameters (such as $\alpha$, $\beta$, $\gamma$, $s_{\rm A}$, $s_{\rm B}$,
$s_{\rm C}$) are to be considered as solved-for quantities along with the debiasing procedure
(Sec.~\ref{res}). Note that the absolute normalization of the population is
to be set by the absolute magnitude distribution modeling discussed in the next section.

Returning to the Trojan families issue, we note that especially the leftmost panel in Fig.~\ref{fig7}
(showing the inclination distribution) is a very useful tool to judge, in a preliminary manner, their potential
role. For instance, we could straightforwardly estimate the population of the most outstanding
Eurybates family in the L4 swarm to be $\simeq 850$ members (this is the level by which it
surpasses the smooth background in its local $A$-bin). Indeed, a more sophisticated clustering
identification of the Eurybates family in the 3D space of the proper orbital elements (see the
Appendix~A, Table~\ref{mb_fams_2023}) reveals a population of 875 currently known members. This
represents $\simeq 11$\% of the the whole L4 swarm. Adding more than 300 members in Hektor and
Arkesilaos families, the fraction in L4 families increases to $\simeq 15$\%, and our analysis
of magnitude distribution of the background population and individual families below shows
that this fraction increases for fainter Trojans. The families share of the net population
seems to be slightly smaller in the L5 swarm, with Ennomos/Deiphobus representing the 
the only standout clan (see the Appendix~A). Its population reaches about $10$\% of the total.  
\smallskip

\noindent{\it The stability zone and the A-B-C correlations.-- }The above-mentioned simple
solution of the libration about the stationary points L4 or L5 of the 1:1 mean-motion resonance
with Jupiter (Eq.~\ref{propera}) neglects eccentricity and inclination terms. Therefore, our
description misses coupling between the libration amplitude or $C$, eccentricity $B$ and inclination
$A$. One may, however, impose such effects after the fact either from theoretical considerations
or from estimates based on numerical simulation and/or data.

In the first step, we represented the known population of Jupiter Trojans using 2D projections
in the space of these parameters, namely $(C,A)$, $(C,B)$, and $(A,B)$ in Fig.~\ref{fig7bis}.
As expected, and long studied \citep[see already][]{r1965,r1967,lev1997}, the strongest
correlation concerns $C$ vs $B$: larger $B$ (eccentricity) values
require smaller $C$ (libration amplitude) values. That helps to avoid close approaches to Jupiter.
They gray line in the middle panels of Fig.~\ref{fig7bis} is a zero order estimate of the stability
limit. There is a weaker effect in the inclination too, but this element is constrained by
a maximum value of $\simeq 43^\circ$; this is likely related to the perturbing effects of
several secular resonances \citep[such as $s_7$ and $s_8$, or even resonances in which secular
frequencies are combined with that of the great inequality; see][]{mar2003,rg2006}.

In the second step, we characterize the stable zone of Jupiter Trojans in a more sophisticated way, namely
using a direct numerical experiment. This work allows us to determine the portion of
the $(A,B,C)$ parameter space where Trojan orbits are long-term stable.  This result will allow us 
to determine the complete Trojan population discussed in Sec.~\ref{complete}. 

We parsed the $(C,B)$ plane with limiting values shown in the middle panel of Fig.~\ref{fig7bis} into
rectangular bins with width $\Delta B=0.01$ and $\Delta C=0.015$, altogether $1040$ such cells. We
considered $28$ variant orbits starting in each of the cells, assigning them $7$ distinct, but fixed,
values of the $A$ parameter (corresponding to the orbital inclination). In order to set up initial
orbital elements, we complemented $(A,B,C)$ values with three angles $(\varphi,\psi,\theta)$,
introduced in Sec.~\ref{secbias}, randomly sampling interval $(0^\circ,360^\circ)$. 

We numerically
propagated this set of synthetic orbits, starting in the libration zone about both the L4 and
L5 stationary points, with the Sun and giant planets for $100$~Myr forward in time (the initial
epoch was MJD 60,000.0). Because the terrestrial planets were neglected, we performed the barycentric
correction to state vectors of all integrated bodies (massive planets and massless Trojans). This means
(i) we added masses of all terrestrial planets to the center (the Sun), and (ii) shifted the Sun to
the barycenter of the inner Solar system. We used the well-tested and efficient package {\tt swift} (specifically
{\tt swift\_rmvs4} code) for our simulation, and used a short timestep of $0.25$~years. During the propagation
we monitored the heliocentric orbits of the test bodies as to their residence in the respective libration
zones about the L4 and L5 points and eliminated those that escaped from their initial regions. At the
end of the simulation, we evaluated statistics of the particle orbital histories and projected them
onto their starting bins in the $(C,B)$ plane. 

The results are shown in Figs.~\ref{fig_stab1a} and
\ref{fig_stab1b}. As expected, the stable region has a triangular shape populated by the observed
Trojans. The region of stability first slightly increases with increasing $A$, but then shrinks
when $A$ becomes large. We note that the integrated timespan of $100$~Myr is rather short, and extending
it to Gyr would lead to additional collapse of the stability zone \citep[e.g.,][]{lev1997,rg2006}.
A small fraction of the observed Trojans (blue symbols on Figs.~\ref{fig_stab1a} and \ref{fig_stab1b})
fall into the unstable bins. This outcome is in agreement with the known marginal stability of the Trojan
swarms, filling the available dynamical zone around the L4 and L5 stationary points to its ``full
capacity'' \citep[see, e.g.,][]{m1993,lev1997,holt2020a}. The resolution given by our large bins is
too poor to identify the very fine secular resonant features discussed by \citet{rg2006} (though some
structures on the stability borderline at the top panels of Figs.~\ref{fig_stab1a} and \ref{fig_stab1b}
might be traced to these effects). Finding this behavior, however, was not the primary motivation of our
experiment. We instead aimed at the zero order definition of the phase space zone available for the
stable Trojan orbits. This information will be used in Sec.~\ref{complete} for a definition
of the orbital zone in the $(A,B,C)$ space of variables over which we shall debias the
Trojan population.

\begin{figure}[t!]
 \plotone{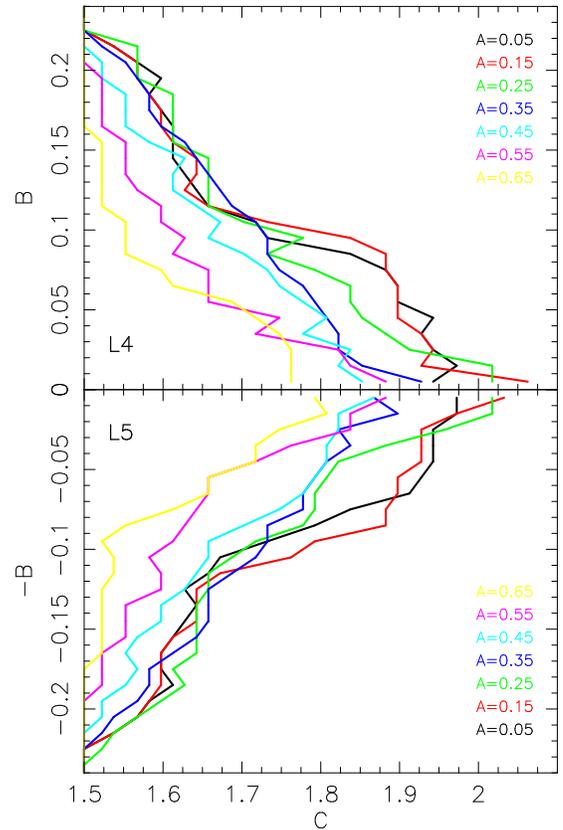} 
 \caption{Comparison of the stability limits for Trojan orbits in the $C$ vs $B$ plane
  determined by our numerical experiment spanning $100$~Myr. Different curves for 7 different
  values of $A$ given by the label and color-identified. The upper panel for L4 Trojans,
  the bottom panel for L5 Trojans; for sake of comparison, we transformed $B\rightarrow -B$
  in this panel. There is no difference in the orbital stability for L4 and L5 clouds
  \citep[see also][]{rg2006,disisto2014}.}
 \label{fig_stab2}
\end{figure}

\begin{deluxetable*}{lccccccccccccccr}[ht]
\caption{\label{fams}
 Parameters of the Trojan families included in our analysis (for a more complete list of the currently
 known Trojan families and a discussion see the Appendix~A, Table~\ref{mb_fams_2023}). }
\tablehead{
 \multicolumn{2}{c}{Family/location} & \colhead{$A_{\rm ref}$} & \colhead{$s^{\rm F}_{\rm A}$} &
  \colhead{$\alpha^{\rm F}$} & \colhead{$B_{\rm ref}$} & \colhead{$s^{\rm F}_{\rm B}$} &
  \colhead{$\beta^{\rm F}$} & \colhead{$C_{\rm ref}$} & \colhead{$s^{\rm F}_{\rm C}$} &
  \colhead{$\gamma^{\rm F}_1$} & \colhead{$\gamma^{\rm F}_2$} &
 \colhead{$\sin I_{\rm P}$} & \colhead{$e_{\rm P}$} & \colhead{$H_{\rm LR}$} & \colhead{$N_{\rm mem}$} 
 }
\startdata
	Eurybates  & L4 & $0.129\phantom{2}$  & $0.004$ & $3.5$ &
		   $0.049\phantom{3}$  & $0.009$ & $2.6$ &
		   $1.553$  & $0.04\phantom{4}$  & $1.1$ & $1.75$ &
                   $0.129$  & $0.044$ & $\phantom{1}9.82$ & 875 \\
	Arkesilaos & L4 & $0.153\phantom{2}$ & $0.004$ & $2.8$ &
		   $0.035\phantom{3}$  & $0.007$ & $3.5$ &
		   $1.5\phantom{00}$    & $0.02\phantom{5}$  & $0.6$ & $1.3\phantom{5}$ &
                   $0.155$  & $0.029$ & $11.97$ & 235 \\
	Hektor     & L4 & $0.323\phantom{2}$  & $0.004$ & $3.5$ &
                   $0.0553$ & $0.007$ & $3.5$ &
		   $1.558$  & $0.05\phantom{5}$  & $1.3$ & $2.1\phantom{5}$ &
                   $0.326$ & $0.054$ & $\phantom{1}7.45$ & 118 \\
 Thronium\tablenotemark{a} & L4 & $0.5252$ & $0.004$ & $4.5$ &
                   $0.0415$ & $0.007$ & $5\phantom{.0}$ &
                   $-$  & $-$  & $-$ & $-$ &
                   $0.526$ & $0.049$ & $\phantom{1}9.69$ & 69 \\
	Teucer     & L4 & $0.3565$ & $0.004$ & $3\phantom{.0}$ &
		   $0.0715$ & $0.01\phantom{0}$ & $4\phantom{.0}$ &
                   $1.5\phantom{00}$ & $0.055$  & $1.7$ & $2\phantom{.05}$ &
                   $0.358$ & $0.073$ & $\phantom{1}8.87$ & 86 \\
	Ennomos\tablenotemark{b} & L5 & $0.464\phantom{2}$ & $0.011$ & $4\phantom{.0}$ &
		   $0.03\phantom{15}$ & $0.01\phantom{7}$ & $3\phantom{.0}$ &
                   $1.5\phantom{00}$    & $0.055$  & $1.4$ & $1.3\phantom{5}$ &
                   $0.459$  & $0.032$ & $\phantom{1}8.68$ & 88 \\
	Deiphobus\tablenotemark{b} & L5 & $0.505\phantom{2}$ & $0.018$ & $4.5$ &
		   $0.032\phantom{3}$ & $0.008$ & $4\phantom{.0}$ &
		   $1.5\phantom{00}$    & $0.05\phantom{5}$  & $0.9$ & $1.6\phantom{5}$ &
                   $0.475$  & $0.029$ & $\phantom{1}8.45$ & 233 \\
	2001~UV209 & L5 & $0.415\phantom{2}$ & $0.007$ & $2.5$ &
		   $0.041\phantom{3}$ & $0.006$ & $3\phantom{.0}$ &
		   $1.5\phantom{00}$    & $0.01\phantom{5}$  & $0.1$ & $3.5\phantom{5}$ &
                   $0.415$  & $0.041$ & $12.62$ & 46 \\
 \enddata
\tablenotetext{a}{Thronium was formerly called 1996~RJ family in most of
  the previous literature. In this case the $dN_{\rm fam}(C)$ was represented using a simple boxcar
  distribution in between $1.5$ and $1.53$.}
\tablenotetext{b}{These two families, overlapping in the space of proper orbital elements, were
  most often denoted just Ennomos family in the previous literature.}
\tablecomments{The second column indicates in which swarm --L4 or L5-- is the family located. The
 third to twelfth columns specify parameter of the family in $A$, $B$ and $C$ parameters distributions
 defined in Eqs.~(\ref{adist_f}), (\ref{bdist_f}) and (\ref{cdist_f}). The
 ninth and tenth columns are proper orbital elements of the largest fragment in the family for sake of
 comparison (we use our new determination of the Trojan proper elements described in the Appendix~A
 and make them available in the electronic form at \url{https://sirrah.troja.mff.cuni.cz/~mira/tmp/trojans/}).
 The last columns give the absolute magnitude of the largest remnant in
 the family $H_{\rm LR}$ and number of members in the nominal family $N_{\rm mem}$.}
\end{deluxetable*}

Figure~\ref{fig_stab2} shows a comparison of the stability limit of the Trojan orbits identified
in our numerical experiments for several values of the proper inclination parameter $A$. We
specifically compare the limiting curves for L4 and L5 libration zones, using a ``mirror representation''
in the upper and lower panels. There are no statistically meaningful differences, implying that the
orbital stability of the L4 and L5 regions is the same. Obviously, this is not a novel result but rather
a confirmation of what has been known previously \citep[e.g.,][]{mar2003,disisto2014,hou2014}. This 
symmetry stems from the stationary configuration of planetary orbits, Jupiter in particular.
It would be broken if the planets were to migrate \citep[e.g.,][]{sd2003,houCMDA2016,pirani2019a,li2023}.
\smallskip

\noindent{\it Assumed orbital distribution of the Jupiter Trojans in families.-- }Several
statistically reliable Trojan families have been identified in the Trojan population
\citep[see, e.g.,][ and references therein]{netal2015}. Since the latest publicly available
identification of the Trojan families dates back to the 2015 release of the catalog at the PDS
node \citep[with some additional data available from][]{vin2015,retal2016,vino2020,holt2020a},
we decided to take the ``bull by the horns'' and update this information. Our work was motivated by the
rapid increase of the known
Trojan population and a need to separate the family members from the background as much
as possible in this work. To that end we (i) determined a new set of Trojan proper elements
using methods described in the Appendix~A of \citet{holt2020b} and applied to the currently
known population of Trojans, and (ii) identified principal families in the population.
Results are briefly summarized in the Appendix~A.

In spite of having this new set of proper elements available to us, we opted to continue using the simplified
variables introduced above for the population analysis in this paper. The reason is because they
have a more straightforward connection to the osculating orbital elements (a step which 
would be more complicated for the proper elements), and this facilitates determination of the
detection probability of CSS discussed in Sec.~\ref{secbias}. We thus use the families identified
in the space of proper elements and locate them in the space of parameters $A$, $B$ and $C$.%
\footnote{We note that \citet{vin2015} and \citet{vin2019} used variables similar to our $A$
 and $B$ to search for Trojan families in the restricted 2D space.}
Note that this step is only approximate and may potentially lead to some inaccuracies, mainly because 
the families are identified in the proper element space as a collection of
individual Trojans. We now need to represent each of the families by a continuous distribution
function
\begin{equation}
 dN_{\rm fam}(A,B,C) = dN_{\rm fam}(A)\,dN_{\rm fam}(B)\,dN_{\rm fam}(C)\; .  \label{abcdist1}
\end{equation}
By itself, the function tells us ``where the family is located'', but does not exclude presence of
possible interloping (background) Trojans in the same zone as well. Therefore $dN_{\rm fam}(A,B,C)$
should have a non-zero value in a correctly-tuned region: (i) not too large (as this would imply
too many interlopers are included), and (ii) not too small (which would mean we undershot the 
contribution of families). There are only approximate solutions between the two limits.

We experimented with different representations of $dN_{\rm fam}(A,B,C)$, starting
with a simple boxcar functions delimiting the corresponding parameter in a certain
interval of values, such as $dN_{\rm fam}(A) \propto \left(H\left[A-A_{\rm min}\right]-H\left[A-
  A_{\rm max}\right]\right)\,dA$, where $H[x]$ is the Heaviside function and $(A_{\rm min},A_{\rm max})$
the corresponding interval of $A$-values. However, tests indicated this definition would
often wrap many interloping objects from the background populations which are unrelated to
a given family. A tighter representation was educated by distribution of the known members
in the families, as identified in the space of proper orbital elements. We used
\begin{equation}
 \frac{dN_{\rm fam}(A)}{dA} \propto \exp\left[-\frac{1}{2}\left(\frac{A-A_{\rm ref}}{s^{\rm F}_{\rm A}}
  \right)^{\alpha^{\rm F}}\right]\; ,  \label{adist_f}
\end{equation}
\begin{equation}
 \frac{dN_{\rm fam}(B)}{dB} \propto \exp\left[-\frac{1}{2}\left(\frac{B-B_{\rm ref}}{s^{\rm F}_{\rm B}}
  \right)^{\beta^{\rm F}}\right] \; , \label{bdist_f}
\end{equation}
and
\begin{equation}
 \frac{dN_{\rm fam}(C)}{dC} \propto \left(C-C_{\rm ref}\right)^{\gamma^{\rm F}_1} \exp\left[-\frac{1}{2}\left(
  \frac{C-C_{\rm ref}}{s^{\rm F}_{\rm C}}\right)^{\gamma^{\rm F}_2}\right]\; . \label{cdist_f}
\end{equation}
Coefficients $(A_{\rm ref},s^{\rm F}_{\rm A},\alpha^{\rm F};B_{\rm ref},s^{\rm F}_{\rm B},\beta^{\rm F};
C_{\rm ref},s^{\rm F}_{\rm C},\gamma^{\rm F}_1,\gamma^{\rm F}_2)$ were determined from distribution of
the identified (although biased) population in the family. They were fixed and not subject to further
refinement, when the complete Trojan population was determined in Sec.~\ref{res}. Their
values for the families considered are given in the Table~\ref{fams}.

\subsubsection{Magnitude distribution}
Apart from the orbital distribution of Trojans in their respective clouds described above,
their absolute magnitude $H$ distribution is the second quantity of major interest here.
In differential form, we denote $dN(H)$ as the number of Trojans in the bin $(H,H+dH)$. We
assume the orbital and magnitude distributions are decoupled.  The reason is because collisional
communication within one Trojan cloud (L4 and/or L5) is large enough that everything hits
everything else,
making the magnitude distribution in each orbital bin equal. As a result, the differential
form $dN(A,B,C;H)$ of the Trojan distribution in the orbital and magnitude space, specifying
the complete Trojan population in a bin $(A,A+dA;B,B+dB;C,C+dC;H,H+dH)$, may be written as
\begin{equation}
 dN_{\rm model}(A,B,C;H) = dN(A,B,C)\, dN(H)\; , \label{abchdist1}
\end{equation}
with $dN(A,B,C)$ from either Eq.~(\ref{abcdist}) or (\ref{abcdist1}). This means we distinguish the
background population and the population in major Trojan families, which linearly superpose to the
total population
\begin{eqnarray}
 dN_{\rm model}(A,B,C;H) &=& dN_{\rm back}(A,B,C;H) + \nonumber \\
                         & & dN_{\rm fam}(A,B,C;H)\; . \label{abchdist2}
\end{eqnarray}
Each term on the right hand side again breaks independently to the orbital and
magnitude parts as in Eq.~(\ref{abchdist1}). Next, we describe the assumed functional
form of the magnitude terms.
\smallskip

\begin{figure*}[t!]
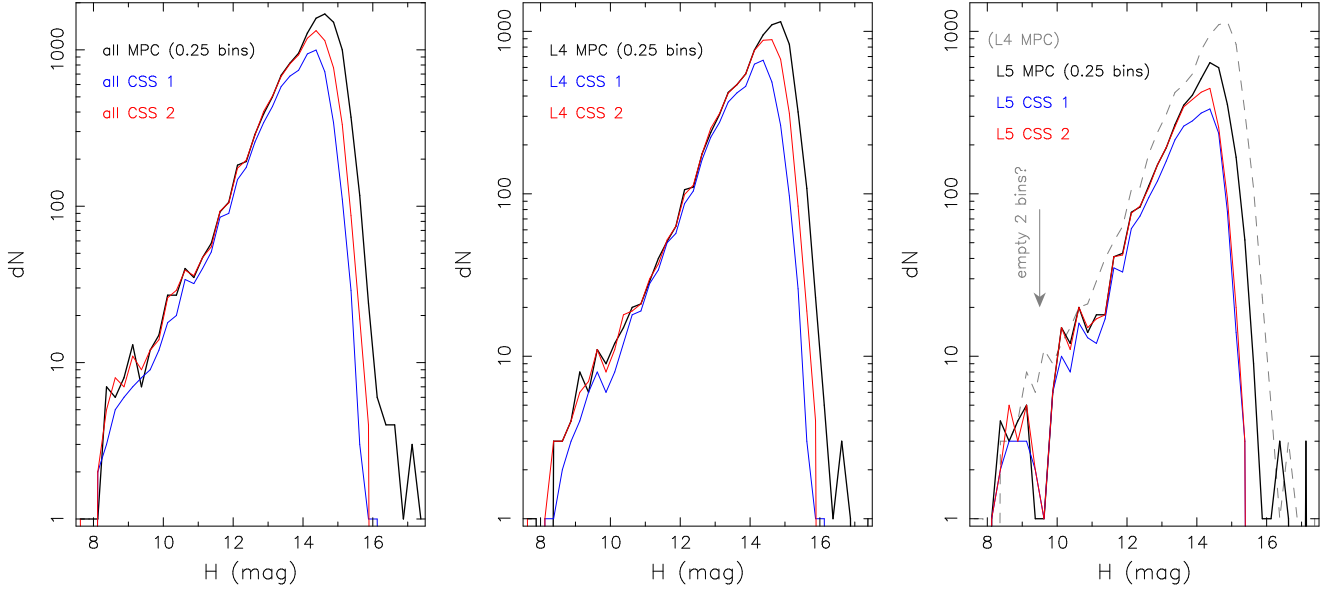

 \begin{center}
 \begin{tabular}{ccc}
  \includegraphics[width=0.31\textwidth]{f15a.eps} &
  \includegraphics[width=0.31\textwidth]{f15b.eps} &   
  \includegraphics[width=0.31\textwidth]{f15c.eps} \\
 \end{tabular}
 \end{center}
 \caption{Differential magnitude distribution of Trojan populations (bin width 0.25 mag):
  (i) left is the complete count (L4 and L5 together), (ii) middle is the L4 cloud,
  and (iii) right is the L5 cloud. Black line is the
  population in the MPC database, blue are detections by CSS during the phase I, red are
  detections by CSS during the phase II. The dashed gray curve on the right panel reproduces
  the MPC L4 population for sake of comparison. We find it curious, that the L5 population
  has nearly empty two bins centered at $\simeq 9.5$ magnitude: there are 46 L4 objects with $H<10$
  vs 26 L5 objects with $H<10$. While this may be an effect of a small number statistics,
  the paucity of the L5 continues beyond the 11 magnitude.}
 \label{fig8}
\end{figure*}

\begin{figure*}[t!]
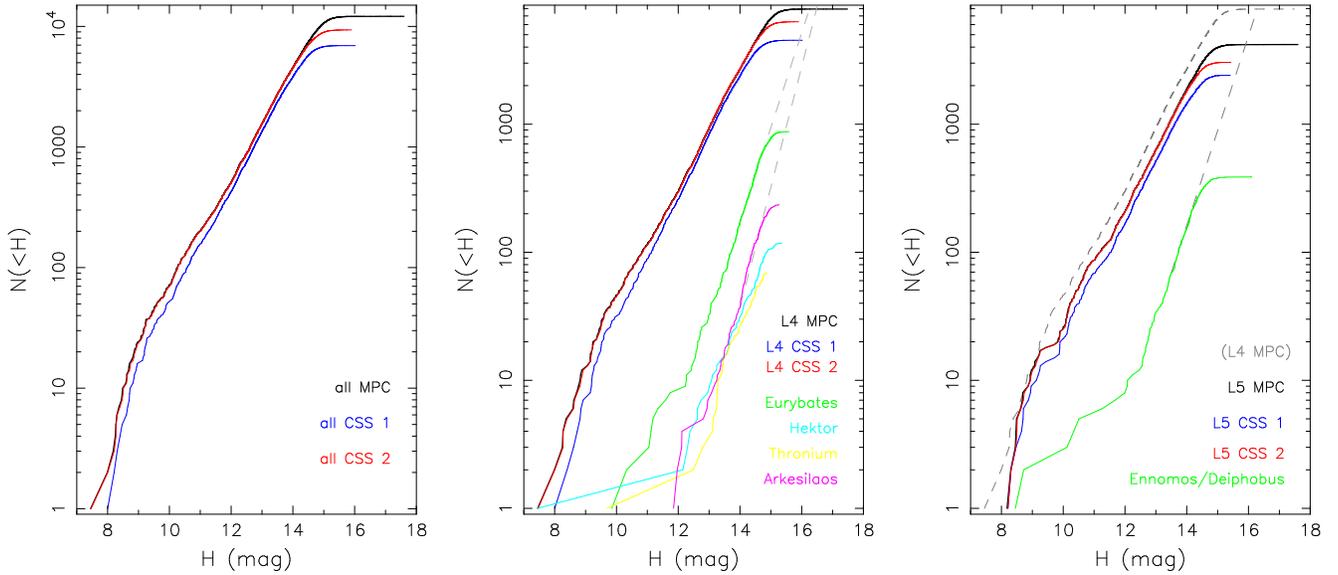

 \begin{center}
 \begin{tabular}{ccc}
  \includegraphics[width=0.31\textwidth]{f16a.eps} &
  \includegraphics[width=0.31\textwidth]{f16b.eps} &   
  \includegraphics[width=0.31\textwidth]{f16c.eps} \\
 \end{tabular}
 \end{center} 
 \caption{The same as in Fig.~\ref{fig8}, but now the cumulative rather than differential
  distribution is shown. The principal Trojan families are also shown: (i) Eurybates (green),
  Hektor (cyan), Arkesilaos (magenta), Thronium (yellow) in the L4 population, and (ii)
  Ennomos/Deiphobus clan in the L5 population. Eurybates has the largest population of 875
  known members, followed by Arkesilaos with 235 known members (see Table~\ref{mb_fams_2023}).
  Even at magnitude 14.5 these two families represent compositely $\simeq 12$\% of the total
  L4 population. Furthermore, the magnitude distribution is much steeper (see the dashed gray
  extrapolation), and their population may reach up to $25-30$\% of the total population
  at magnitudes $16.5-17$, if continued with a straight power-law. The Ennomos/Deiphobus is
  equally abundant among the L5 population: at magnitude $\simeq 14.25$ their population
  represents $\simeq 10$\% of the total and this fraction increases towards larger magnitude.}
 \label{fig9}
\end{figure*}

\noindent{\it Background population.-- }Traditionally, the Trojan magnitude distribution has been
represented using a broken power law with a steeper leg (characterized by an exponent $\alpha_1$)
below a break at magnitude $H_{\rm b}$ and a shallower slope (characterized by an exponent $\alpha_2$)
above, such that%
\footnote{Yet another, more complicated model, would divide the background population into two
 sub-populations, namely the spectrally ``very red'' and ``red'' groups \citep{wb2015}. As shown by
 these authors, both have different magnitude distributions. We do not pursue this interesting
 pathway in this paper because we do not use color data in the optimization procedure. However, our ability
 to include the most prominent Trojan families, which distinctly show in the orbital element space,
 partly substitutes the missing ``red'' population that \citet{wb2015} attribute to collisional
 fragments}
\citep[see, e.g.,][]{jewitt2000,wetal2014,wb2015}
\begin{eqnarray}
 dN_{\rm back}(H) &=& 10^{\alpha_1 (H-H_0)}\,dH\qquad\qquad\quad\;\;\, {\rm for}\, H<H_{\rm b}\; , \nonumber \\
  & &  \label{hbak1} \\
 dN_{\rm back}(H) &=& 10^{\alpha_2 H +(\alpha_1-\alpha_2)H_{\rm b}-\alpha_1 H_0}\,dH\;\;\, {\rm for}\, H>H_{\rm b}\; .
  \nonumber \\ & &  \label{hbak2}
\end{eqnarray}
There are four independent parameters in this model, three of which --$(H_{\rm b},\alpha_1,\alpha_2)$--
characterize the shape of the magnitude distribution, and $H_0$ sets the absolute normalization
(see Figs.~\ref{fig8} and \ref{fig9}). Two more
parameters are to be added, when a second break at a few kilometer size has
been identified beyond magnitude $H_{\rm b2}\simeq 15$ \citep{wb2015}; for $H>H_{\rm b2}$ the magnitude distribution
becomes even shallower \citep[likely because the population has experienced $4.5$~Gyr of collisional grinding,
e.g.,][]{mar2022,bottke2023}.
Further studies obtained slightly different values, shifting the second break-point down to $H_{\rm b2}\simeq
13.6$ \citep{yt2017,u2022}, but the changes in the slope are quite subtle. Apriori, the 1.5 m aperture of the
CSS/G96 telescope at Mt. Lemmon cannot rival results of Subaru-scale telescopes in reaching
very faint Trojans (see Figs.~\ref{fig3a} and \ref{fig3b}). As a consequence, our ability to accurately
determine the location of this second breakpoint, and magnitude distribution properties just below and beyond it, is
limited. At best, we could eliminate implausibly low values of $H_{\rm b2}$ (see Sec.~\ref{res}).

Given the significantly larger number of known Trojans today, and the level of detail at which
we want to describe their population, we considered
an analytical description of $dN_{\rm back}(H)$, such as given in Eqs.~(\ref{hbak1}) and (\ref{hbak2}),
would be too rigid. Instead, we used a  more flexible approach introduced to study the magnitude
distribution of the near-Earth population by \citet{neomod1,nes2023}. In particular, we represent
the cumulative magnitude distribution ${\rm log}_{10} N_{\rm back}(<H)$ using cubic splines in the
interval $7<H<15$ in the initial analysis, extending to a larger interval $7<H<15.5$ in further
tests: (i) there are no Trojans with $H<7$ (the largest being 624 Hektor with MPC
$H=7.45$), and (ii) there are too few G96 detections beyond the magnitude $15.5$, such that
no reliable results may be obtained for a population of these faintest objects from the data
we have. In practice, we split this interval into a certain number of segments. In each of those
we consider a mean slope parameter to be a free parameter (for $n$ segments this implies $n$
independent slopes to be fitted). Additionally, we add $(n+1)$-th parameter representing the absolute
normalization $N_{\rm back}(<H_{\rm r})$ of the the population at a certain reference magnitude $H_{\rm r}$; we
use $H_{\rm r}=14.5$. There no additional parameters to be fitted, since the solution for
${\rm log}_{10} N(<H)$ is then fixed by enforcing continuity (including the first derivative) at the
segment boundaries. After a brief phase of experimenting with the model, we set up at $6$ segments,
implying $7$ free parameters in the initial analysis. Comparing the degrees-of-freedom in our
approach to the traditional method, we find that our work does not over-parametrize the problem, yet
it does gain a significant benefit in generality.

In Sec.~\ref{solution} we find support for
statistically robust features of the magnitude distribution which cannot be described by the
simple broken power-law model. In addition to this aspect, we point out a conceptual issue.
The piecewise spline approximation is in fact very close to piecewise power-law, with different
power slopes in different segments, but splines connect different slopes smoothly (i.e., without
discontinuities, including the derivatives). This conforms the real size distributions which are
continuous and smooth. On the other hand, the broken power-law approach violates the natural
smoothness requirement.

\smallskip

\noindent{\it Population in families.-- }Similar to the situation in the asteroid belt,
studies of Trojan families (clusters of fragments all related to the disruption of a 
parent body) flourished when a large population of Trojans was 
known and accurate proper elements became available \citep[see the pioneering
works of][]{m1993,br2001}. As the number of Trojans has rapidly increased over the past
decade or so, the potential of reliably identifying families in the Trojan population
has increased \citep[see, e.g.,][]{br2011,retal2016}. This advance has also motivated a number
of photometric and spectroscopic studies focused on family members, with the goal of
seeking possible differences with respect to the background Trojan population. 

In our
simple description of the Trojan orbits, the principal families are most easily
discerned in inclination (Figs.~\ref{fig4}, \ref{fig7} and \ref{fig7bis}). The left
panel on Fig.~\ref{fig7} clearly demonstrates that families represent a surplus over the
background population that may account for as much as $10-15$\% of the whole population.
In order to compare the background populations in the L4 and L5 clouds, we need to treat the
families as separate populations.

Several studies \citep[see][ and our Appendix for the most recent updates]{retal2016} indicate
that (i) the magnitude distribution of the family members is steeper than the background population
in the $H\simeq 12-15$ range (e.g., Fig.~\ref{fig9}), and (ii) except for the few largest
members it can be reasonably approximated by a single power-law
\begin{equation}
 dN_{\rm fam}(H) \simeq 10^{\beta_1 (H-H_0)}\,dH\; . \label{hfam}
\end{equation}
For instance, approximating the observed (biased) population we obtain $\beta_1\simeq 0.74$ for the
Eurybates and $\beta_1\simeq 0.94$ for the Arkesilaos families in the L4 swarm,
and $\beta_1\simeq 0.74$ for the Ennomos/Deiphobus clan in the L5 swarm (see Fig.~\ref{fig9}).
These values are significantly larger than the slope $\simeq 0.46$ of the
background population in the same magnitude range, signaling the families may be an important
population partner to the background at smaller Trojan sizes. Obviously, this depends on their
age, as collisional communication with the background population will grind the families
and change the power law slope of their size distributions from a certain size down \citep[e.g.,][]{mar2022}.

In order to prevent over-simplistic analytical assumptions about the magnitude
distribution of the families, such as in (\ref{hfam}), we use again the cubic spline representation of
${\rm log}_{10} N_{\rm fam}(<H)$ as above for the background population. While keeping
$H_{\rm r}=14.5$ for the absolute normalization parameter $N_{\rm fam}(<H_{\rm r})$, we can
use less segments now. There exact number has been adopted for each of the families
separately, with four segments used in the case of prominent families like Eurybates and
Arkesilaos, and only three segments other, less populous families.

\section{Complete Trojan population via model calibration on the CSS data}\label{complete}
After presenting (i) the available observations (Sec.~\ref{datag96}), and (ii) the Trojan
orbital and magnitude model (Sec.~\ref{model}), we now describe how we connect
them via adjustments to the free parameters of the model. Note that the model itself does not
allow us to describe the observations directly because we need to account for and
quantify incompleteness (bias) in the CSS survey. It is only the convolution of the bias and the model
that may be compared to the observations. So the last element to be formulated is the bias
description.

\subsection{Detection probability defined on the space of model parameters}\label{secbias}
Given our limited dataset, we have chosen here to compromise complexity for computational efficiency. 
Consider that every single Trojan orbit is described by six orbital elements, and that this information is 
complemented by the body's absolute magnitude.  Together, these seven parameters have their own probability of detection 
during the time-constrained survey, which in turn is limited by its photometric sensitivity. To avoid small number statistics in
our method, we choose to downgrade the available information for each orbit by averaging. 

The scheme is driven by the model, whose intrinsic parameters are
the proper elements $(A,B,C)$ and the absolute magnitude $H$. To make things simple, we
shall assume the detection probability ${\cal P}$ is given as a function of the same parameters,
namely ${\cal P}={\cal P}(A,B,C;H)$. The Trojan population, predicted to be seen by the survey,
is then given by the distribution function
\begin{eqnarray}
 dN_{\rm pred}(A,B,C;H) &=& {\cal P}(A,B,C;H)\times \nonumber \\ & &  dN_{\rm model}(A,B,C;H) \; , \label{abchdist3}
\end{eqnarray}
which can be finally compared to the observations. Before we get to the method of this comparison
(Sec.~\ref{opti}), we deal with the detection probability ${\cal P}$.

As to the orbital part, having ${\cal P}$ dependent on $(A,B,C)$ only, implies averaging over the
complementary three angular variables. Uniformity of the observed Trojan population in these
angles appears as a reasonable justification for this procedure.%
\footnote{Further justification is provided by our numerical test in the Appendix~B.}
The most straightforward situation
concerns the inclination ($A$) and eccentricity ($B$), for which the suppressed angles are directly
$\varphi$ and $\psi$ in Eqs.~(\ref{properi}) and (\ref{propere}). These equations also provide a
simple mapping
of $(A,B,\varphi,\psi)$ onto osculating orbital elements $(e,I,\Omega,\varpi)$. The reader should 
keep in mind that neither of the resonant
variables, $da/a_{\rm J}$ and $\sigma=\lambda-\lambda_{\rm J}$, is suitable for the intended
averaging. Rather, it would be a (polar) angular variable $\theta$ with the origin shifted to the
respective libration center (L4 or L5), representing an affine parameter on the chosen 
$C=$~const. isoline. Put more simply, in the process of letting $\theta$ span the whole interval
of values $0^\circ$ to $360^\circ$, one completes one turn about the libration center along the
chosen $C-$isoline. 

The definition of $\theta$ still depends on how we scale the ordinate and
abscissa in Fig.~\ref{fig6}. Here we follow \cite{m1993}, who proposed a relation
\begin{equation}
 da = 0.2783\, \left(\sigma - \chi_\star\right)\,\tan\theta  \; , \label{theta}
\end{equation}
where $\chi_\star=\pm \pi/3$ corresponds to the respective libration center for L4 or L5 Trojans.
Inputting (\ref{theta}) into Eq.~(\ref{propera}), one finds a correspondence of $\theta$ to 
$\sigma$ and $da$ on the given $C-$isoline. Unlike the $da$ and $\sigma$ minimum and maximum values,%
\footnote{Given a certain value of the $C$ parameter ($C>C_{\rm min}=1.5$), we can determine the limiting
 values of $da$ and the resonant angle $\sigma$ on the integral curve (\ref{propera}) analytically. 
 The case of semimajor axis is trivial because the minimum and maximum values $\pm a_{\rm J}
 \sqrt{8\mu\,(C-C_{\rm min})/3}$ are readily obtained by plugging $\sigma=\chi_\star$ into 
 Eq.~(\ref{propera}). A more interesting problem is to find extremes, say $\sigma_1$ and $\sigma_2$, 
 of the resonant angle $\sigma$. This is a nice exercise of the ``casus irreducibilis'' within the
 Cardano family of the cubic equation solutions. Intending to express the solution using real
 variables, we first define an auxiliary angle $\xi\in (\pi/2,\pi)$, such that
 \begin{equation}
   \sin \xi = \sqrt{1-(3/2C)^3}\; , \quad \cos \xi = -(3/2C)^{3/2}\; .
 \end{equation}
 Then, we have
 \begin{eqnarray}
   \sin(\sigma_1/2) & = & \pm \sqrt{\frac{2C}{3}}\,\cos \xi/3\; , \\
   \sin(\sigma_2/2) & = & \pm \sqrt{\frac{2C}{3}}\,\cos (\xi-2\pi)/3\; ,
 \end{eqnarray}
 where the $\pm$ sign on the right hand side corresponds to L4, resp. L5, case (thus $\sigma$ either
 positive or negative).}
this needs to be solved by numerical iterations, and it provides a one-to-one mapping of $(C,\theta)$
onto the remaining osculating orbital elements $(a,\lambda)$.

With this preliminary information, we now outline the procedure how we determine the detection
probability ${\cal P}$ by the survey: 
\begin{itemize}
\item Consider a small bin about the central value $(A,B,C)$, namely 
 $(A-\frac{1}{2}dA,A+\frac{1}{2}dA,B-\frac{1}{2}dB,B+\frac{1}{2}dB,C-\frac{1}{2}dC,C+\frac{1}{2}dC)$, and
 choose $(A',B',C')$ within this bin assuming a uniform distribution. Complement it with randomly chosen
 angles $(\varphi,\psi,\theta)$ having again a uniform distribution in the $0^\circ$ and $360^\circ$
 interval. The set $(A',B',C',\varphi,\psi,\theta)$ uniquely transforms onto Keplerian orbital elements
 as described above. In order to sample a sufficient number of configurations, we generate $N_{\rm orb}=5000$
 such orbits within each bin. The epoch of osculation is MJD 60,000.0, or Feb~25, 2023, corresponding to
 the epoch of data in the MPC catalog we used for identification of the presently known population of
 Jupiter Trojans. Each of these orbits is numerically propagated backwards in time over little more
 than a decade, namely to Jan~1, 2013 which precedes the first available frame of the phase~I observations
 of CSS. 
\item Next, we consider information about the CSS operations during phases~I and II (Sec.~\ref{datag96}).
 This includes a full list of frames with specific field-of-view obtained at certain specific time: there
 are (i) $N_{\rm FoV}=61,585$
 such frames in the phase~I, and (ii) $N_{\rm FoV}=162,280$ such frames in the phase~II operations of 
 CSS (altogether making $223,865$ frames available). \citet{nes2023} conducted
 a detailed analysis of moving objects identified on each of these frames and determined their detection
 probability $\epsilon(m,w)$ as a function of the apparent magnitude $m$ and the rate of motion $w$.
\item We then analyze the orbital history of the synthetic trajectories propagated from the orbital
 bins $(A,B,C)$ and consider whether it geometrically appears in the field-of-view of any of the CSS
 frame. To that end we use the publicly available {\tt objectsInField} code (oIF) from the Asteroid
 Survey Simulator package \citep{naidu2017}
 For frames not crossed by the particular orbit we assign a detection probability $\epsilon=0$,
 when the body appears in a certain frame, we use $\epsilon(m,w)$ to determine its detection probability.
 We test a sufficient range of absolute magnitudes $H$ in between $7$ and $19$ (using $0.25$ magnitude bins)
 and compute the apparent magnitude $m$ on the frame using Pogson's relation:
 \begin{equation}
  m=H+5\log(R\Delta)-P(\alpha)\; . \label{pogson}
\end{equation}
 As expected, $m$ is directly related to the absolute magnitude $H$ --thus the dependence of ${\cal P}$
 on $H$-- plus several correction factors. The second term on the right hand side of (\ref{pogson})
 simply follows from the flux dilution with heliocentric distance $R$ and observer distance $\Delta$.
 The last term is the phase function, dependent on the phase angle $\alpha$ (i.e., Sun-asteroid-observer);
 $P$ describes the fraction of the asteroid's hemisphere seen by the observer that is illuminated by the Sun,
 but is also captures the complicated effects of the sunlight reflection and scattering in the surface
 layer. In accord with previous Trojan studies, we adopt a simple $H-G$ magnitude phase function system
 \begin{equation}
  P(\alpha)=2.5\log\left[\left(1-G\right)\Phi_1+G\Phi_2\right]\; , \label{phase1}
 \end{equation}
 relying on two base functions $\Phi_1$ and $\Phi_2$, whose contribution is weighted by the
 slope parameter $G$ \citep[see, e.g.,][]{bow1989,mui2010}, which read
 \begin{equation}
  \Phi_i = \exp\left(-A_i\left[\tan\left(\alpha/2\right)\right]^{B_i}\right)\qquad (i=1,2)
   \; , \label{phase2}
 \end{equation}
 with empirical constants $A_1=3.33$, $A_2=1.87$, $B_1=0.63$, $B_2=1.22$. The partitioning slope
 factor $G$ has to be found from observations. The previous Trojan literature \citep[e.g.,][]{grav2011,yt2017,u2022}
 used $G=0.15$, and this was also our first choice as well. \citet{shev2012} conducted a careful
 study of the phase function for three bright Trojans and found a mean magnitude-phase slope of
 $0.042$ mag~deg$^{-1}$ up to $\simeq 10^\circ$ phase angle. This is in a reasonable accord with
 the formulation in (\ref{phase1}) and (\ref{phase2}), and $G=0.15$. While popular so far, we
 note that the canonical $G=0.15$ has been found most appropriate for the taxonomic class C of
 asteroids. The more common D- and P-types among Trojans, however, have a mean-magnitude-phase slope
 even steeper than the C's, making the effective $G$ parameter lower, namely $G\simeq 0.09$ \citep[e.g.][]{ver2015}.
 For that reason we consider this lower $G$ value nominal in our work, saving the previously used
 higher value for simulations whose purpose is to calculate realistic uncertainties in our results. This
 procedure is especially important for the faintest detectable Trojans, for which the apparent
 magnitude $m$ determined using Eq.~(\ref{pogson}) is near the photometric detection limit of the
 survey.

 Modeling of additional factors of uncertainty in $m$, such as the role of rotation phase at
 the detection for small, irregularly shaped Trojans with possibly large amplitude of the rotation curve
 \citep[e.g.,][]{kal2021,chang2021}, represents a difficult problem and we do not include it in
 our analysis. For Trojans near the detection limit of the telescope, the corresponding noise element
 in the $m(H)$ relation (\ref{pogson}) may not contribute with zero average. As a result, faint Trojans
 will be preferentially detected near the peak of their apparent magnitude variation, and thus deemed
 brighter. But we find this issue more of the interpretation problem (once the reader knows these effects
 have not been taken into account). Indeed, the inferred population of small Trojans will appear
 volumetrically larger, than it really is, but this only means the resulting absolute magnitude
 distribution we obtain for Trojans in Sec.~\ref{solution} should not be overinterpreted towards the
 physical parameters of the population.
\end{itemize}
As a result, we have for (i) each of the synthetic orbits $j=1,\ldots,N_{\rm orb}$, (ii) each of the
frames $k=1,\ldots,N_{\rm FoV}$ and (iii) all the $H$ magnitudes within the needed range, the corresponding
detection probability $\epsilon_{j,k}$ computed. Finally, the representative detection probability
of a given orbit in the $(A,B,C)$ bin and a magnitude $H$ over the whole duration of the survey is
expressed as
\begin{equation}
 {\cal P}\left(A,B,C;H\right) = \frac{1}{N_{\rm orb}}\sum_{j=1}^{N_{\rm orb}}\left\{1-\prod_{k=1}^{N_{\rm FoV}}
  \left[1-\epsilon_{j,k}\right]\right\}\; . \label{bias}
\end{equation}
Note, that ${\cal P}$ is actually evaluated as a complementary value to the non-detection of the body,
which on each frame reads $1-\epsilon_{j,k}$. Similarly, we define the representative rate of detection
\begin{equation}
 {\cal R}\left(A,B,C;H\right) = \frac{1}{N_{\rm orb}}\sum_{j=1}^{N_{\rm orb}} \sum_{k=1}^{N_{\rm FoV}}
  \epsilon_{j,k}\; . \label{rate}
\end{equation}
providing the mean number of frames in which a Trojan in the bin around $(A,B,C;H)$ should be detected
by the survey.
\begin{deluxetable}{ccccc}[t!]
\tablecaption{\label{bins}
 Orbital parameters and absolute magnitudes considered in our model.}
\tablehead{
 \colhead{parameter} & \colhead{minimum} & \colhead{maximum} & \colhead{bin width} &
 \colhead{number of bins} }
\startdata
    $A$ & $0$   & $0.7$  & $0.025$ &  28 \\
 $\; B$ & $0$   & $0.25$ & $0.01$  &  25 \\
 $\; C$ & $1.5$ & $2.05$ & $0.01$  &  55 \\
 $\; H$ & $7$   & $17$   & $0.10$  & 100 \\ 
\enddata
\tablecomments{The second and third columns give the minimum and maximum values, the fourth column is
 the width of the bin, and the fifth column gives number of bins.}
\end{deluxetable}
\begin{figure*}[t!]
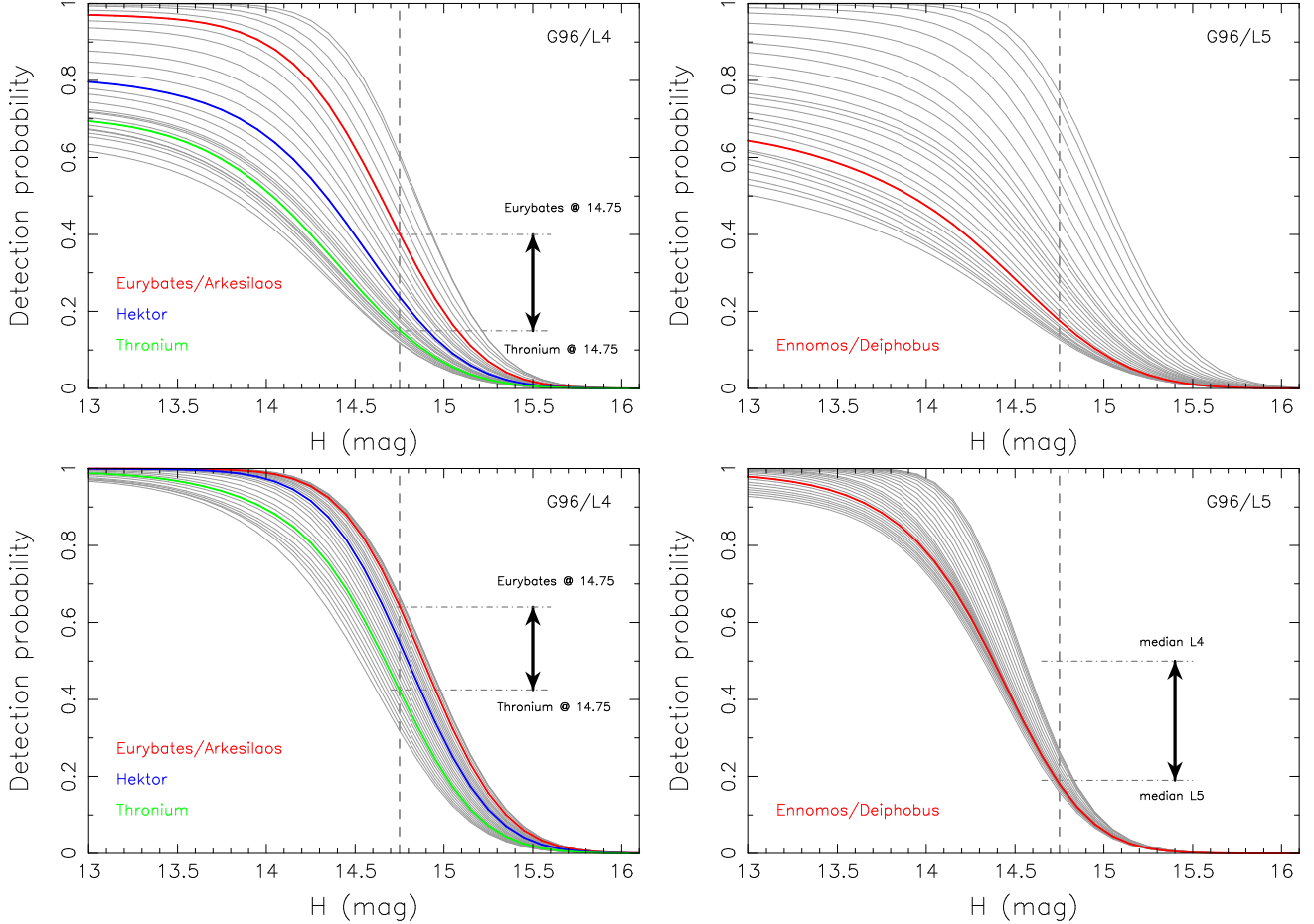

 \begin{center}
 \begin{tabular}{cc}
  \includegraphics[width=0.47\textwidth]{f17a.eps} &
  \includegraphics[width=0.47\textwidth]{f17b.eps} \\  
  \includegraphics[width=0.47\textwidth]{f17c.eps} &
  \includegraphics[width=0.47\textwidth]{f17d.eps} \\  
 \end{tabular}
 \end{center}
 \caption{Detection probability ${\cal P}(A,B,C;H)$ of the L4 (left panels) and L5 (right panels) population
  Trojans in a specific set of the orbital bins centered about $B=0.055$ and $C=1.605$ values: various curves for
  a full range of values of the $A$ parameter from $0$ to $0.7$ (the upper curves for the smallest $A$ value,
  progressing to the lowest curve for the largest $A$ values). The abscissa is the absolute magnitude
  $H$. The top panels for the phase~I of the CSS operations, the bottom panels for the phase~II of the CSS
  operations (Sec.~\ref{datag96}). The five major Trojan L4 and L5 families are intersected by the chosen
  string of bins at different values of $A$ (i.e., orbital inclination). The respective detection probability
  curves are color-highlighted: (i) Eurybates centered at $A\simeq 0.125$ and Arkesilaos at $A\simeq 0.15$
  (both represented in red, L4), (ii) Hektor at $A\simeq 0.325$ (green, L4), (iii) Thronium at
  $A\simeq 0.525$ (magenta, L4), and (iv) Ennomos/Deiphobus at $A\simeq 0.475$ (red, L5). Bins/families
  at higher inclinations have progressively smaller detection probability. For instance, at  magnitude $14.75$
  the detection probability spread from smallest to the largest inclination bins in the L4 population is
  $\simeq 0.4$ to $\simeq 0.72$ (phase~II), and $\simeq 0.12$ to $\simeq 0.65$ (phase~I). Detection probability
  in the phase~I and high inclination bins is systematically much lower, not reaching unity even for large
  Trojans (small $H$ values). A more subtle effect concerns the phase~II, in which detection probability for
  the same bins in the L5 swarm levels-off the unity earlier by about $0.3$~magnitude.}
 \label{fig_bias1}
\end{figure*}
\begin{figure*}[t!]
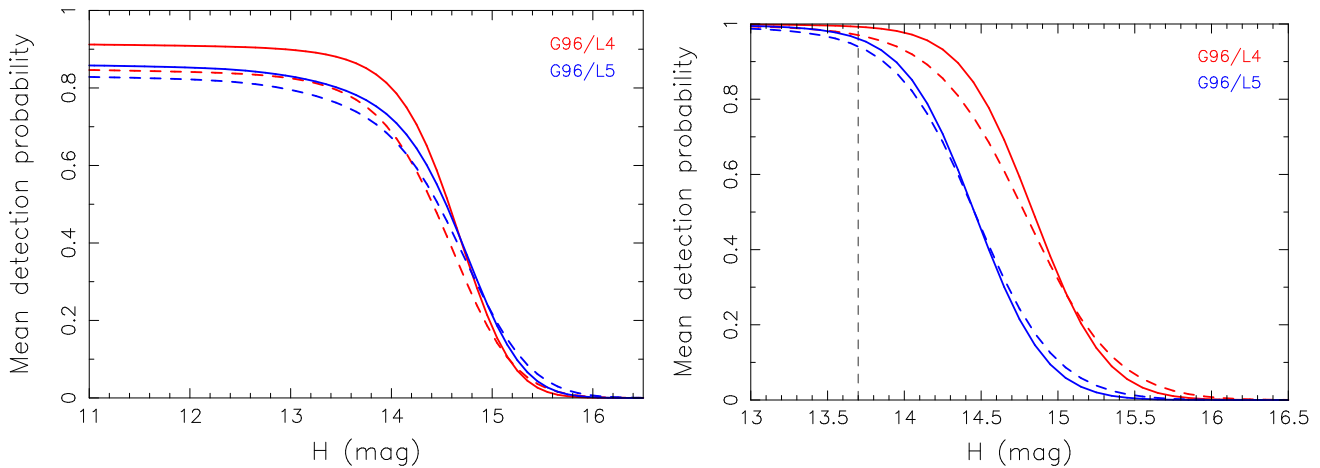

 \begin{center}
 \begin{tabular}{cc}
   \includegraphics[width=0.47\textwidth]{f18a.eps} &   
   \includegraphics[width=0.47\textwidth]{f18b.eps} \\
 \end{tabular}
 \end{center} 
 \caption{Detection probability ${\bar{\cal P}}(H)$ of the L4 (red curve) and L5 (blue curve) population
  Trojans averaged over all orbital bins. Two variants of weighting defined in Eq.~(\ref{weig}) are shown:
  (i) a uniform weight $w=1$ (dashed lines), and (ii) a weight based on Eqs.~(\ref{adist}) to (\ref{cdist})
  (solid lines). The abscissa is the absolute magnitude $H$. The left panel is for the phase~I of the
  CSS operations, the right panel is for the phase~II of the CSS operations (Sec.~\ref{datag96}). The
  vertical dashed line at $13.7$ magnitude in the right panel is an approximate formal completeness limit.
  This limit is not reached during the phase~I, because of its shorter timespan and fields-of-view pointing
  near the  ecliptic (missing thus a fraction of high-inclination Trojans).}
 \label{fig_bias3}
\end{figure*}

\subsection{Optimization of model parameters}\label{opti}
The orbital-magnitude space is the arena in which the target function of the optimization is
evaluated. In practice,
we discretize it into a large number of bins about the central values $(A,B,C;H)$. Table~\ref{bins}
provides the information about the range of their values and bin width. Altogether, there is
more than $3.8$ million bins (in reality, though, nearly half of the orbital bins are
excluded, as they would lead to long-term unstable motion as determined in Sec.~\ref{model}).
This is far larger than number of observations by CSS, which implies
that the orbital-magnitude space is very sparsely populated by the data.

Consider a certain bin indexed $j$ in which CSS detected $n_j (\geq 0)$ individual Trojans and, at the same bin,
our model predicts $\lambda_j=dN_{\rm pred}(A,B,C;H)$ such objects (see Eq.~(\ref{abchdist3})). Assuming 
conditions of the Poisson statistics are satisfied, drawing $n_j$ objects out of $\lambda_j$ expected 
obeys a probability distribution
\begin{equation}
 p_j(n_j) = \frac{\lambda_j^{n_j}\exp(-\lambda_j)}{n_j!}\; . \label{prob0}
\end{equation}
Combining the information from all bins in the orbital-magnitude parameter space, and assuming no
correlations among the bins, we have a joint probability of the model prediction versus data
\begin{equation}
 P = \prod_{j} \frac{\lambda_j^{n_j}\exp(-\lambda_j)}{n_j!}\; . \label{prob1}
\end{equation}
Finally, the target function (log-likelihood) is defined as
\begin{equation}
 {\cal L} = \ln P = -\sum_j \lambda_j + \sum_j n_j \ln \lambda_j\; , \label{prob2}
\end{equation}
where a constant term $-\sum_j \ln(n_j!)$ has been dropped (representing just an absolute
normalization of ${\cal L}$ that cannot affect the optimization of the model parameters). The
parameter optimization procedure seeks maximization of ${\cal L}$ on the space of free 
parameters of the model. The role of the two terms on the right hand side of (\ref{prob2})
is complementary: (i) the first does not depend on $n_j$ and thus this is the single 
contribution for those bins, where CCS has $n_j=0$ detected Trojans; clearly, the situation
when the model predicts a large number $\lambda_j$ of objects needs to be penalized and thus
the negative sign of this term; and (ii) the second term, which includes number of detected
Trojans $n_j$, combines in these bins with the first term to maximize ${\cal L}$ when
$\lambda_j\simeq n_j$; clearly, the match of the model with data is the preferred configuration.
\begin{figure*}[t!]
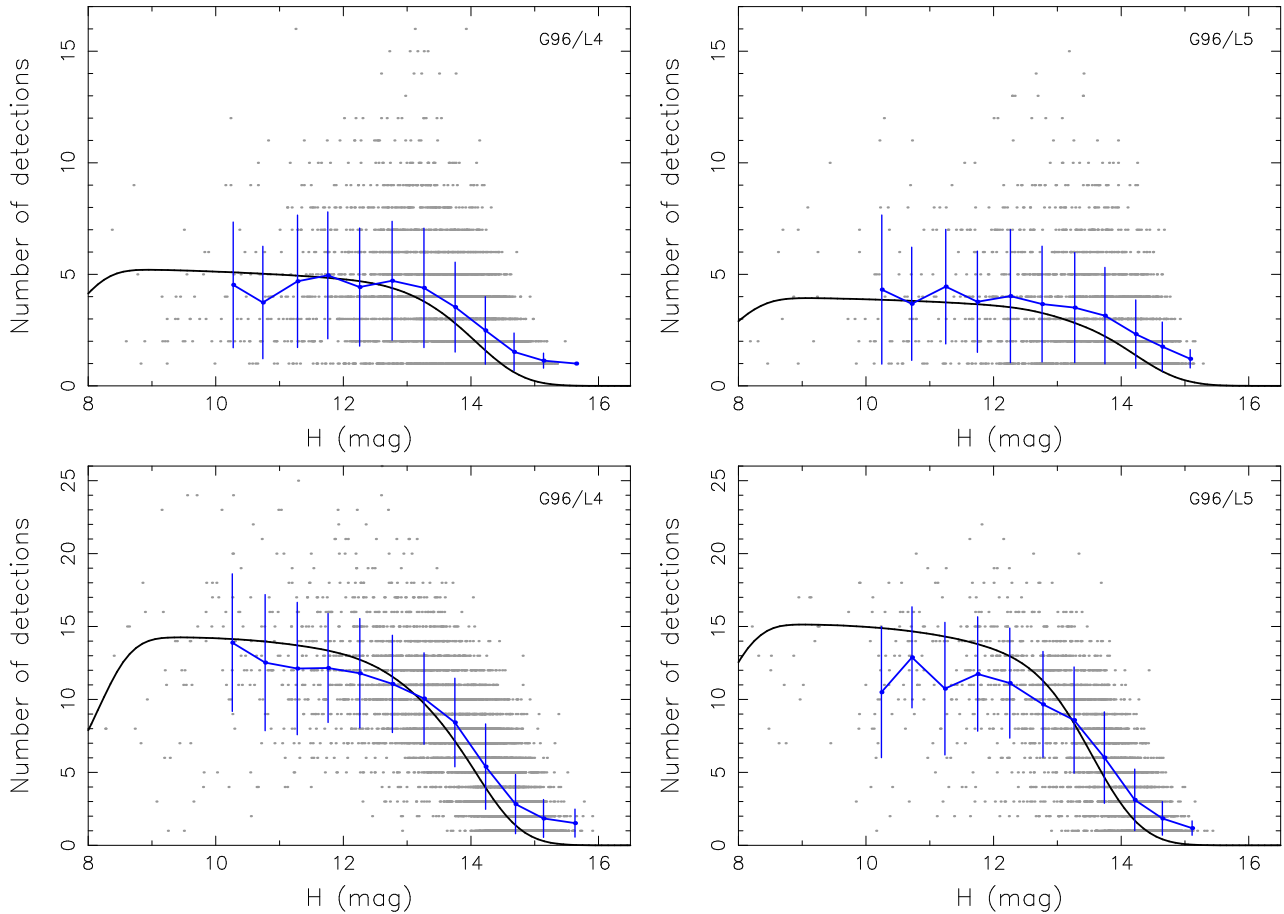

 \begin{center}
 \begin{tabular}{cc}
  \includegraphics[width=0.46\textwidth]{f19a.eps} &
  \includegraphics[width=0.46\textwidth]{f19b.eps} \\   
  \includegraphics[width=0.46\textwidth]{f19c.eps} &
  \includegraphics[width=0.46\textwidth]{f19d.eps} \\  
 \end{tabular}
 \end{center}
 \caption{Number of independent detections (gray symbols) for Jupiter Trojans observed by CSS station
  G96 during the phase~I (top panels) and phase~II (bottom panels) period (Sec.~\ref{datag96}). Left panels
  for the L4 swarm, right panels for the L5 swarm. The abscissa is the absolute magnitude $H$. The black
  curves provide the predicted detection rate ${\bar{\cal R}}(H)$ for G96 determined by averaging over all
  long-term stable $(A,B,C)$ bins using weights based on Eqs.~(\ref{adist}) to (\ref{cdist}). The blue
  curves are simply mean values of detection-counts grouped in $0.5$ wide bins in absolute
  magnitude, and the vertical bars represent the standard deviation within the respective magnitude
  bin. By their definition, the blue curves cannot decrease below one detection, while the predicted rates
  may. This produces an apparent mismatch beyond $H\simeq 14.5-15$.}
 \label{fig_bias2}
\end{figure*}

The optimization procedure seeks maximum of ${\cal L}$ over the space of free parameters of the
model. Because the new camera significantly improved CSS performance in May 2016, we formally consider
phases~I and II described in Sec.~\ref{datag96} as two independent surveys. The target
function to be maximized is a linear combination of the individual parts, namely ${\cal L}=
{\cal L}_{\rm I}+{\cal L}_{\rm II}$. As for parameter set, there are 8 parameters of the $dN_{\rm back}(A)$, 
$dN_{\rm back}(B)$ and $dN_{\rm back}(C)$ distribution functions given by Eqs.~(\ref{adist}),
(\ref{bdist}) and (\ref{cdist}), defining the orbital distribution. The magnitude component has 7 parameters
of the background population
$dN_{\rm back}(H)$ in Eqs.~(\ref{hbak1}) and (\ref{hbak2}), and each of the Trojan families
add additional parameters (typically 4 to 5). Therefore, with up to five families, the L4 population
requires altogether 38 parameters. The L5 population has only one significant cluster (the 
Ennomos/Deiphobus clan) and a smaller family about 2001~UV209, making altogether 27 parameters when
we represent each part in the Ennomos/Deiphobus clan as an individual family (Table~\ref{fams}).

We use powerful and well tested {\tt MultiNest}%
\footnote{\url{https://github.com/JohannesBuchner/MultiNest}}
code to perform optimization
procedure, namely parameter estimation and error analysis \citep[e.g.,][]{feroz2008,feroz2009}.
The popularity and power of {\tt MultiNest} stems from it ability to efficiently deal with
complex parameter space, endowed with degeneracies in high-dimensions. For sake of brevity
we refer to the above quoted literature to learn more about this versatile package.

\section{Results}\label{res}

\subsection{Detection probability of CSS Trojan observations}

As an introductory step, we first
describe behavior of the detection probability ${\cal P}(A,B,C;H)$. We also define a population-averaged
detection probability
\begin{equation}
 {\bar{\cal P}}(H)=\frac{\int_{\cal D} dD\,w(A,B,C)\, {\cal P}(A,B,C;H)}{\int_{\cal D} dD\, w(A,B,C)}
  \; , \label{weig}
\end{equation}
where the integration is performed over a stable domain ${\cal D}$ of the orbital parameter space
($dD=dA\, dB\, dC$), and $w(A,B,C)$ is some weighting function. A uniform weight $w=1$ is the simplest
choice, but this likely assigns too much statistical significance to the extreme borders of the
orbital stability zone. We also used $w$ defined by a product of the one-dimensional distribution
functions given in Eqs.~(\ref{adist}) to (\ref{cdist}) as a refined alternative. At this moment, we use
the set of coefficients $(\alpha_1,\alpha_2,\ldots,\gamma)$ given in the caption of Fig.~\ref{fig7},
which matches the biased populations of Trojans. The reason is that we only want to illustrate
the difference to the uniform weight case. In the same way, we also define the population mean rate
of detection ${\bar{\cal R}}(H)$ with ${\cal R}(A,B,C;H)$ from Eq.~(\ref{rate}).

Figure~\ref{fig_bias1} shows detection probability ${\cal P}(A,B,C;H)$ for fixed values $B=0.055$ and $C=1.605$,
centers of the respective bins, as a function of the absolute magnitude $H$ at the abscissa. Different
gray curves are for $28$ values of $A$, namely mid-values in the bins covering the definition interval
$(0,0.7)$ (see Table~\ref{bins}). The $B$ and $C$ values were purposely chosen to reach some of the
principal Trojan families for a particular $A$ value (see Table~\ref{fams}). As expected, the overall performance
during phase~II is much better than in phase~I of the CSS operations (Sec.~\ref{datag96}). In the
former part, the detection probability does not reach unity at mid- and high-inclinations even for bright
Trojans. This is because this phase was shorter, but mainly the fields-of-view pointed near the ecliptic
plane, not reaching to large ecliptic latitudes (Fig.~\ref{fig0}). At low-inclinations the phase~I detection
probability is nearly as good as that during the phase~II for L5 Trojans, but still worse for L4 Trojans.
The family-corresponding detection probability profiles are highlighted by colors and indicated by labels. The
lowest-inclination families (Eurybates and Arkesilaos) have a detection advantage over the highest-inclination
family (Thronium). The effect is not too large for bright objects during phase~II, but at the faint 
end ($H\geq 14.5$, say) the detection probability in the Thronium family is nearly twice as small as the
Eurybates family. The similar trend is also seen in the L5 swarm (right panel). The sensitivity of the
observations for the high-inclination zone of the Ennomos/Deiphobus clan is poor at the faint end: at $H\geq 14.75$,
shown by the dashed line, the detection probability reaches barely 30\% during phase~II, and only 20\% during
phase~I. Indeed, the right panel on Fig.~\ref{fig9} shows that the detected population of the Ennomos/Deiphobus  
clan levels-off the straight power-law at $\simeq 14.1$ magnitude. A subtler effect is seen in comparison
of the L4 and L5 detection probabilities during phase~II. The L4 detection probability is systematically higher 
than the L5 detection probability (at $H\simeq 14.75$ magnitude the difference in mean detection
probability may only reach values modestly smaller than a factor of $2$).

Figure~\ref{fig_bias3} shows the population-averaged detection probability ${\bar{\cal P}}(H)$ from
Eq.~(\ref{weig}) for both L4 and L5 swarms and for both phases~I (left panel) and II (right panel). This
is global information for the swarms from the respective phases, such that many different factors
related to fields-of-view pointing, galactic plane intersection, and phase of Jupiter's motion in
declination contribute non-trivially to these curves. The comparison of the sensitivity during
phase~I and II therefore exhibits different behavior, whose main aspects can be straightforwardly
understood.

During phase~I the detection probability for the L4 and L5 swarms reach nearly the same
$H$ limit (with ${\bar{\cal P}}\simeq 0$ at $H\simeq 16$), but never reach unity even for bright
Trojans. Additionally, the effect is more serious for the L5 swarm, for which ${\bar{\cal P}}\simeq
0.87$ at $H\simeq 11$, while ${\bar{\cal P}}\simeq 0.92$ for the same magnitude for the L4 swarm.
This stems from the fact that G96 operations during phase~I was limited to fields-of-view near the ecliptic
and did not reach large latitude zones. This means that segments of high-inclination orbits
among Trojans escaped observations, and the effect is independent of object brightness.

Another type of behavior is observed for phase~II. Here, we find that ${\bar{\cal P}}(H)\simeq 1$ up to
$\simeq 13.7$ magnitude for both swarms, beyond which the detection probability levels off. Interestingly,
the L5 population is more biased, as already noted in Fig.~\ref{fig_bias1}. The relation between the two population
curves on Fig.~\ref{fig_bias3} may be approximately characterized as a $\simeq 0.35$ magnitude shift. Overall,
this finding confirms earlier estimates of the completion limit of Jupiter Trojans \citep[e.g.][]{hm2020}.
Indeed, the largest Trojans discovered within the past decade are 2013~BD8 in L4 (with $H=13.7$), and
2013~SP62 in L5 (with $H=13.6$), and there are only $4$/$9$ discovered Trojans during this interval of
time brighter than $H=14$ in L4/L5. 
At magnitudes $\simeq 16$ and $\simeq 15.5$, for L4 and L5 swarms, ${\bar{\cal P}}(H)\simeq 0$ because the
Trojans become too faint even at perihelia of their orbit to overcome the apparent magnitude limit for
the G96 detection. The reason for the magnitude shift was discussed at the end of
Sec.~\ref{datag96} (see Fig.~\ref{fig5}), and has to do with a systematically larger heliocentric
and geocentric distances of the L5 Trojans during phase~II of the G96 operations.

This shift is confirmed on Fig.~\ref{fig_bias2},
where we compare the number of detections for the observed Trojans with the population-averaged 
prediction ${\bar{\cal R}}(H)$ (using the simple weighting based on based on Eqs.~(\ref{adist}) to
(\ref{cdist}) and preliminary parameters from matching the biased population in Fig.~\ref{fig7}).
The spread in number of detections for a given absolute magnitude, shown by blue vertical intervals,
indicates pricipally the expected dependence of this quantity in the orbital inclination. There
are generally less detections during phase~I, and detection
of L4 Trojans during phase~I reaches up to magnitude $16$, while only $\simeq 15.6$ for L5 Trojans.
The theoretical curves roughly match the trend seen in the observations,
whose scatter is large. This is partly because we have combined observations of Trojans at all
possible orbits, but also because it is likely that ${\bar{\cal R}}(H)$ is more sensitive to
the averaging over the angular variables $(\varphi,\psi,\theta)$ associated with the elements
$(A,B,C)$ (Sec.~\ref{secbias}). At this moment, we satisfy ourselves with the fair agreement, but
do not use information about the detection rate in the model fitting procedure.

\begin{deluxetable}{l|rrcc}[t] 
 \tablecaption{\label{model_back}
  Median and uncertainties of Jupiter Trojan background population parameters in the nominal model.}
 \tablehead{
  \colhead{} & \colhead{median} & \colhead{$\pm\sigma$} 
	      & \colhead{median} & \colhead{$\pm\sigma$} }
\startdata
 \rule{0pt}{3ex}
 & \multicolumn{4}{c}{{\it -- Orbital distribution parameters --}} \\ [2pt]
  & \multicolumn{2}{c}{{\it -- L4 parameters --}} & \multicolumn{2}{c}{{\it -- L5 parameters --}} \\ [1pt]
 $s_A$      & $0.0533$ & $0.0105$ & $0.3352$ & $0.0178$ \\
 $\alpha_1$ & $1.274\phantom{0}$ & $0.091\phantom{0}$ & $0.644\phantom{0}$ & $0.046\phantom{0}$ \\
	$\alpha_2$ & $1.008\phantom{0}$ & $0.063\phantom{0}$ & $3.012\phantom{0}$ & $0.247\phantom{0}$ \\
 $s_B$      & $0.0191$ & $0.0028$ & $0.0307$ & $0.0041$ \\
 $\beta_1$  & $1.156\phantom{0}$ & $0.077\phantom{0}$ & $1.106\phantom{0}$ & $0.086\phantom{0}$ \\
 $\beta_2$  & $1.144\phantom{0}$ & $0.061\phantom{0}$ & $1.140\phantom{0}$ & $0.088\phantom{0}$ \\  
 $s_C$      & $0.0507$ & $0.0031$ & $0.0443$ & $0.0036$ \\
 $\gamma$   & $0.941\phantom{0}$ & $0.030\phantom{0}$ & $0.929\phantom{0}$ & $0.038\phantom{0}$ \\ [3pt] 
 & \multicolumn{4}{c}{{\it -- Magnitude distribution parameters --}} \\ [2pt]
  & \multicolumn{2}{c}{{\it -- L4 parameters --}} & \multicolumn{2}{c}{{\it -- L5 parameters --}} \\ [1pt]
 $N_{\rm back}$ & $3951\phantom{.0000}$ & $44\phantom{.0000}$ & $2664\phantom{.0000}$ & $39\phantom{.0000}$    \\ 
 $\gamma_1$ & $1.236\phantom{0}$ & $0.448\phantom{0}$ & $1.404\phantom{0}$ & $0.386\phantom{0}$ \\
 $\gamma_2$ & $0.656\phantom{0}$ & $0.094\phantom{0}$ & $0.638\phantom{0}$ & $0.090\phantom{0}$ \\
 $\gamma_3$ & $0.429\phantom{0}$ & $0.021\phantom{0}$ & $0.431\phantom{0}$ & $0.025\phantom{0}$ \\
 $\gamma_4$ & $0.497\phantom{0}$ & $0.014\phantom{0}$ & $0.468\phantom{0}$ & $0.016\phantom{0}$ \\
 $\gamma_5$ & $0.447\phantom{0}$ & $0.008\phantom{0}$ & $0.473\phantom{0}$ & $0.010\phantom{0}$ \\
 $\gamma_6$ & $0.388\phantom{0}$ & $0.055\phantom{0}$ & $0.344\phantom{0}$ & $0.009\phantom{0}$ \\ [2pt]
\enddata

\tablecomments{The L4/L5 cloud parameters in the
 highlighted columns. The seven parameters at the bottom part of the Table specify the
 cumulative magnitude distribution: (i) $\gamma_i$ are mean slopes on seven magnitude
 segments defined by intervals ($7-8.5$, $8.5-10$, $10-12$, $12-13$, $13-14$, $14-15$),
 and (ii) $N_{\rm back}=N_{\rm back}(<H_{\rm r})$ is the complete population up to $H_{\rm r}=14.5$ magnitude. The eight
 parameters at the top part of the Table specify parameters of the orbital distribution
 in the $(A,B,C)$ space, Eqs.~(\ref{adist}), (\ref{bdist}) and (\ref{cdist}).}
\end{deluxetable}

\begin{figure*}[t!]
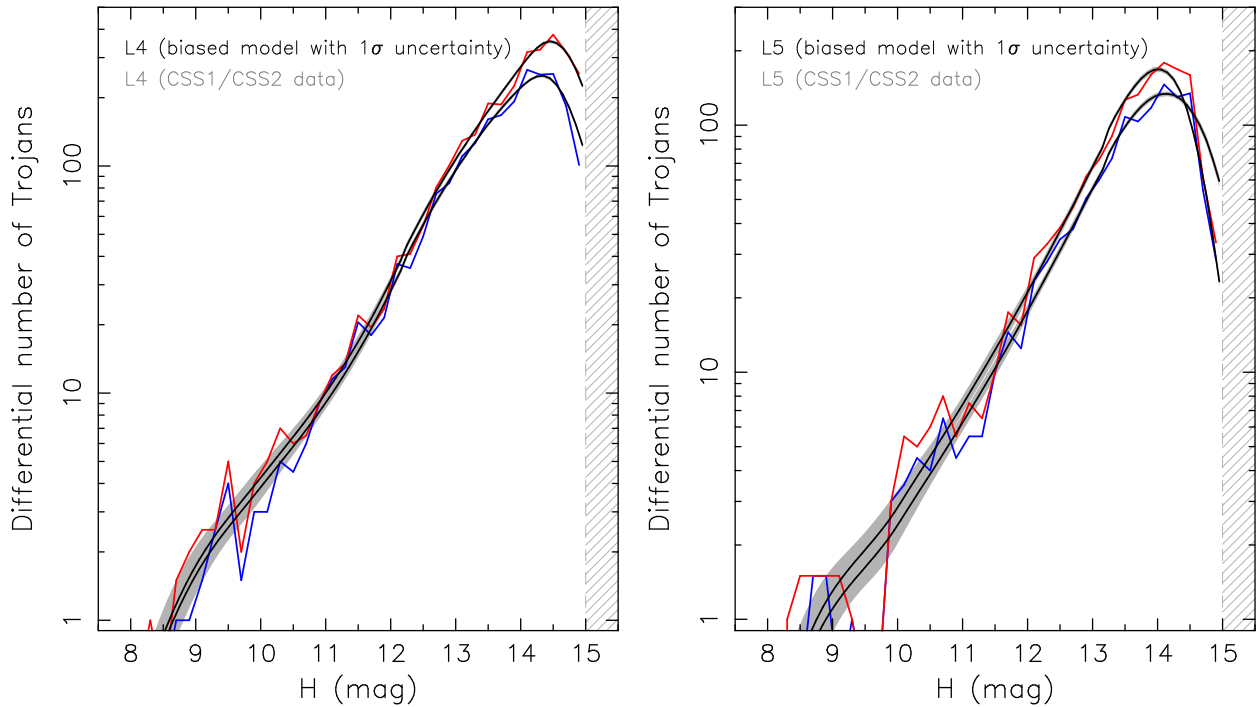

 \begin{center}
 \begin{tabular}{cc}
  \includegraphics[width=0.45\textwidth]{f20a.eps} &
  \includegraphics[width=0.45\textwidth]{f20b.eps} \\
 \end{tabular}
 \end{center}  
 \caption{Differential magnitude distribution of the G96 observations (blue for phase~I and red
  for phase~II) compared with the best-fitting biased model (black line with a $\sigma$ interval,
  based on analysis of $10,000$ posterior random samples of the model, depicted by the gray zone).
  Left panel for L4 Trojans, right panel for L5 Trojans. Magnitude bin has a width of $0.1$ magnitude
  for the model, and $0.2$ magnitude for the data.}
 \label{model_h_biased}
\end{figure*}
\begin{figure*}[t!]
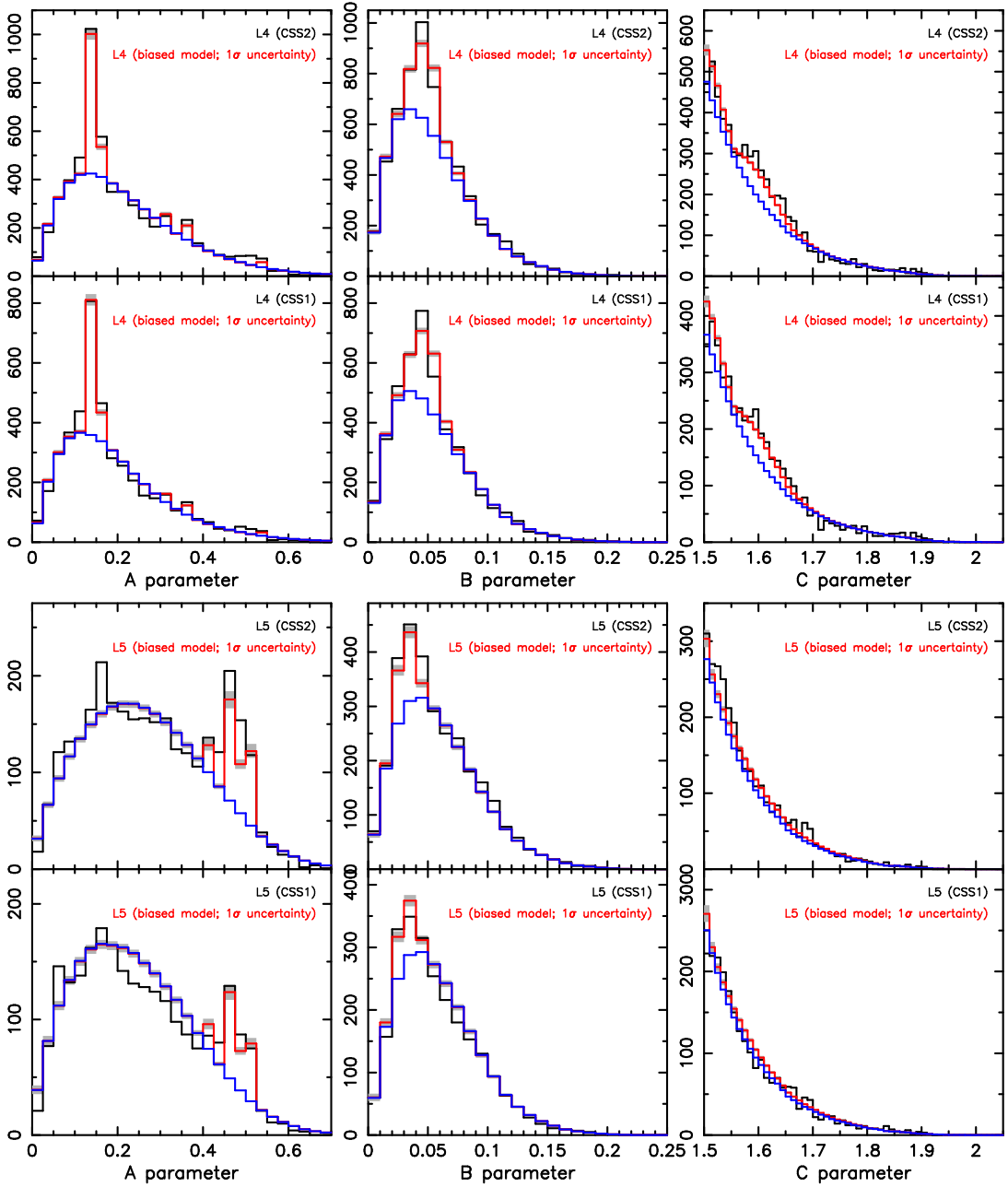

 \begin{center}
 \begin{tabular}{c}
  \includegraphics[width=0.8\textwidth]{f21a.eps} \\
  \includegraphics[width=0.8\textwidth]{f21b.eps} \\
 \end{tabular}
 \end{center}
 \caption{Projected distributions of the orbital parameters $A$ (left), $B$ (middle) and $C$ (right) of the
  biased populations from our model (red lines) and the G96 observations (black histogram) using the
  phase~I (bottom sectors) and phase~II (top sectors) data: L4 data at the top panels, L5 data at the bottom
  panels. The shaded gray area delimits the $\sigma$ region of the solution (created by $10,000$ posterior
  random samples of the model). The blue histogram is the best-fitting model of the background population.}
 \label{model_abc_biased}
\end{figure*}
\begin{figure*}[t!]
 \begin{center}
  \includegraphics[width=0.95\textwidth]{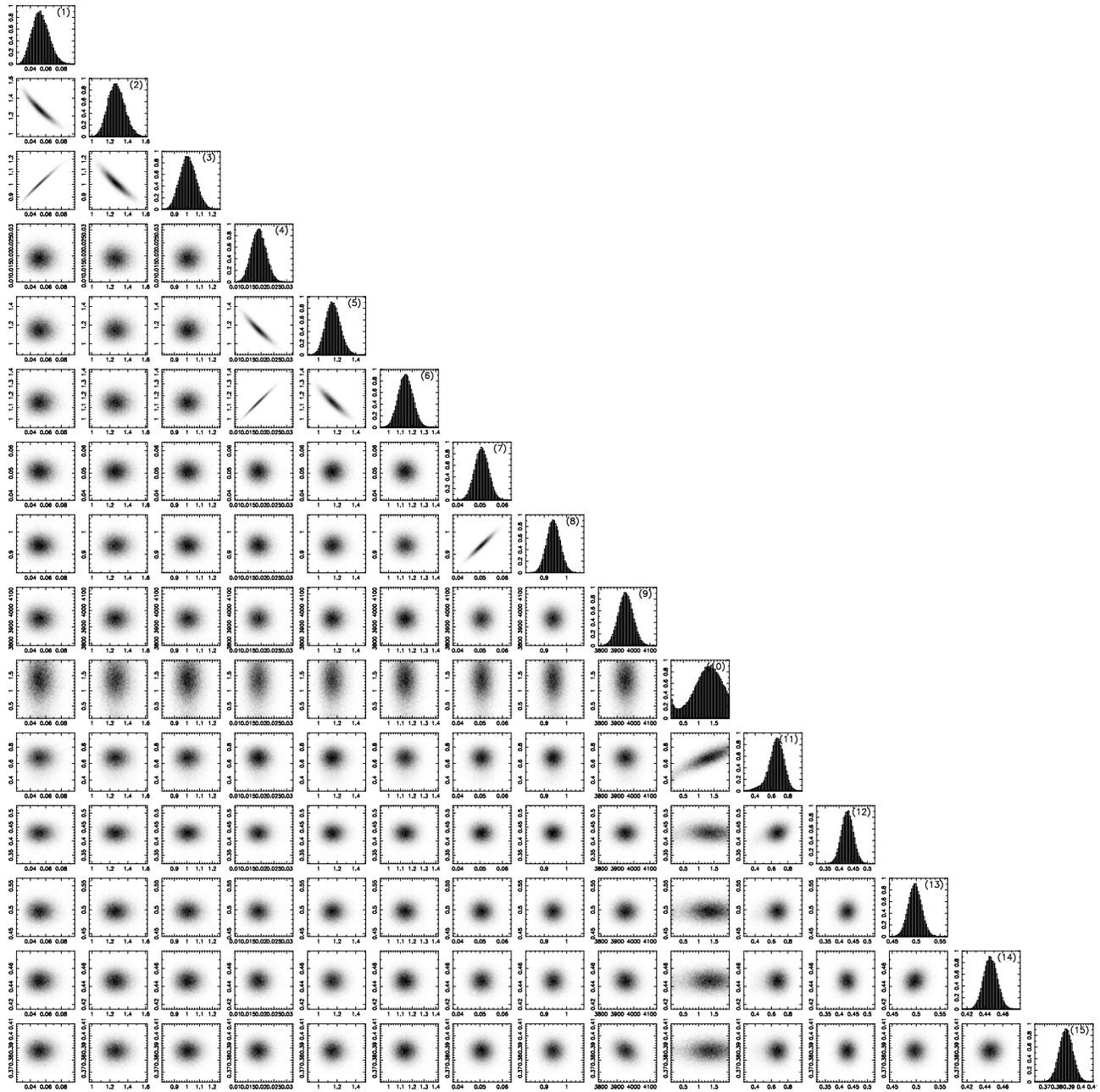} 
 \end{center}  
 \caption{The posterior distribution of $15$ model parameters characterizing the background population of the
  L4 Trojans from our nominal {\tt MultiNest} fit to G96 observations (sometimes called the ``corner
  diagram''; L5 solution shows a fairly similar
  pattern). The individual plots are labeled (1) to (15) following the model parameter sequence given in
  Table~\ref{model_back}; the first eight parameters determine the orbital distribution in $A$, $B$
  and $C$ elements, the last seven parameters determine the absolute magnitude cumulative distribution (with
  the ninth parameter being the normalization at $14.5$ magnitude), and the last six parameters the mean
  slope values in the chosen segments (Table~\ref{model_back}).}
 \label{model_corr}
\end{figure*}

\subsection{Bias-corrected L4/L5 populations} \label{solution}
We now describe the main results, namely properties of the bias-corrected population of the L4 and L5 
clouds of Jupiter Trojans. We start with a nominal model that has the 
maximum absolute magnitude set to $15$. We find there are too few observations of L5 Trojans by G96 
beyond this limit (see Fig.~\ref{fig8}). As a result we explore the solution beyond $H=15$ only for
L4 Trojans in an extension of the nominal solution at the end of this section.

\begin{figure*}[t!]
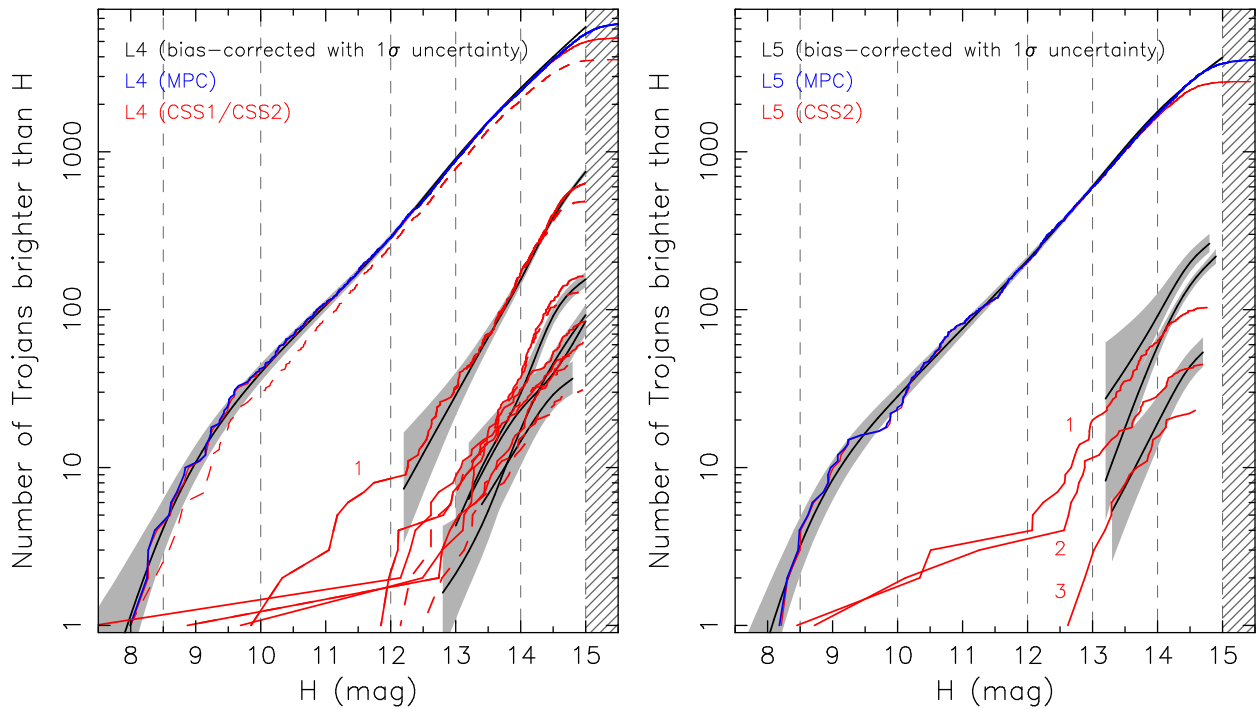

 \begin{center}
 \begin{tabular}{cc}
  \includegraphics[width=0.45\textwidth]{f23a.eps} &
  \includegraphics[width=0.45\textwidth]{f23b.eps} \\
 \end{tabular}
 \end{center}  
 \caption{Bias-corrected cumulative magnitude distribution of the L4 (left panel) and
  L5 (right panel) Jupiter Trojan populations based on G96 observations. The background population,
  shown by the upper curves, is separated from the major families: (i) Eurybates (label~1),
  Arkesilaos, Hektor, Thronium and Teucer in the L4 population, and (ii) Deiphobus (label~1),
  Ennomos (label~2) and 2001~UV209 (label~3) in the L5 population. The red lines are the CSS observations
  (solid are the data in the phase~II, dashed are the data in the phase~I), the black line is the
  bias-corrected model with $\sigma$ interval (gray zone). The blue line is the whole background
  population from the MPC catalog. The vertical dashed lines show the magnitude segments used
  for representation of the background population (Table~\ref{model_back}).}
 \label{model_h_unbiased}
\end{figure*}

We combine full sets of CSS phase~I and phase II detections for both L4 and L5 observations. In each
of the survey phases we use unique detections only (i.e., we do not account for Trojan re-detections).
However, considering the two CSS phases separately, a number of objects may be detected in both of
them. The observations are mapped onto the bins in $(A,B,C;H)$ space described in Table~\ref{bins}.
Next we run simulations%
\footnote{These simulations were run on the NASA Pleia\-des Super\-computer.} 
using {\tt MultiNest} to optimize parameters of the global model described by Eq.~(\ref{abchdist2}),
with the observation prediction given by convolution of the model with the detection probability 
as in Eq.~(\ref{abchdist3}). Recall the model has a (i) smooth, background component, and (ii)
contribution from selected Trojan families. In the case of the background population, we adjust
parameters of the orbital distribution described in Eqs.~(\ref{adist}) to (\ref{cdist}) and the
absolute magnitude distribution (in the nominal setup in between magnitudes $7$ and $15$ using six
segments; Table~\ref{model_back} and Fig.~\ref{model_h_unbiased}). In the case of the families,
we use distribution functions in Eqs.~(\ref{adist_f}) to
(\ref{cdist_f}) with fixed parameters listed in Table~\ref{fams} to make them identified in the
orbital space $(A,B,C)$, and we adjust only parameters of their absolute magnitude range of values
$(H_1,H_2)$ individual to each family. The range for each of the families is shown by the
gray polygons in Fig.~\ref{model_h_unbiased}. The largest family members typically have an irregular
distribution of magnitudes that is not suitable for any analytic representation. This sets our
individual choice of $H_1$ for each of the families. Having avoided modeling the largest few
members in the families is not a problem for the global model because we make sure that the
population is complete to $H_1$. The maximum value $H_2$ expresses a limit beyond which the
nominal family, determined by the clustering method in the proper element space (see the Appendix~A),
has already too few members detected by CSS. 

The typical value of $H_1$ ranges from $12.2$ to $13.3$,
while the $H_2$ values span an interval between $14.7$ and $15$. Note that the location in orbital
space is the only information about the families we provide to the optimization procedure.
We do not separate the background and family populations in the observations, and let the split to
be decided by the {\tt MultiNest} algorithm uniquely.
\begin{figure}[t!]
 \plotone{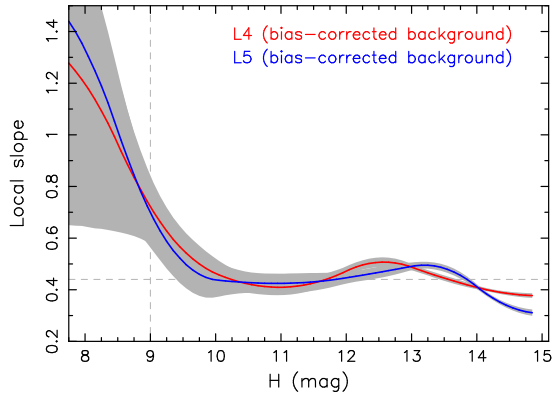}
 \caption{Local slope of the cumulative absolute magnitude distribution for the bias-corrected,
  background population of L4 (red) and L5 (blue) Trojans. The gray zone is the $\sigma$ interval of
  the solution. The vertical dashed line indicates the often-quoted break at $\simeq 9$ magnitude
  delimiting the transition from a very steep distribution of larger Trojans to a shallower
  distribution of smaller Trojans. The horizontal dashed line at $0.44$ slope is the characteristic
  value for $H>9$ magnitude Trojans in the previous literature \citep[see, e.g.,][]{u2022}. In
  the last segment of our solution, magnitude range $14-15$, the slope of the L4 population becomes
  gradually shallower reaching $\simeq 0.38$ at $H=15$ magnitude. The solution of the L5 population
  in the same magnitude segment is even shallower, but we consider it unreliable for reasons
  discussed in the main text.}
 \label{gamaloc}
\end{figure}

{\tt MultiNest} also provides posterior distributions of model parameters. These help us to
determine the uncertainties of the parameter values (the best-fit and uncertainty values for the
background populations parameters for the nominal model are listed in the Table~\ref{model_back})
as well as higher-dimensional statistical quantities, such as parameter correlations. They
also serve to construct posterior models, useful to visualize check on the validity of the results, as they are
confronted with observations using different parametric projections.
\smallskip

\noindent{\it Comparison of the biased model to the observations.-- }
Figures~\ref{model_h_biased} and \ref{model_abc_biased} show the biased model predictions
(Eq.~\ref{abchdist3}) compared to the observations using 1-D projections in all parameters $(A,B,C;H)$
for both L4 and L5 clouds.
The solution for the L4 population is markedly superior over the solution for the L5 population.
This is because of two factors: (i) the available observations of L5 Trojans are less numerous and only
a few of them reach the last magnitude segment between $14.5$ and $15$ (see the right panel of Fig.~\ref{fig8}),
and (ii) the Trojan families in the L5 cloud are extended and diffuse, which makes their precise identification 
more difficult to determine (see Fig.~\ref{fig7bis}). Finally, the mask functions for the largest L5 families,
Ennomos and Deiphobus, partly overlap (see Table~\ref{fams}), which brings additional confusion to
the solution of the L5 population. 

As a result, the biased model of the L5 population exhibits several drawbacks;
for instance (i) it does not decrease very fast in the last segment of the CSS phase~I observations
(Fig.~\ref{model_h_biased}), or (ii) it falls short in a compromise between the background and family signal
in orbital space (see especially the $A$ coordinate on Fig.~\ref{model_abc_biased}). While not perfect,
the correspondence between the biased model prediction and the CSS data is much better for the L4 population.
More compact Trojan families in the L4 cloud reveal their identity more clearly in the fitting
procedure. Figure~\ref{model_abc_biased} confirms that the smooth background population matches the expected
trend over which the principal families contribute with the foreseen punctual signal. The small differences may
be due to the large bins we used. The fit of the magnitude distribution of both the phase~I and the
phase~II data in Fig.~\ref{model_h_biased} is fairly satisfactory in this case too.
\begin{figure*}[t!]
 \plotone{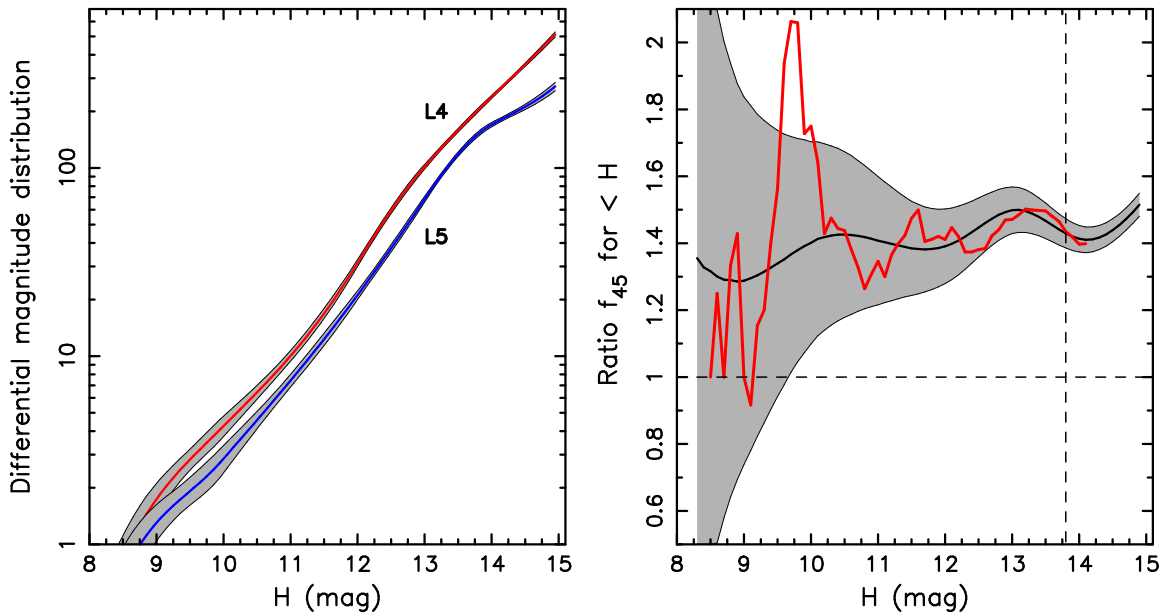}
 \caption{Left panel: Differential magnitude distribution of the bias-corrected background Trojan
  populations: solid line is the median solution (red for L4 and blue for L5), and the gray zone is
  the $\sigma$ uncertainty ($0.1$ magnitude bins wide). The Trojan populations in the families is
  not contributing. Right panel:
  The asymmetry factor $f_{45}=N_4/N_5$ of the L4 and L5 Jupiter Trojans. The background population counts
  $N_4$ and $N_5$ are (i) their respective cumulative values at a certain magnitude limit (the abscissa),
  and (ii) correspond to the bias-corrected, background population in our model. The solid black line is the
  median value and the gray area is the $\sigma$ envelope from the $10,000$ posterior solutions by
  {\tt MultiNest}. The vertical dashed line is the approximate observational completeness to-date.
  The solid red line is the ratio of the observed L4/L5 populations with the Trojans in families
  removed. Up to the completion limit we expect this fraction should reside inside the uncertainty
  limits of the model. The group of the largest Trojans, below magnitude $\simeq 9-9.5$, exhibit
  symmetry, since $f_{45}=1$ is within the statistical uncertainty. Interestingly, the observed-population
  fraction exceeds the $\sigma$ interval of the model in the $\simeq 9.5-10$ range. This feature is
  due to the anomalous vacancy of the L5 Trojans in this magnitude range (see Figs.~\ref{fig8} and
  \ref{fig9}). Trojan populations fainter than $\simeq 10$ deviate from symmetry, with the
  observed fraction returning to the $\sigma$ interval of the model, and the
  statistical robustness increases with increasing $H$. At the approximate observational completeness
  limit, $H\simeq 13.8$, $f_{45}=1.43\pm 0.05$ ($\sigma$ solution), with $f_{45}=1$ excluded at
  $\simeq 7.9\sigma$ level.}
 \label{asymetry}
\end{figure*}

Figure~\ref{model_corr} shows posterior distribution for the 15 parameters of the background L4 population
organized in the triangular form which allows to see possible mutual correlations. A well-behaved model
is perceived to have no, or minimum number of, correlations between the parameters. In our case, this
is the situation of the magnitude sector of the model (parameters 9 to 15). Certain groups of the
orbital distribution parameters are found to be strongly correlated between each other: $(s_{\rm A},
\alpha_1,\alpha_2)$ of the $A$ distribution function $dN_{\rm back}(A)$, $(s_{\rm B},\beta_1,\beta_2)$ of the
$B$ distribution function $dN_{\rm back}(B)$, and $(s_{\rm C},\gamma)$ of the $C$ distribution function
$dN_{\rm back}(C)$ (see the sub-diagonal plots of the first 8 parameters in Fig.~\ref{model_corr}).
This behavior is understood by inspecting the formulae (\ref{adist}), (\ref{bdist}) and (\ref{cdist});
for instance, $s_{\rm A}$ and $\alpha_2$ literally form a single parameter $s_{\rm A}^\prime = s_{\rm A}^{\alpha_2}$
in Eq.~(\ref{adist}), and thus their values must be highly correlated. Nevertheless, robustness of
{\tt MultiNest} algorithm makes the result well-behaved. Re-parametrization of the orbital
distribution functions may lead to a set of parameters with lower correlations, but as long as the
optimization scheme does not fail this is not a serious problem.
\smallskip

\noindent{\it Properties of the bias-corrected model.-- }Figure~\ref{model_h_unbiased} shows the cumulative
distribution of the bias-corrected population of the L4 and L5 clouds. The model naturally provides the
background population separated from the population in the families. To more easily make a one-to-one comparison, we also 
show CSS observations which were, for the purpose of this figure, divided into background and family
components; both are shown by red lines, those from phase~I are dashed and those from the phase~II are solid.
We also show the complete background population of L4 and L5 Trojans identified in the MPC database (in blue).
This population contains a surplus of faint Trojans with $H\geq 14.5$ that were not detected by CSS but instead by 
larger-aperture surveys (e.g., Pan-STARRS) or the occasional dedicated efforts using large telescopes
(e.g., Subaru). We find it useful to display this dataset, even though it is obviously a biased sample, 
as it represents a reference basis. The bias-corrected population determined by our model must at every magnitude
exceed this total population. As before, the solution of the L5 population barely passes this criterion,
while that of the L4 population easily meets the test. This confirms the abundant nature of the L4 population 
over the L5 solution for reasons outlined above. 

The solution for the family populations may appear disappointing
at the first sight. Considering the limitations of our approach, however, we find it satisfactory how the major
families are reasonably reconstructed in both clouds. In particular, the Eurybates family in the L4 population
is well resolved and has a population of $385\pm 20$ members with $H<14.5$ as well as a cumulative magnitude slope
changing at the same reference magnitude from $0.78\pm 0.05$ to $0.60\pm 0.04$. Here, just the last segment
solution with a shallower slope is not exact. Similarly, the solution for the Deiphobus family in the L5
cloud indicates $210\pm 10$ members with $H<14.5$ and a change of the cumulative slope from $0.65\pm 0.12$
to $0.40\pm 0.09$. As in the Eurybates family case, the slope in the last segment is obviously again
underestimated. Apart from scarce set of observational data beyond $14.5$ in the high-inclination zone
of the L5 cloud, the additional complications are due to the Deiphobus family, which partially overlaps with the
Ennomos family. The {\tt MultiNest} solution tends to assign some of the Deiphobus members to the Ennomos
part. The solution of smaller families, with only tens to a hundreds of members, is less accurate
(Fig.~\ref{model_h_unbiased}). Recall that {\tt MultiNest} is only provided general information about
the position of the families in orbital space. It then objectively decides how to sort
the observations into the background and family populations. When the families are large and diffuse,
and when they contain only small to limited numbers of members, the solution is necessarily approximate.
Yet, it contributes some value to the model (note, for instance, the contribution of the small
L4 families to the orbital parameter distribution on the top panels of Fig.~\ref{model_abc_biased}).

Figures~\ref{model_h_unbiased}, and parameters listed in Table~\ref{model_back}, generally confirm
previous results concerning the magnitude distribution of Jupiter Trojans: (i) a steep segment
with a power-law exponent $\gamma\simeq 0.91$ to magnitude $\simeq (8.5-9)$, (ii) followed with a shallower
part characterized by a power-law exponent $\gamma \simeq (0.45-0.5)$ to a certain break-point $H_{\rm b}$,
(iii) beyond which the magnitude distribution becomes even shallower with $\gamma\simeq 0.36$
\citep[see, e.g.,][]{wetal2014,wb2015,yt2017,u2022}. The literature differs at $H_{\rm b}$, with values ranging
from $13.56^{+0.04}_{-0.06}$ \citep{yt2017} to $14.93^{+0.73}_{-0.88}$ \citep{wb2015}. In previous works, 
a rigid broken power law model has often been applied to the observations. We believe our spline representation
of the magnitude distribution is more flexible and accurate at the expense of only small increase in the number 
of free parameters (with an additional quality of fitting the background population). 

Results
in Table~\ref{model_back} indicate the reality is more complicated than a simple change of the power-law
slope at a certain break-point. The magnitude distribution in both Trojan clouds exhibits subtle
features, such as a slight change from shallower to steeper slope values in the third and fourth magnitude
segments.%
\footnote{We performed a quantitative test supporting the need to represent the magnitude distribution
 of the background population with cubic-splines in the following way. L4/L5 populations restricted to
 absolute magnitude $13.5$ were fitted using our core model and, alternatively, by a test approach in which
 the cubic-spline magnitude representation was replaced by the traditional broken power-law recipe 
 (Eqs.~\ref{hbak1} and \ref{hbak2}). Upon convergence we compared Bayesian-based evidence factors $\ln Z$
 of the best-fit solutions provided by {\tt MultiNest}. The relative preference of one over the other
 models is directly by $\exp\left(\Delta \ln Z\right)$ value. We obtained $5\times 10^{-3}$ for L4 and
 $10^{-5}$ for L5 in favor of the cubic spline model.}
Figure~\ref{gamaloc} shows the power law slope variations at the highest resolution, namely
$0.1$ magnitude of the bins used. The switch between steep and shallow regimes at $\simeq 9$ magnitude
is still the dominant feature. We can also conclude that the slope values of the L4 and L5 populations
are statistically equivalent up to $\simeq 14$ magnitude. Beyond this value we has less confidence in
the L5 solution. The slope exponent of the L4 population indicates a small but steady decrease
beyond magnitude $13$. This gradual change is perhaps at the origin of different $H_{\rm b}$ values
mentioned above. The local slope at $H=15$ is formally $\gamma=0.38\pm 0.01$. There is a small slope
difference of the L4 and L5 clouds in between magnitudes $9$ and $10$ with the L4 population having
a steeper trend.
\begin{figure*}[t!]
 \plotone{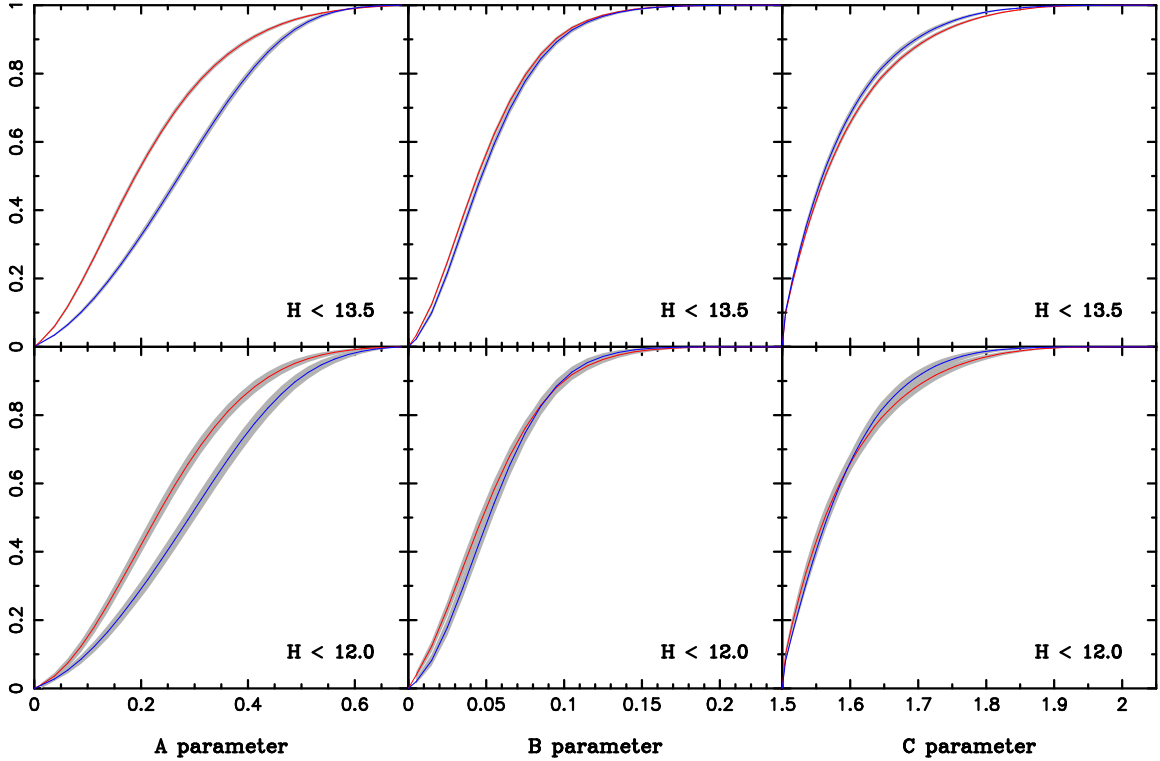}
 \caption{Cumulative distribution of the orbital parameters $(A,B,C)$ (left to right) based on our
  biased-corrected solution of the Trojan background populations: L4 in red and L5 in blue. Because
  the model parameters of the faint L5 Trojans are not reliable enough, we use a maximum $13.5$ magnitude limit
  in computing the upper plots; the lower plots provide the same information but for a population
  limited to absolute magnitude $12$. The color-coded curves are the best-fit solution, the gray
  region is a $\sigma$ interval. The $B$ and $C$ distributions appear to be statistically identical,
  but the $A$ (inclination) distribution is different.}
 \label{abc_cumu}
\end{figure*}
\begin{figure*}[t!]
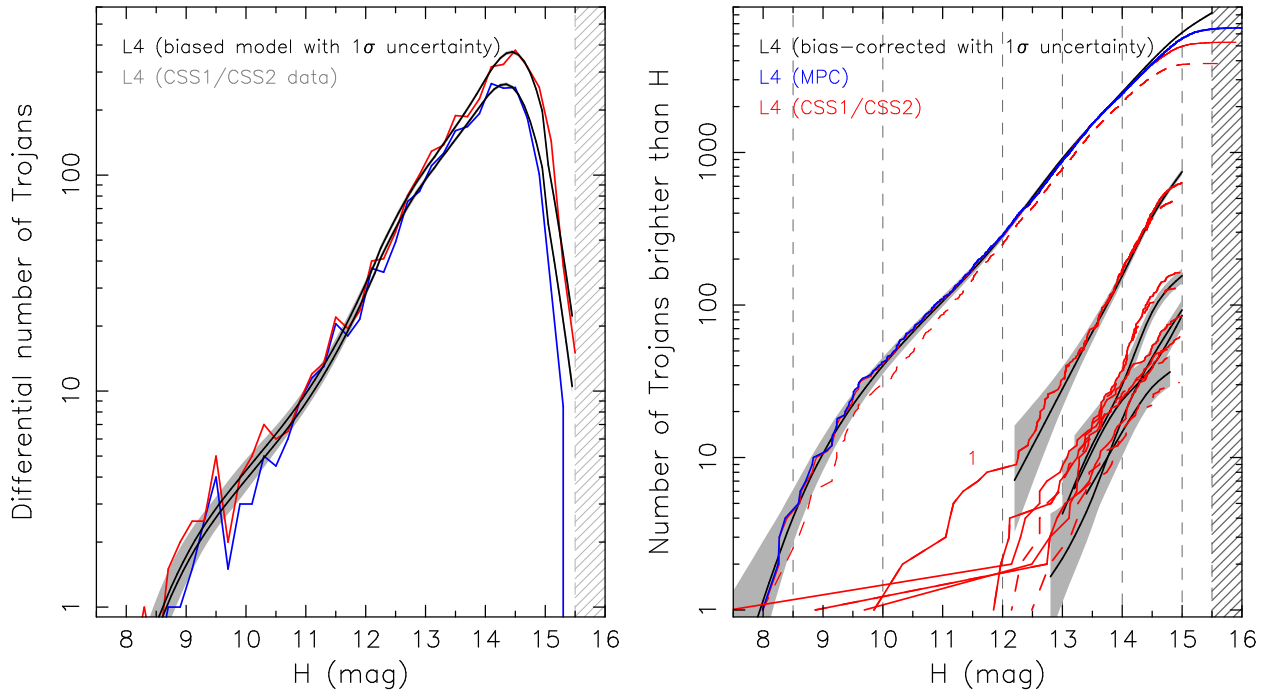

 \begin{center}
 \begin{tabular}{cc}
  \includegraphics[width=0.45\textwidth]{f27a.eps} &
  \includegraphics[width=0.45\textwidth]{f27b.eps} \\
 \end{tabular} 
 \end{center}  
 \caption{Solution of the L4 Trojan population using a model in which the maximum absolute magnitude
  has been extended to $15.5$ (compare with left panels on Figs.~\ref{model_h_biased} and
  \ref{model_h_unbiased}).
  Left panel: Differential magnitude distribution of the G96 observations (blue for phase~I and red
  for phase~II) compared with the best-fitting biased model (black line with a $\sigma$ interval depicted
  by the gray zone).
  Right panel: Bias-corrected cumulative magnitude distribution based on G96 observations. The background
  population, shown by the upper curves, is separated from the major families: (i) Eurybates (label~1),
  Arkesilaos, Hektor, Thronium and Teucer. The red lines are the CSS observations
  (solid are the data in the phase~II, dashed are the data in the phase~I), the black line is the
  bias-corrected model with $\sigma$ interval (gray zone). The blue line is the whole background
  population from the MPC catalog. The vertical dashed lines show the magnitude segments used
  for representation of the background population.}
 \label{model_ext}
\end{figure*}

We find that this behavior is formally at the origin of the population L4/L5 asymmetry in our model.
This is best seen on Fig.~\ref{asymetry}, where the left panel shows the bias-corrected differential
magnitude distributions of the L4 and L5 Trojan background populations. They are statistically
different beyond $H\simeq 11$. In particular, the L4 distribution appears to be shifted by $\Delta H
\simeq 0.4$ magnitude towards smaller $H$ if compared to the L5 distribution (the break of the
L5 population to a shallow trend beyond $H\simeq 14$ is an artifact of scarce dataset in this case
and should not be trusted). 

The origin of the population difference in magnitude limited samples is
unknown, leaving it open for speculation and hypothesis. For instance, assuming the underlying size
distributions
are the same for L4 and L5 Trojans, the inferred magnitude distribution shift may be caused by
$\simeq 1 - 10^{-0.4\Delta H}\simeq 0.44$ fractional difference in the mean albedo value, such that
small L4 Trojans would have a mean albedo of $\simeq 0.05$ and small L5 Trojans would have a mean
albedo of $\simeq 0.07$, as an example. While apparently not huge, such an albedo difference would
have been significant. There is no clear reason for this to happen. Data in \citet{grav2011} and
\citet{grav2012} indicate the albedo distribution for the Trojans becomes much wider below 
$\simeq 30$~km (approximately magnitude $11$ for $0.07$ albedo value), but there is no statistically
substantial difference between the L4 and L5 populations at these smaller sizes. Additionally,
recent observations focused on Lucy mission targets \citep[e.g.,][]{buie2018, buie2021, mot2023}
found that WISE albedo determinations for Trojans could be systematically too high (an issue
possibly related to the shape uncertainty).

Another possibility,
or rather a congruent effect, would be a slight difference in shape statistics of the L4 and L5 Trojans:
more irregular shapes tend to have a larger cross-section (and thus brightness) on average than
less irregular shapes. If the L4 population has undergone slightly more collisional evolution
than the L5 population \citep[see, e.g.,][]{obm2008}, resulting in a systematically more irregular shapes
of small objects, it would appear on average slightly brighter. \citet{neill2021} use the observations 
from ATLAS survey to support the case.

The asymmetry of the L4/L5 clouds in magnitude-limited population has been discussed
for more a decade. Possible evidence for some kind of asymmetry of the
orbital architecture of the L4 and L5 stability zones has yet to be identified. The above mentioned anomalies
in the inclination distribution \citep[e.g.,][]{jewitt2000,sb2014} are readily explained by
the population of the principal Trojan families. Separation of the families from the background has been
one of the major goals of this work. 

Therefore, we can now revisit the issue of a possible orbital asymmetry between the L4 and L5
populations more substantially by limiting our studies to the background populations
in both clouds. Upper panels at Fig.~\ref{abc_cumu} show the cumulative distribution of the
three orbital elements $A$, $B$ and $C$ for background populations with a magnitude cut at
$H=13.5$. We used this limit because we previously argued that the L5 solution is less reliable
for faint Trojans. These results have been obtained by specifically fitting the Trojan
populations restricted to this magnitude, but they appear fairly similar to those obtained
from the complete fit to limiting magnitude $15$ and downgrading to magnitude $13.5$.
Applying the Kolmogorov-Smirnov (KS) test to compare L4 and L5 data for distributions of the
three elements, we obtain $D=0.221$, $D=0.051$ and $D=0.043$ as a KS distance in the $A$, $B$ and $C$
elements. With the background population numbers from Table~\ref{model_back}, $N_4=1587$ and $N_5=1083$, 
which have been rescaled to $H=13.5$ limit, we obtain the probability of the distributions
being drawn from the same sample to be $\leq 10^{-20}$, $0.08$ and $0.17$ for $A$, $B$ and $C$
distributions. While not entirely robust, the distributions of the $B$ and $C$ elements appear
to be statistically similar, while the $A$ distributions are distinct. The extent of the formal
$\sigma$ areas about the best fit distributions, constructed from the posteriori variants of the
fit, conform the conclusion. To make sure these results are not affected by the choice of the
limiting magnitude, we repeated the test, restricting now the Trojan populations to magnitude
$H=12$ (bottom panels at Fig.~\ref{abc_cumu}). The cumulative distributions in all orbital
elements exhibit the same behavior as before. The KS distances of the best-fit solutions are only
slightly changed to $D=0.193$, $D=0.093$ and $D=0.084$ for the $A$, $B$ and $C$ distributions. 
The smaller populations at $12$ magnitude limit imply the KS probabilities change to
$2.5\times 10^{-4}$, $0.20$ and $0.35$ for the $A$, $B$ and $C$ distributions (see also wider
$\sigma$ regions, now basically overlapping in $B$ and $C$ cases). We therefore conclude that the
L4 and L5 distributions in eccentricity and semimajor axis (or the libration amplitude) appear to
be compatible with one another, while the distribution in inclination still appears
to be statistically inconsistent. Here again we leave a more thorough analysis and a possible
explanation for future studies.
\smallskip

\noindent{\it Variants of the solution.-- }We now explore the possibility of constraining the Jupiter Trojan
magnitude distributions beyond the limit $H=15$ adopted in the nominal model. For the reasons discussed
above, we will discard the L5 population and focus on observations from the L4 population. Keeping our
nominal setup, 
we only add one more magnitude segment in the spline representation between $15$ and $15.5$ magnitudes.

Figure~\ref{model_ext} shows (i) the biased model compared to the differential magnitude distribution of
the G96 observation in phases~I and II, and (ii) the bias-corrected cumulative magnitude distribution
for both the background and family components in the L4 cloud (compare with Figs.~\ref{model_h_biased} and
\ref{model_h_unbiased}). While visually the solution appears satisfactory to us, we believe it should be treated
with caution.
As expected from previous studies, beyond the $15$ magnitude limit the cumulative distribution becomes
shallow. The local magnitude slope at $H=15$ is now $0.30\pm 0.01$ and at $H=15.5$ becomes even $0.24\pm 
0.02$. Here we raise a flag of warning because the slope at $H=15$ is not statistically compatible with
that of the nominal model. Apparently, the {\tt MultiNest} algorithm has decided to choose a very shallow slope at
$H=15.5$, and this affects the solution at $H=15$ due to a rather small, half-a-magnitude last
segment. 

The left panel on Fig.~\ref{model_ext} shows that beyond magnitude $15$, the number of CSS detections
become limited: only a few tens of detections in phase~I and slightly more than 200 in phase~II. In this
situation, the fidelity of the detection probability determination is a critical element of the
model. Figure~\ref{fig_bias3} then shows that the nominal probabilities are $\leq 0.3$ even in phase~II
(and three times smaller in phase~I), where they depend on various details of the model. These could include
more in-depth analysis of the detection probability at very low values of the apparent motion, phase
function uncertainties, or further considerations about the choice of variables and detection probability
computation
(as an example, analysis in the Appendix~B allows us to consider our approach reasonable, but it could
certainly be improved). For that reason, we hold a study of a very faint end of the Trojan population
until further efforts have been made in the detection probability determination and more data has been acquired.

\section{Conclusions}\label{concl}
The main results of our work may be summarized as follows:
\begin{enumerate}
\item We developed a novel approach for description of the orbital architecture and magnitude
 distribution of Jupiter Trojan populations. The orbital part uses quasi-proper elements
 corresponding to the semimajor axis (or libration amplitude), eccentricity and inclination.
 The model takes a simple averaging over the complementary three orbital angles. The orbital
 space is segmented into 38,500 bins. Probability density distribution used by the model in the
 orbital space is represented by a simple convolution of three
 1-D analytical functions, and the long-term stability zone is determined using numerical
 integration. The magnitude part is described by cubic spline representation of the
 cumulative distribution on 6 segments, and it is parsed to 80 magnitude bins. In addition to
 the simultaneous analysis of the orbital and magnitude Trojan parameters, another novel aspect
 of or model is a separation of the Trojan population in clustered families and the smooth
 background.
\item New determination of Jupiter Trojan proper orbital elements and identification of Trojan
 families is a side result to our principal goal (see Appendix~A). We find evidence of 9 statistically
 robust families among L4 Trojans, and 4 statistically robust families among L5 Trojans. In the
 L5 population we helped to resolve the confusion that existed about the high-inclination Trojan clan (most often
 called the Ennomos family in the previous literature). Here we show it consists of two partially
 overlapping families about the largest members (1867) Deiphobus at higher inclination and
 (4709) Ennomos at lower inclination. This separation nicely matches conclusions of \citet{wb2023}
 where multiband photometric observations are used to investigate members of the Ennomos-Deiphobus
 clan. The morphology of the L4 and L5 families is surprisingly
 different: the L4 families are sharp and highly concentrated clusters, with possibly very low
 populations of interlopers, while the L5 families are diffuse and broad clans, with possibly a
 substantial population of interlopers. The origin of this difference is unknown, but possibly
 bears important clues about the past collisional evolution of these populations as well as
 their origin.
\item We applied our model to observations of CSS station G96 taken between January 2013 and
 June 2022. More than 220,000 documented fields-of-view were well characterized in terms of
 detection probability as a function of apparent magnitude and rate of motion. G96 camera upgrades
 in May~2016 made us split the survey period into phases I and II, which we consider as
 two independent surveys. During part~I, G96 detected 4,551 and 2,460 individual
 Jupiter Trojans in the L4 and L5 populations. These numbers increased during part~II
 to 6,307 and 3,041 individual Jupiter Trojans detected in the L4 and L5 populations. By
 June~2022, G96 had detected more than $85$\% of the known populations in the two Trojan clouds.
 We developed a method to determine the detection probability defined over the orbital element 
 and absolute magnitude range used by our model. This probability is a convolution of the aforementioned
 detection probabilities within a given field-of-view and the probabilities that occur in any of the
 survey fields-of-view. 
\item We applied the {\tt MultiNest} code to determine which parameters within the model are needed to 
 reach a maximum likelihood match to the CSS observations. This outcome leads to a suitably robust algorithm
 to tackle
 the multi-parametric task with existing correlations. Conveniently, {\tt MultiNest} also provides
 posterior distributions of the solved-for parameters, enabling us to evaluate the statistical significance
 of the model and its components.
\item Our results are in a good agreement with previous inferences about the magnitude distribution
 of the Trojan population. Traditionally a simple broken power-law model with one or two breakpoints has
 been used. Thanks to the cubic spline representation of the magnitude distribution, our analysis is
 more detailed and shows evidence of fine variations of the slope exponent (e.g., a shallow local
 maximum between magnitudes $12$ and $13$; Fig.~\ref{gamaloc}). Between magnitudes $14.5$ and $15$ the
 power-law slope 
 becomes shallower, reaching $0.38\pm 0.01$ at $H=15$. The available CSS observations of the L5 population
 are less numerous, making our ability to clearly separate the family populations less certain. 
 For those reasons the L5 solution should be considered less reliable beyond magnitude $14-14.5$. We
 tried to extend the
 L4 population solution beyond $H=15$, but the CSS data to-date are insufficient to characterize the
 very faint end of the population. 
\item Our results confirm the previously debated asymmetry in magnitude limited populations of the
 L4 and L5 Trojans. However, unlike before, we evaluated the asymmetry measure using the background
 population of Trojans, objectively separating the population in the Trojan families. At magnitude
 $H=15$ we find $f_{45}=N_4(<H)/N_5(<H)=1.43\pm 0.05$. The origin of this asymmetry is unknown. 
 The end-member approach in this respect is to associate it with the possibility of a large-scale
 inward migration of Jupiter and in-situ formation of Trojans \citep[e.g.,][]{pirani2019a,pirani2019b}
 \cite[there are, however, arguments against such a model, e.g.,][]{dei2022}. While effects of a more
 limited Jupiter migration (or a jump from Jupiter via an encounter with an ice giant) could play 
 some role \citep[e.g.,][]{houCMDA2016,li2023,li2023b},
 there are also additional possibilities, such as a slight albedo and/or shape irregularity difference between
 the small Trojans in the L4 and L5 populations (e.g., Sec.~\ref{solution}). The population asymmetry
 therefore requires further study in the future.
\item Our analysis suggests the possibility of an asymmetry in the orbital inclination of the Jupiter Trojans,
 with the L4 population having a tighter inclination distribution and the L5 population having a broader one.
 However, since the L5 population solution is less reliable (for reasons mentioned above), we consider
 the inclination distribution asymmetry an interesting hypothesis that needs to be verified when problems
 of the L5 population modelling are improved in the future.
\end{enumerate}

There were fewer and smaller magnitude-reaching observations of the Jupiter Trojan L5 population in
the G96 dataset between 2013 and 2022 (see Sec.~\ref{datag96}). As a result, our ability to characterize
this cloud was limited in this paper.
The observational conditions of L5 populations significantly improved from the 2023 season, which is 
motivation to revisit our Jupiter Trojan analysis in a few years time. Continuing operations
of CSS could provide the basis of this effort. They could be readily complemented with data from other,
well characterized surveys, such as the powerful Vera C. Rubin Observatory \citep[e.g.,][]{schwa2023} or
the upcoming NASA mission NEO Surveyor \citep[e.g.,][]{surv}.  


\acknowledgments
 We thank the referee whose insightful comments helped us to improve the original version of this paper.
 CSS operations are currently funded under grant 80NSSC21K0893-NEOO. The simulations were performed on
 the NASA Pleiades Supercomputer. We thank the NASA NAS computing division for continued support. The
 work of DV and MB was partially supported by the Czech Science Foundation (grant~21-11058S). DN's work
 was supported by the NASA Solar System Workings program.  WB's work in this paper was supported by
 NASA’s Solar System Workings program through Grant 80NSSC18K0186 and NASA’s Lucy mission through
 contract NNM16AA08C.
 
\bibliographystyle{aasjournal}

\appendix
\begin{figure*}[t!]
 \plottwo{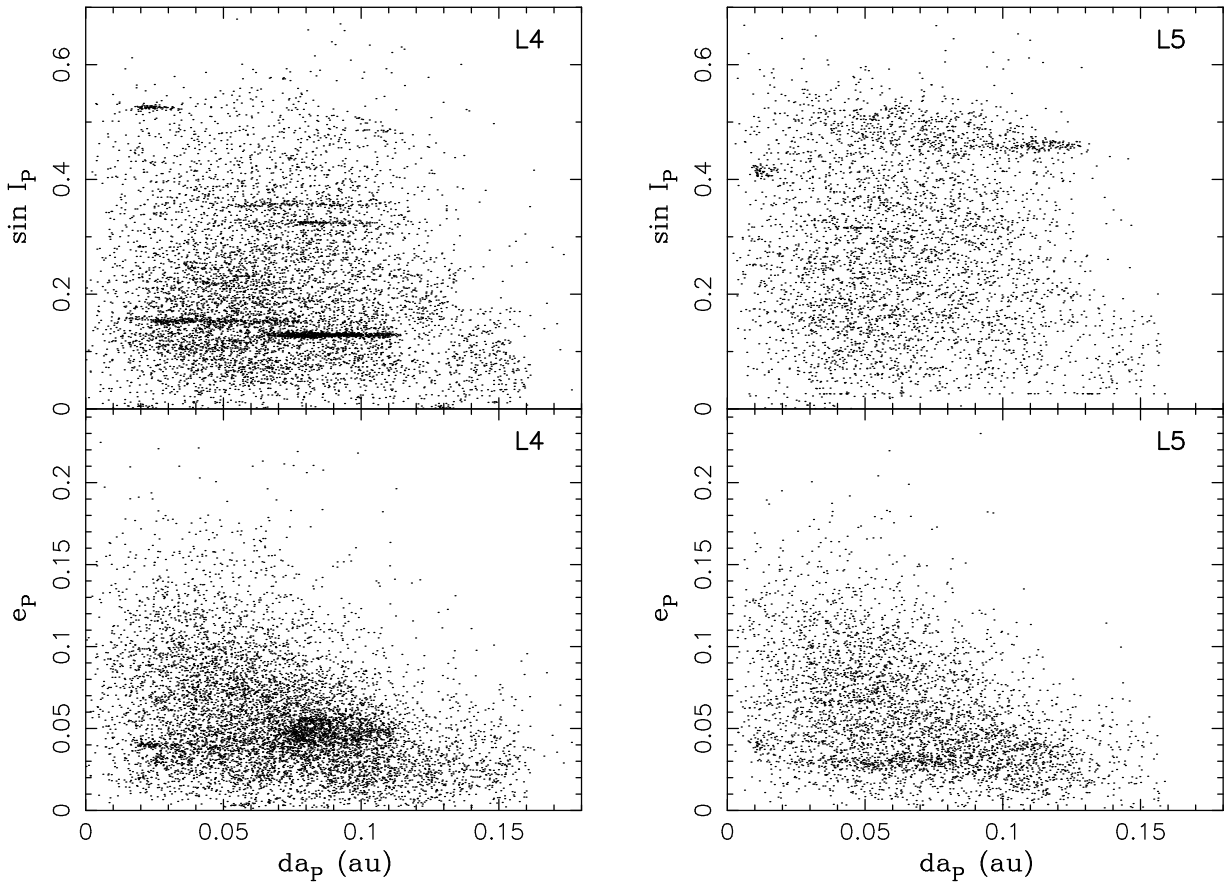}{fAppA1b.eps}
 \caption{The newly determined proper elements of L4 (left) and L5 (right) Jupiter Trojans with the
  input orbital information as of April 2023. Top panels show a projection onto the $da_{\rm P}$ vs $\sin I_{\rm P}$
  plane, bottom panels show a projection onto the $da_{\rm P}$ vs $e_{\rm P}$ plane. Distinct clusters,
  especially among the L4 Trojans, are subject of our analysis resulting in a new catalog of the
  Trojan families.}
 \label{fig_prop_mb}
\end{figure*}

\section{Appendix: New catalog of Trojan proper elements and identification of the families}
\label{properel}
Our approach to determine proper elements of Jupiter Trojans closely follows the synthetic
theory of \cite{m1993} (see also technical steps outlined in the Appendix~A of \citet{holt2020b})
and applies it to the currently known population of objects.%
\footnote{For sake of brevity, we thus relegate an interested reader to \cite{m1993} and \citet{holt2020b}
 to learn about technical details, while here we discuss the results. The final products, namely (i) the
 catalog of the Trojan proper elements, and (ii) identification of the Trojan families, are available at
 \url{https://sirrah.troja.mff.cuni.cz/~mira/tmp/trojans/}.}
In this case, we used the MPC catalog of minor planet orbits as of April~15, 2023 containing about 1,275,000
entries, and selected Jupiter Trojans using the same steps as described in Sec.~\ref{datag96}. We
obtained 8,144 orbits in the L4 swarm, and 4,252 orbits in the L5 swarm. The slight increase with respect to
the numbers given in Sec.~\ref{datag96} reflects (i) partly the new detections during the spring 2023, but
(ii) more importantly the fact that now we did not drop the orbits with short observation arcs (except
the really poorly determined ones with arcs less than a few days).

\begin{figure*}[t!]
 \begin{center}
 \begin{tabular}{cc}
  \includegraphics[width=0.47\textwidth]{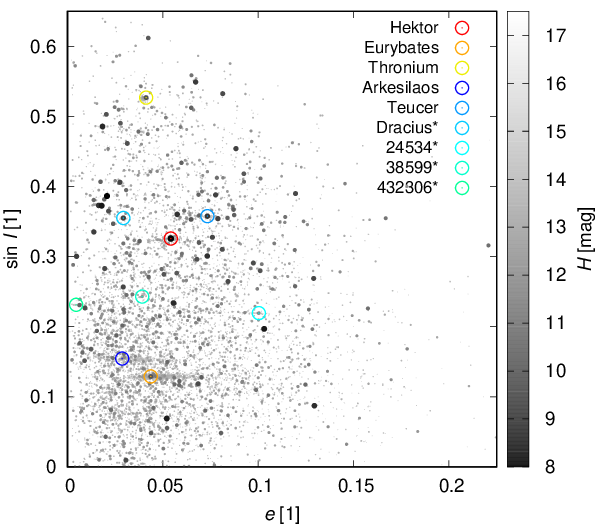} &
  \includegraphics[width=0.47\textwidth]{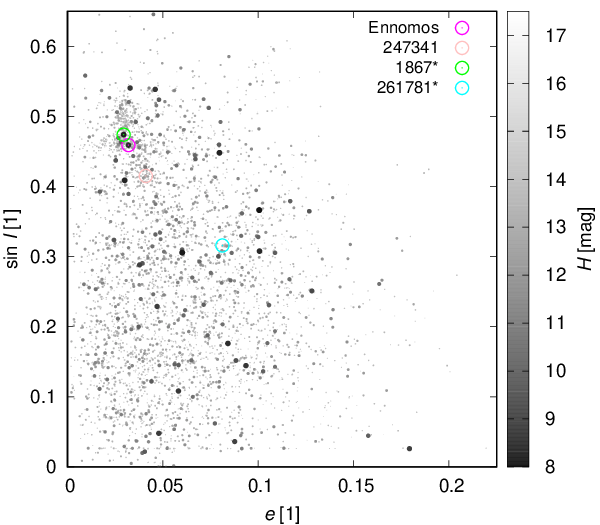} \\
 \end{tabular}
 \end{center}  
 \caption{Trojan families highlighted in the plane of the proper eccentricity $e_{\rm P}$ and the proper
  sine of inclination $\sin i_{\rm P}$ (compare with the rightmost panel on Fig.~\ref{fig7bis});
  the objects librating about the leading point L4 on the left
  panel, those librating about the trailing point L5 on the right panel. The scale of gray and size of 
  symbols are proportional to the absolute magnitude $H$ (see the vertical bar). The family location is
  indicated by the largest remnant (color-coded open circles and labels); the newly discovered families
  are identified by the asterisk (*).}
 \label{proper_ei_2023}
\end{figure*}

Figure~\ref{fig_prop_mb} shows projection of the Trojan proper elements onto the $(da_{\rm P},e_{\rm P})$
and $(da_{\rm P},\sin I_{\rm P})$ planes. The principal features to be noticed include (i) the
correlation between the maximum proper eccentricity and $da_{\rm P}$, which has been known
since the works of E.~Rabe \citep[see, e.g.][]{r1965,r1967,lev1997} (see also Figs.~\ref{fig_stab1a}
and \ref{fig_stab1b}), and (ii) statistically
distinct orbital clusters, principally among the L4 population. In order to characterize the
latter in a quantitative way, we follow the traditional steps \citep[see, e.g.,][]{m1993,br2001}.
First, we adopt the $d_3$ metrics to define distances of the orbits in the proper elements space
and we use the Hierarchical Clustering Method (HCM) to discern these principal families in the
population. Our results confirm and extend those in \citet{br2011}, \citet{retal2016},
\cite{vin2019} and \cite{vino2020}.
In particular, we identify the previously known families and extend their membership towards
Trojans with smaller size (obviously, this is the harvest of the powerful sky surveys from the
last several years). Additionally, we see evidence for new, previously not known families
(see the Table~\ref{mb_fams_2023} for a summarizing information). Figure~\ref{proper_ei_2023} highlights
location of the identified families in the projection onto the proper element $(e_{\rm P},\sin I_{\rm P})$
plane, suitable for their visualization (this is because they are often most extended in
the $da_{\rm P}$ coordinate --or the libration amplitude-- which is the least stable of the proper
elements). Pushing the knowledge of membership in Trojan families towards smaller objects provides
important hints about their contribution to the total Trojan population as a function of the absolute
magnitude. As already mentioned in the main text, we selected the prominent families in each of
the Trojan swarms, and remapped their location from the proper element space to the space
of $(A,B,C)$ orbital parameters (see data in the Table~\ref{fams}).
\begin{figure}[t!]
 \plotone{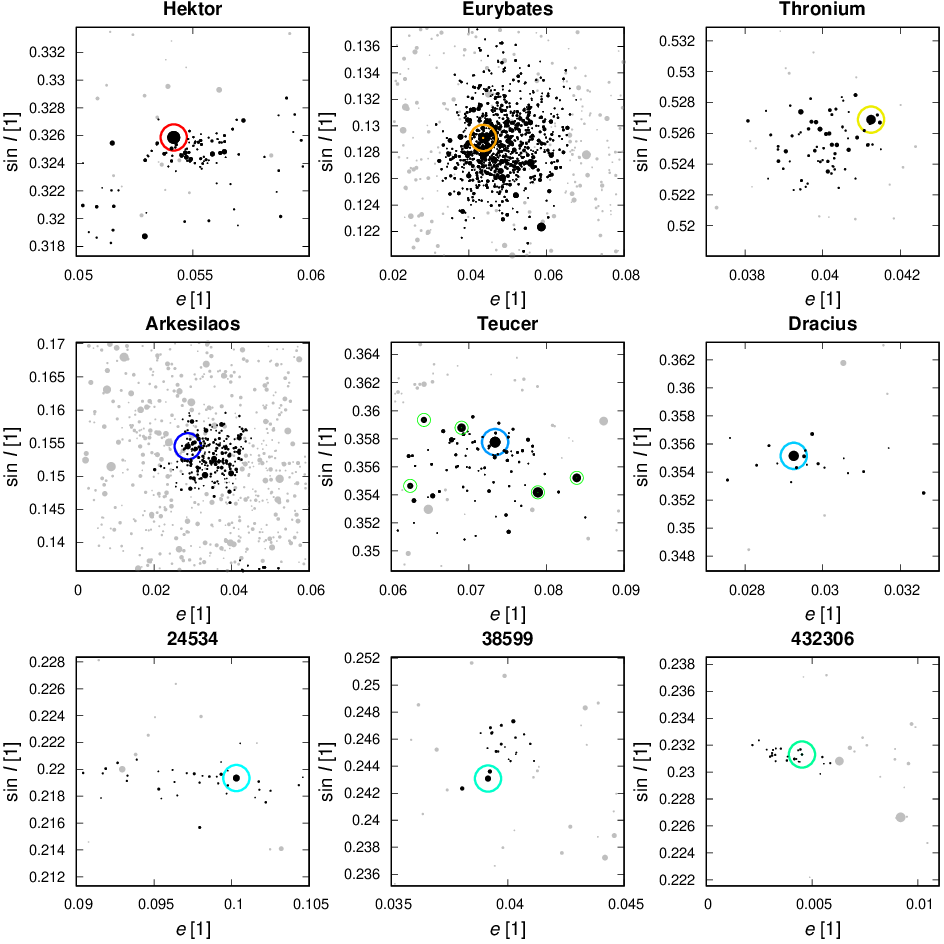}
 \caption{The same as Fig.~\ref{proper_ei_2023} (left panel), but now zoomed on the individual families
  among the L4 Trojans. The family members are black (the symbol size proportional to the physical
  size of the body), non-members projecting onto the same $(e_{\rm P},\sin I_{\rm P})$ zone (but having
  offset the $da_{\rm P}$ third coordinate) are shown in gray. The family identification is at the
  top label, position of the largest remnant is highlighted by the open circle. The low-inclination
  families ($\sin I_{\rm P}<0.2$) are typically resolved by having a large population of members, the
  high-inclination ones may be resolved more easily, because of low background population of Trojans.
  In the case of Hektor family (top and right panel), we show its dense core consisting of only
  very small fragments, while previous studies considered its more extended version
  \citep[e.g.,][]{retal2016}.}
 \label{proper_ei_2023_fam_l4}
\end{figure}
\begin{figure}[t!]
 \plotone{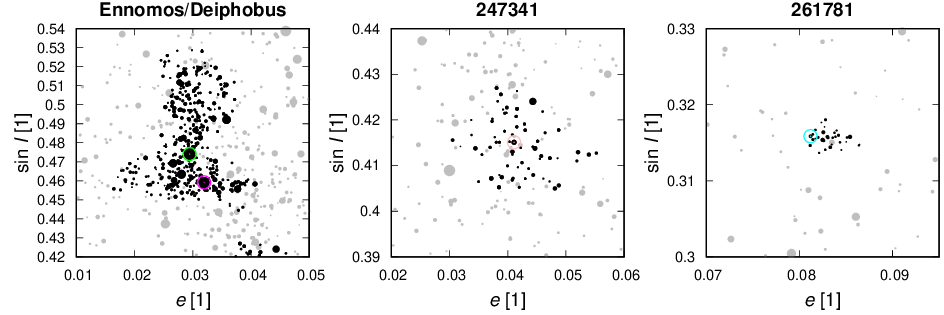}
 \caption{The same as in Fig.~\ref{proper_ei_2023_fam_l4}, but now for the three families identified among
  the L5 Trojans. We argue that the former Ennomos family, shown on the leftmost panel, is in fact
  composed of two overlapping families: (i) Ennomos, at the right-bottom corner, and (ii) Deiphobus,
  extending towards the high-inclinations.}
 \label{proper_ei_2023_fam_l5}
\end{figure}
\begin{deluxetable*}{rlrrrrr}[t] 
 \tablecaption{\label{mb_fams_2023}
  Statistically significant families in the L4 and L5 Trojan populations determined using the HCM and our
  new catalog of synthetic proper elements. Newly discovered clusters are indicated by asterisk symbols (*).}
 \tablehead{
  \colhead{number} & \colhead{designation} & \colhead{$v_{\rm cutoff}$} &
  \colhead{$N_{\rm mem}$} & \colhead{$D_{\rm LR}$} & \colhead{$D_{\rm LF}$} & \colhead{$D_{\rm Durda}$} \\ [-2pt]
  \colhead{} & \colhead{} & \colhead{(m s$^{-1}$)} & \colhead{} & \colhead{(km)} &
  \colhead{(km)} & \colhead{(km)}
 }
\startdata
 \multicolumn{7}{c}{{\it -- Families in the L4 cloud --}} \\ [2pt]
   624$\phantom{*}$ & Hektor\tablenotemark{a}       & 40 & 118 & 230  &  21  & 250 \\
  3548$\phantom{*}$ & Eurybates                     & 40 & 875 &  68  &  53  & 140 \\
  9799$\phantom{*}$ & Thronium                      & 20 &  69 &  72  &  22  & 120 \\
 20961$\phantom{*}$ & Arkesilaos\tablenotemark{b}   & 40 & 235 &  23  &  21  & 110 \\
  2797$\phantom{*}$ & Teucer\tablenotemark{c}       & 60 &  86 & 113  &  17  & 130 \\
  4489*             & Dracius                       & 60 &  24 &  95  &  15  &     \\
 24534*             & 2001 CX27                     & 60 &  45 &  29  &  15  &     \\
 38599*             & 1999 XC210                    & 60 &  22 &  20  &  12  &     \\
432306*             & 2009 SQ357\tablenotemark{d}   & 40 &  24 &   7.3&   6.8&     \\ [3pt]
  \multicolumn{7}{c}{{\it -- Families in the L5 cloud --}} \\ [2pt]                                                 
  4709$\phantom{*}$ & Ennomos\tablenotemark{e}      & 50 &  88 &  80  &  20  & 110  \\
  1867*             & Deiphobus\tablenotemark{e}    & 60 & 233 & 131  &  24  & 180  \\
247341$\phantom{*}$ & 2001 UV209\tablenotemark{f}   & 55 &  46 &  18  &  16  &  90? \\
261781*             & 2006 BG132\tablenotemark{g}   & 60 &  31 &  14  &  13  &      \\
\enddata
\tablenotetext{a}{Alternatively, only a tight cluster of small bodies at the HCM velocity of
 $20\,{\rm m}\,{\rm s}^{-1}$ around (624) Hektor.}
\tablenotetext{b}{\cite{vino2020} opted to associate this cluster to (2148) Epeios, formerly also by
 \citet{br2001}.}
\tablenotetext{c}{Identified by \cite{vin2015}, formerly also by \citet{m1993}.}
\tablenotetext{d}{This cluster contains only small objects; large Trojans (4035) Thestor and (6545) Leitus are
 nearby, but offset in proper semimajor axis (see also Fig.~\ref{proper_ei_2023_fam_l4}).}
\tablenotetext{e}{The earlier studies \citep[e.g.,][]{br2011,retal2016} had Ennomos as a single large,
 but diffuse and spectrally controversial family-candidate in the L5 cloud. \cite{vino2020} associated
 this cluster with (1867) Deiphobus. We propose the former Ennomos-cluster consists of two overlapping
 families associated with (4709) Ennomos and (1867) Deiphobus largest objects; this is also nicely supported by
 the detailed color analysis in \citet{wb2023}.}
\tablenotetext{f}{This family has been associated with (37519) Amphios in \cite{vino2020}, but
 (247341) 2001 UV209 resides closer to the family center.}
\tablenotetext{g}{\citet{vino2020} mentions a cluster around (1172) Aneas, which is close to this cluster,
 but offset in the proper eccentricity.}
\tablecomments{The number and designation correspond to the central body, $v_{\rm cutoff}$ is the velocity
 cutoff used in the HCM identification, $N_{\rm mem}$ is the number of associated members, $D_{\rm LR}$ is the size
 of the largest remnant, $D_{\rm LF}$ is the size of the largest fragment, and $D_{\rm Durda}$ is the size of the
 parent body estimated using the method of \cite{durda2007}.}
\end{deluxetable*}

In what follows, we provide a brief overview of the novel findings about the Trojan families.
Similarly to the case of the total observed populations, the gradual increase in the asymmetry
of the Trojan families among the L4 and L5 swarms arose during the recent decade. Here we report nine
cases in the L4 population versus only four in the L5 population. Additionally, their morphology
seems different: sharp and concentrated families among the L4 Trojans, versus rather broad
and diffuse families among the L5 Trojans.

The Eurybates family in the L4 cloud remains to be the most pronounced of them, itself representing
more then $\simeq 10$\% of the population at magnitude $H\simeq 14.5$. This is because, as shown on
Fig.~\ref{fig9}, the Eurybates family magnitude distribution keeps to be steeper than the
background population till that magnitude limit. \citet{retal2016} payed a close attention to a
broad cluster of Trojans about the historically third discovered Trojan object (624) Hektor
\citep[e.g.,][]{wolf1907a,wolf1907b,strom1907}.
Their motivation was multiple: (i) Hektor has a satellite, possibly formed along with the family,
(ii) Hektor, and a few smaller fragments, have a rare D-type taxonomy among the asteroid families,
and (iii) Hektor family was the second most populous at the time. Here, we find that the newly
discovered small Trojans in the Hektor family zone preferentially occupy an immediate vicinity
of the largest fragment, constituting thus its
dense core. It is not clear whether this structure follows from the original family formation
event, or if it arises from a secondary cratering on (624) Hektor more recently. Further studies
of this interesting cluster are warranted. New data about the Arkesilaos family small members continue
the trend of a steep size distribution predicted in \citet{retal2016}. In our identification, it is
already the second most populous cluster in the L4 swarm (see also Fig.~\ref{fig9}). When plainly 
extrapolated to even smaller fragments, this family may rival the Eurybates family at $H\simeq 
(16.5-17)$. Finally, the cluster about the L4 Trojan (9799) 1996~RJ, first discovered by \citet{br2011},
remains to be one of the fairly distinct families (the low background population at its very
high inclination helping its identification). Because this object obtained a final designation, we call
it the Thronium family.

\citet{br2011} reported a discovery a broad cluster of L5 Trojans about the largest object (4709)
Ennomos. Yet, its diffuse nature and spectral/albedo peculiarity immediately rose questions about
its legitimacy and the interloper density. They have been recently carefully tested by dedicated
broadband photometric observations of \citet{wb2023}. These authors found that many of the proposed
members in the Ennomos family are spectrally indistinguishable from the overall background
population, except a limited number of Trojans in the Ennomos vicinity. \citet{vino2020}, also
observing a the extended nature of this cluster, took the liberty to associate it with another
large L5 Trojan (1867) Deiphobus. Our results support this shifting point of view, or rather make it
more accurate. When picking Ennomos and Deiphobus
as the largest remnants in their respective families, we observe they keep distinct at small HCM
$v_{\rm cutoff}$ velocities and merge only when it exceed $60$ m~s$^{-1}$. The Ennomos family occupies
smaller $\sin i_{\rm P}$ and larger $e_{\rm P}$ zone, the Deiphobus family extends to higher $\sin i_{\rm P}$ 
zone (Fig.~\ref{proper_ei_2023_fam_l5}). Additionally, the Ennomos part is more concentrated at
larger libration amplitude (or $da_{\rm P}$), helping it separate from the Deiphobus part at smaller
libration amplitude in the 3D proper element space. This nicely matches results of the photometric
observations of \citet{wb2023}.

The newly discovered families are typically compact clusters with a limited number of known and small
Trojans. See, for instance, families about (38599) 1999~XC210 or (432306) 2009~SQ357 in the L4 swarm, or
the family about (261781) 2006~BG132 in the L5 swarm (Figs.~\ref{proper_ei_2023_fam_l4} and
\ref{proper_ei_2023_fam_l5}).
\begin{figure*}[t!]
 \plottwo{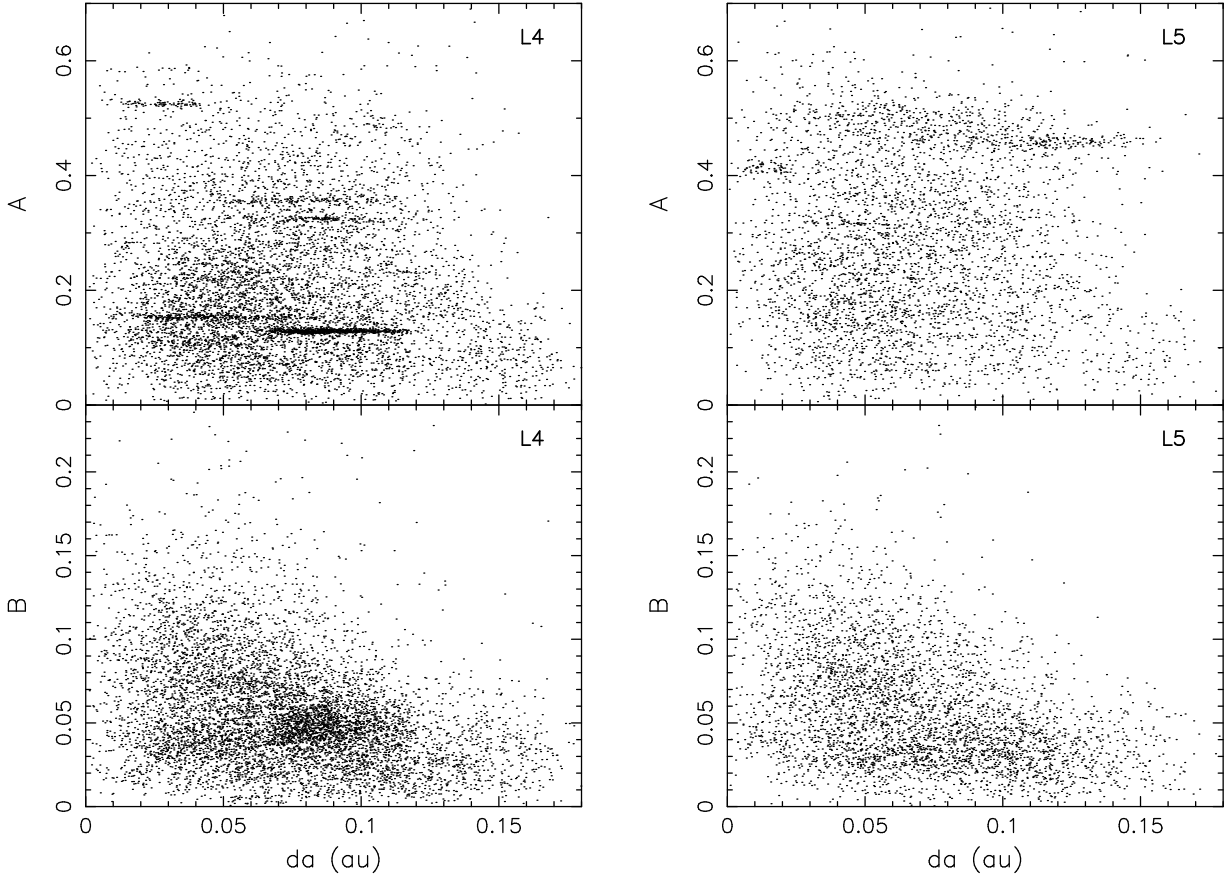}{fAppA5b.eps}
 \caption{Population of L4 (left) and L5 (right) Jupiter Trojans from the MPC catalog (as of February~22, 2023)
  projected onto the ``quasi-proper'' elements used in our debiasing procedure: (i) $da_{\rm P}(C)=a_{\rm J}
  \sqrt{8\mu\,(C-1.5)/3}$ vs $A$ (top), and (ii) $da_{\rm P}(C)=a_{\rm J}\sqrt{8\mu\,(C-1.5)/3}$ vs $B$
  (bottom). While these variables are much simpler than the true proper elements, they provide a picture
  nearly as accurate (compare with Fig.~\ref{fig_prop_mb}).}
 \label{fig_abc}
\end{figure*}

It is interesting to compare architecture of the Trojan orbits in the space of proper orbital elements
described above to that described by the ``quasi-proper'' variables $(A,B,C)$. This is similar to
Fig.~\ref{fig7bis}, but here --to allow a more direct correspondence to Fig.~\ref{fig_prop_mb}--  we use
$da_{\rm P}(C)=a_{\rm J}\sqrt{8\mu\,(C-1.5)/3}$ to replace $C$. The result is shown in Fig.~\ref{fig_abc}. While the
proper elements are certainly more accurate quasi-integrals of motion, we note that Trojan representation
in the space of $(A,B,C)$ parameters captures are important features of the population. Most importantly,
Trojan families are very well reproduced in these simplified variables. This justifies their
applicability for the population debiasing efforts.
\begin{figure*}[t!]
 \plotone{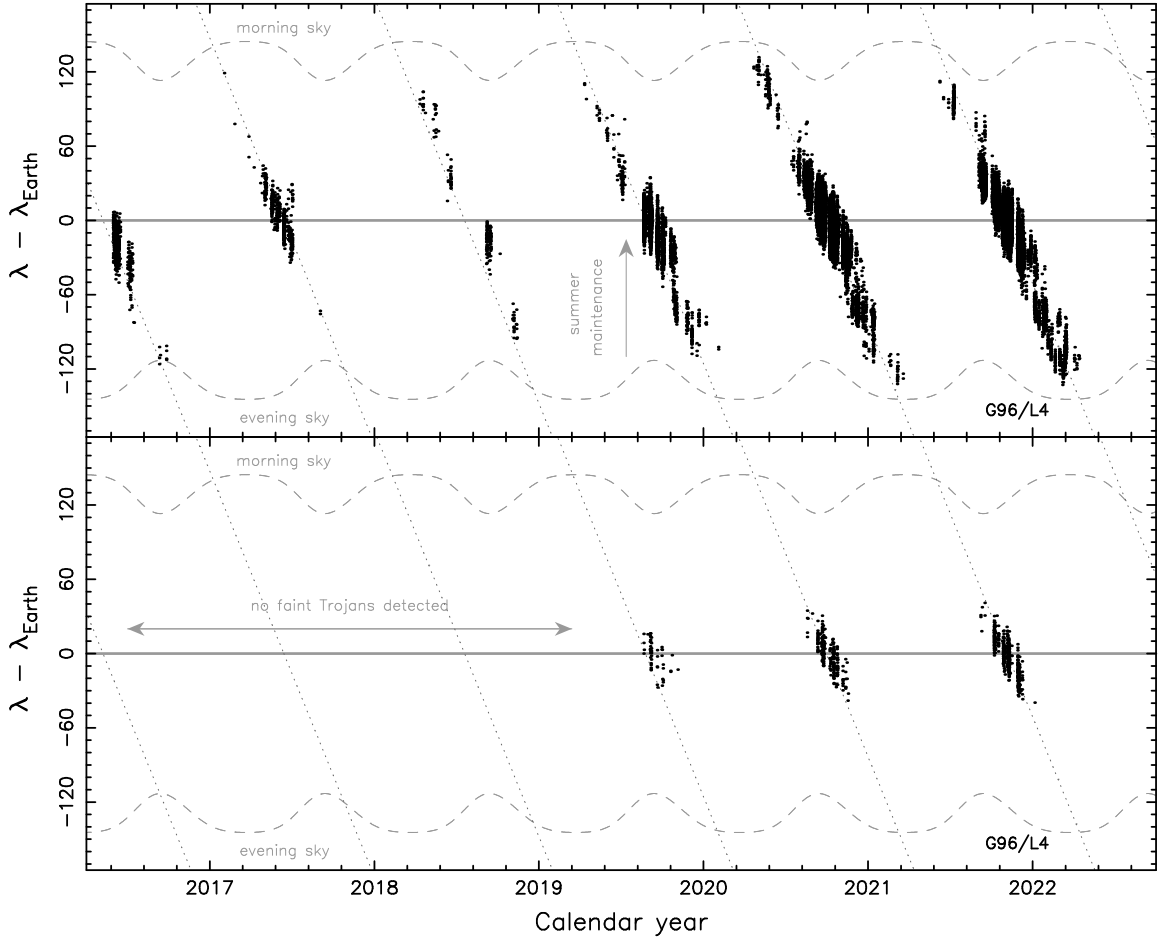} 
 \caption{Detections of the L4 Trojans during the phase~II of the CSS observations (calendar
  date at the abscissa; see also Fig.~\ref{fig2}). The ordinate shows the Trojan ecliptic longitude
  $\lambda$ referred to the ecliptic longitude of the Earth $\lambda_{\rm Earth}$ at the moment of
  detection. The upper panel for
  $H\leq 14$ Trojans (population of large and bright objects), the lower panel for $H\geq 15$ Trojans
  (population of small and faint objects). The dashed curves are the extremal configurations defined by
  (i) the Sun being $10^\circ$ below the local horizon, (ii) the Trojan being at the ecliptic and
  $15^\circ$ above the local horizon (the cyclic nature is due to longer nights as well as higher
  maximum position of the ecliptic above the horizon in winter). At each season the Trojans start to
  be observable on the
  extreme morning sky, and over a period of little less than a year they move to the extreme evening
  sky. This is due to the Earth faster mean motion about the Sun, such that the L4 swarm near the
  Jupiter's libration center is being caught by the Earth-bound observed; the dotted line has the
  relevant synodic frequency $n-n_{\rm J}$. The scatter of detections in $\lambda-\lambda_{\rm Earth}$
  at any given moment corresponds to Trojans at different libration amplitude about L4. A combination
  of geometric, CSS-operation dependent and photometric selection effects impact detections. Bright
  Trojans are detectable during the whole season, unless galactic plane interferes (pre-2019 observations;
  see Figs.~\ref{fig1} and \ref{fig2}), or the observations are interrupted by summer maintenance (see the example
  in 2019 shown by the arrow). Faint objects are observed selectively near the opposition ($\lambda\simeq
  \lambda_{\rm Earth}$) which decreases their apparent magnitude in both factors on the right hand side of
  Eq.~(\ref{pogson}): (i) the product $R\Delta$ is minimum, as well as (ii) the magnitude-phase correction
  $P(\alpha)$. Due to the galactic plane disturbance no faint Trojans near L4 were detected during the 
  phase~II until the opposition in the late 2019.}
 \label{fig_dlam_l4}
\end{figure*}
\begin{figure*}[t!]
 \plotone{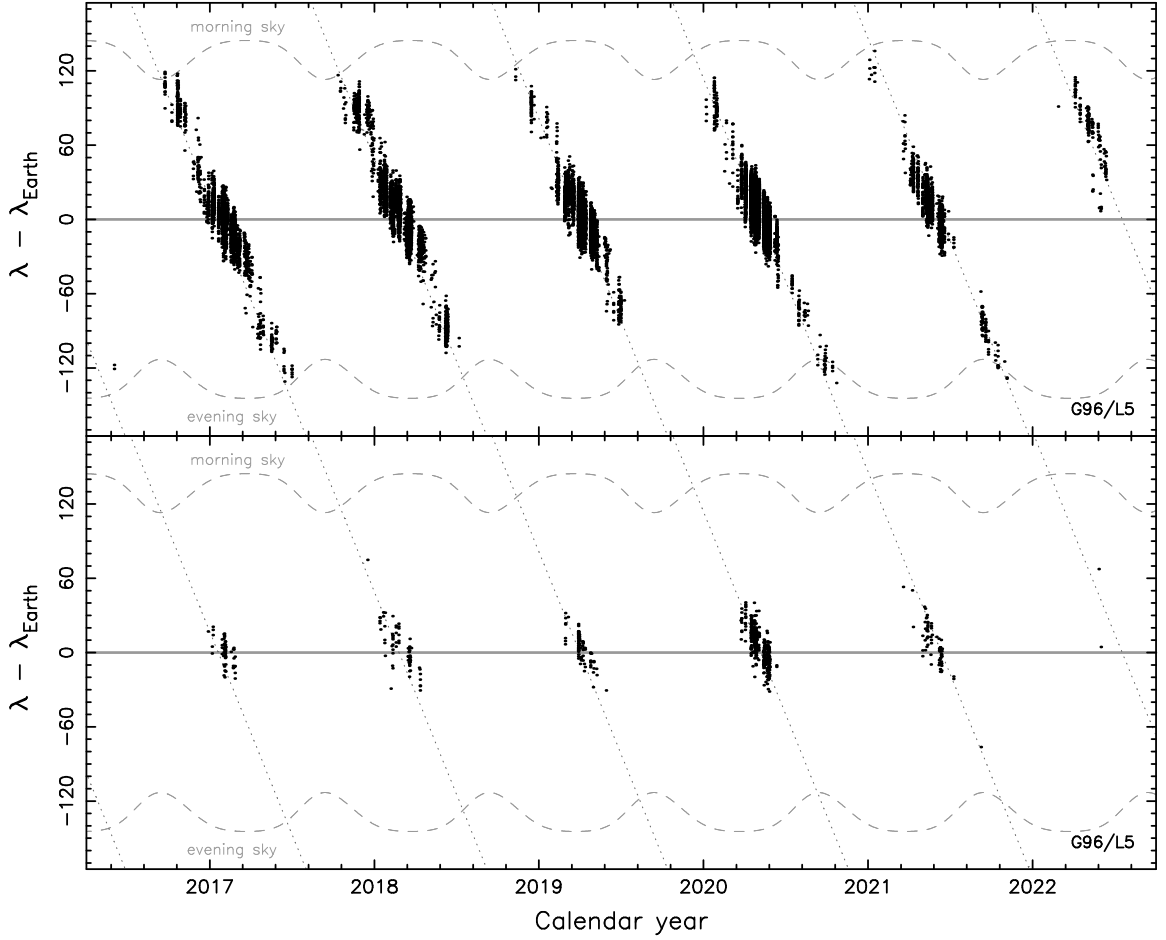} 
 \caption{The same as in Fig.~\ref{fig_dlam_l4}, but now for the L5 Trojans. The upper panel for detections
  of the $H\leq 14$ Trojans, the lower panel for detections of the $H>14.5$ Trojans.}
 \label{fig_dlam_l5}
\end{figure*}

\section{Inferences of the detection bias from the observations}\label{appbias}
Following information in Sec.~\ref{datag96}, we provide few more examples of the observation
biases affecting Trojan detections by CSS that need to be accounted for and removed when
estimating the complete populations near L4 and L5 libration centers. We also test our
formulation of the detection probability ${\cal P}(A,B,C;H)$ introduced in Sec.~\ref{secbias},
seeking justification of averaging over the proper angle $\psi$ associated with the
eccentricity parameter $B$.
\begin{figure*}[t!]
 \plottwo{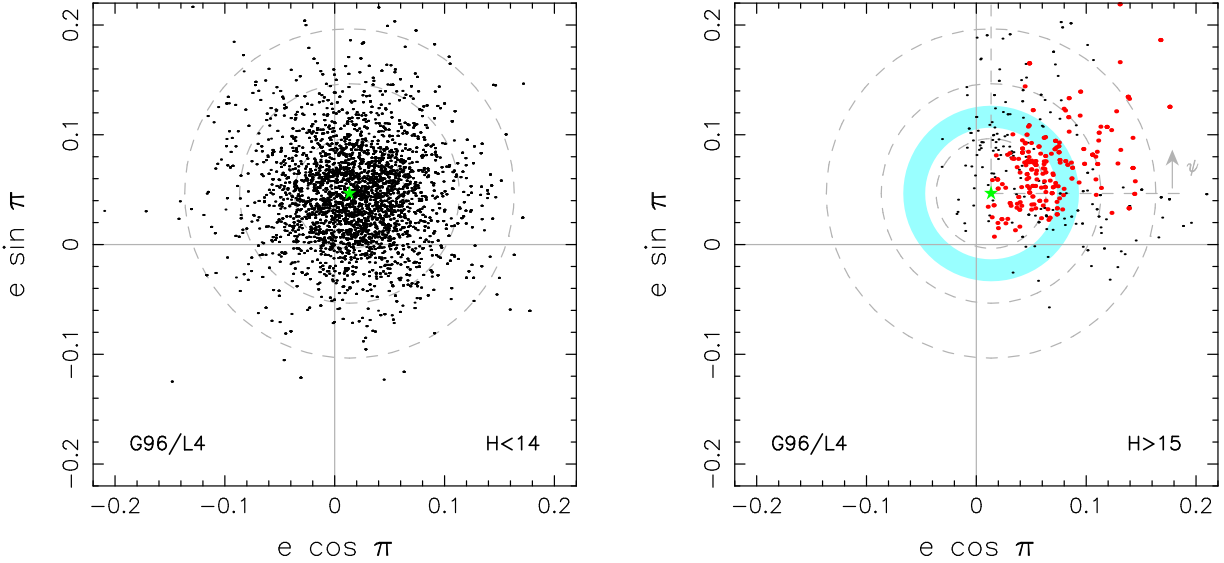}{fAppB3b.eps}
 \caption{Trojans in the L4 swarm detected by the CSS in the April 2021 and April 2022 season
  (see Fig.~\ref{fig2} and \ref{fig_dlam_l4}), projected on the plane of osculating non-singular
  elements $(e\cos\varpi,e\sin\varpi)$. Left panel for bright objects with $H<14$, right panel
  for faint objects with $H>15$. The green star is the forced center of the quasi-proper elements
  $(B,\psi)$, namely at $e_{\rm J}$ polar distance from the center and polar angle $60^\circ+
  \varpi_{\rm J}$ (Eq.~(\ref{propere})); several constant $B$ circles are shown for sake of
  illustration. The bright objects are distributed symmetrically about the forced center. The faint
  objects are mostly detected in the first quadrant, thought clearly their complete population
  should follow the symmetric distribution shown by the bright objects. The symbols on the right
  panel are colored in red, when the detection occurred at $20^\circ$ angular distance from the
  pericenter; this is the case of nearly all detected faint objects. The light blue annulus
  depicts $B\in (0.06,0.08)$ interval of values that we have chosen for testing the validity of
  the detection probability ${\cal P}(A,B,C;H)$ averaging over the polar angle $\psi$ (defined
  from the shifted axis as shown by the arrow).}
 \label{fig_zb_l4}
\end{figure*}

Figures~\ref{fig_dlam_l4} and \ref{fig_dlam_l5} show the angular difference between the longitude
in orbit $\lambda$ of the Trojan and the longitude in orbit $\lambda_{\rm Earth}$ of the Earth at the
epoch of detection for the L4- and L5-swarm objects by CSS during the phase~II (May~31, 2016 to
June~15, 2022). The opposition geometry has $\lambda-\lambda_{\rm Earth}\simeq 0$ (neglecting Trojan's
orbital inclination), while positive/negative values of the difference shift the detection towards
the morning/evening sky of the night. Detections proceed in ``seasons'' with the first objects
in the swarm are detected on the extreme morning sky and drift toward the extreme evening sky
during nearly a year interval. They are affected not only by the galactic-plane interference
(Figs.~\ref{fig1} and \ref{fig2}), but also by practical aspects of the CSS operations. For instance,
system maintenance is typically scheduled in summer, and this produces a gap in the observations
(see the summer interruptions in detections of even the bright Trojans on the top panels of
Figs.~\ref{fig_dlam_l4} and \ref{fig_dlam_l5}. Small Trojans, which naturally appear faint, are
detected only during the opposition (bottom panels). This helps to overcome the detection bar
set by the limiting apparent magnitude on the image ($\leq 21.5$ typically), by minimizing
magnitude-phase correction on the right hand side of Eq.~(\ref{pogson}). All the effects of the
viewing geometry and CSS operations are presumably correctly accounted for in our determination of
the detection probability ${\cal P}$ (Sec.~\ref{secbias}). However, the photometric limit related issues
are slightly more delicate and warrant a further testing.

This is because in determination of ${\cal P}$ we downgraded the full dependence on all orbital
elements to only three ``quasi-proper elements'' $(A,B,C)$, thus ${\cal P}(A,B,C;H)$. The fully
detailed detection probability has been averaged over the associated ``quasi-proper angles''
$(\phi,\psi,\theta)$ (Sec.~\ref{secbias}). Consulting the statistics of the detected Trojans
of different absolute magnitude $H$ reveals that the averaging over $\phi$ and $\theta$ angles
is well justified. The case of the $\psi$ angle, associated with the eccentricity $B$, is slightly
more tricky, and requires a numerical check.

Figure~\ref{fig_zb_l4} shows L4-swarm Trojans detected by CSS from April~2021 to April~2022 (the
last season shown in Fig.~\ref{fig_dlam_l4}). During this period of time galactic latitude of the detected Trojans
was large, such that no losses due to dense stellar fields near the galactic plane occurred
(Figs.~\ref{fig1} and \ref{fig2}). The summer maintenance interruption affected only a small
portion of the possible detections on the morning sky, such that the conditions are favorable to
test the remaining photometric effects. The left panel on Fig.~\ref{fig_zb_l4} shows position
of the detected large Trojans, namely those with $H<14$ magnitude. Their distribution in the
plane of osculating heliocentric elements $(e\cos\varpi,e\sin\varpi)$ is fairly symmetrically
distributed about the origin shifted to $(e_{\rm J}\cos(\pi/3+\varpi_{\rm J}),e_{\rm J}\sin(\pi/3+
\varpi_{\rm J})$ (this is the principal forcing effect by Jupiter; Eq.~(\ref{propere})). When switching
to the population of small L4 Trojans, with $H>15$ magnitude on the right panel of Fig.~\ref{fig_zb_l4},
things are different. Objects are preferentially detected only in the first quadrant characterized
by $\varpi\in (0,\pi/2)$. Obviously, this only means that the complementary objects on different
values of $\varpi$ were not detected and we need to check that the proposed simplified scheme
for ${\cal P}(A,B,C;H)$ describes the situation correctly.
\begin{figure*}[t!]
 \plottwo{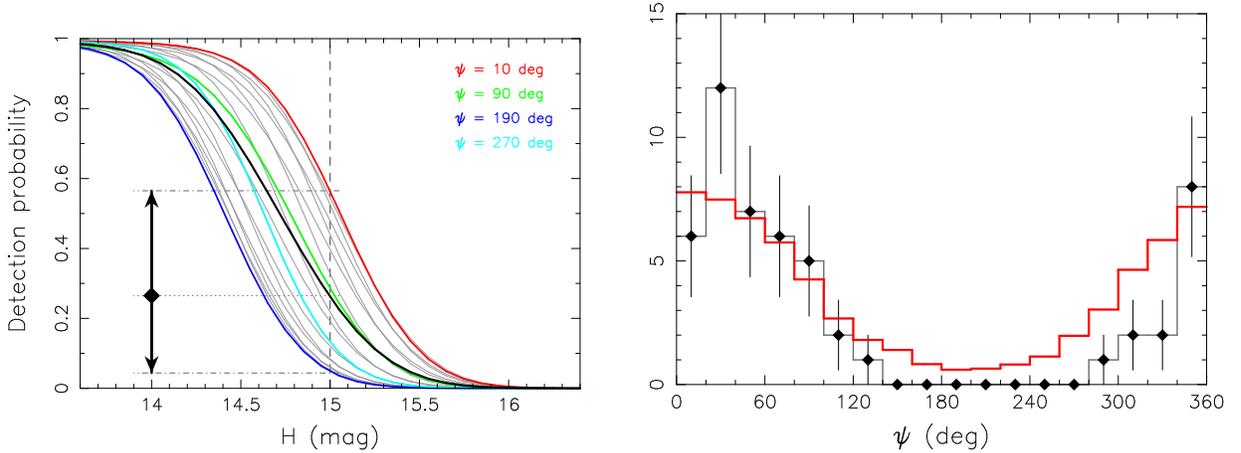}{fAppB4b.eps}
 \caption{Left panel: Detection probability of L4 Trojans in the quasi-proper orbital element
  bin defined by (i) $A\in (0.2,0.225)$, (ii) $B\in (0.06, 0.08)$, and (iii) $C\in (1.58,1.60)$.
  Here we use only CCS observations in between April~2021 and April~2022. The bold black line
  is the $\psi$-averaged detection probability ${\cal P}_{\rm ave}={\cal P}(A,B,C;H)$,
  the gray curves show the $\psi$-resolved detection probabilities ${\cal P}_i$ in $20^\circ$
  wide bins. Four individual values are color-highlighted. At $H=15$ magnitude the minimum and
  maximum detection probabilities ${\cal P}_i$ are more than an order-of-magnitude different
  (see the arrow). The diamond symbol is the ${\cal P}_{\rm ave}$ value. Right panel: The
  appropriately scaled ${\cal P}_i$ values (red histogram), compared to actual number $N_i$ of
  detected Trojans in the $(B,\psi)$ annulus with $B\in (0.06, 0.08)$, but all $(A,\varphi,C,\theta)$
  values (in order to increase statistics); for simplicity, the $N_i$ are assumed to have
  $\sqrt{N_i}$ uncertainties shown by vertical intervals.}
 \label{fig_psi_test}
\end{figure*}

Inquiring little more about parameters of the detected faint Trojans we analyzed the phase of their
heliocentric motion at the detection epoch. In particular, we color-coded in red those Trojans,
whose mean anomaly $M$ was less than $20^\circ$ from the perihelion. As suspected, most of the
detections of the large-$H$ Trojans occurs when they are near perihelion (i.e., $\lambda\simeq \varpi$).
From the bottom panel on Fig.~\ref{fig_dlam_l4} we already know that they are also detected near
opposition (i.e., $\lambda\simeq \lambda_{\rm Earth}$). Combining both conditions together we obtain
$\varpi\simeq \lambda_{\rm Earth}\simeq 50^\circ$ in the middle of the 2021 season, and this is where
we see them. What is special about both conditions? They both help minimizing the apparent
magnitude $m$ of the Trojan on the image by making both the second and the third terms on the
right hand size of Pogson's relation Eq.~(\ref{pogson}) as small as possible: (i) the opposition and
perihelion geometry make both $R$ and $\Delta$ small, and (ii) the opposition geometry makes
the magnitude-phase correction $P(\alpha)$ small as well. These are the optimum conditions, and
the smallest detectable Trojans require this situation to occur.

In order to proceed quantitatively, we selected the following bin of the quasi-proper elements:
(i) $A\in (0.2,0.225)$, (ii) $B\in (0.06, 0.08)$, and (iii) $C\in (1.58,1.60)$. This is still a
fairly stable Trojan configuration (Fig.~\ref{fig_stab1a}), and in $A$ and $C$ parameters quite
representative of the population. When determining ${\cal P}$ we averaged over $\phi$ (associated
with $A$) and $\theta$ (associated with $C$). In the case of $\psi$ angle we proceeded as follows.
First, we determined the proposed ${\cal P}_{\rm ave}={\cal P}(A,B,C;H)$ by averaging over $\psi$ too. 
This is the approach we use for debiasing Trojan population in this paper. In order to check its 
validity, we next evaluated ${\cal P}$ using an extended formulation, where $\psi$ angle has been 
also explicitly considered, thus ${\cal P}(A,B,C,\psi;H)$ and we used $20^\circ$ bins in $\psi$. 
The bin indexed $i$ has $\psi=\psi_i\in [(i-1)\,20^\circ,i\,20^\circ]$, and we denote the locally evaluated
detection probability ${\cal P}_i={\cal P}(A,B,C,\psi_i;H)$. Left panel on Fig.~\ref{fig_psi_test} shows
the individual ${\cal P}_i$ values as a function of $H$, highlighting those in the bins with $i=1$ 
($\psi\in(0^\circ,20^\circ$)), $i=5$ ($\psi\in(80^\circ,100^\circ$)), $i=10$ ($\psi\in(180^\circ,200^\circ$)), 
and $i=14$ ($\psi\in(260^\circ,280^\circ$)). For sake of comparison, we also show the averaged
probability ${\cal P}_{\rm ave}$. At absolute magnitude $\simeq 13.75$ all probabilities are basically
unity, indicating these are bright enough objects to be detected. At absolute magnitude $\simeq 16.1$
all probabilities are basically zero, indicating these are too faint objects to be detected. The
interesting transition values are at $H\simeq 15$. The maximum ${\cal P}_i$ values for small $\psi$
angles reach $\simeq 0.68$, while the minimum ${\cal P}_i$ values at about $\psi=190^\circ$ as
$\simeq 0.06$. This is more than a magnitude difference indicating that even these small Trojans
may be detected if $\psi\simeq 10^\circ$, but they are unlikely to be detected if $\psi\simeq 190^\circ$.
The averaged ${\cal P}_{\rm ave}\simeq 0.27$ is intermediate. This is the qualitative picture we
get from the right panel on Fig.~\ref{fig_zb_l4}. The right panel on Fig.~\ref{fig_psi_test} makes
this comparison quantitative. Here we show $N_i$, namely the number of Trojans in the $\psi_i$ bin,
and compare it with appropriately scaled ${\cal P}_i$ detection probability in each of the bins
(red histogram). While the observed population in this particular bin is already low, such that
the distribution of $N_i$ is subject to considerable fluctuations,%
\footnote{We also consider $H>15$ Trojans having all possible $(A,\varphi,C,\theta)$ values in the chosen
 annulus in $B$ and different $\psi$ bins. This is because number of detections of such faint Trojans
 are already low.}
the trend is reasonably well 
explained by the distribution of ${\cal P}_i$. Moreover, if we define ${\cal P}_{\rm eff}=(\sum_i N_i)/
(\sum_i N_i/{\cal P}_i)$, an effective detection probability for Trojans in the tested blue annulus
on Fig.~\ref{fig_zb_l4}, we obtain ${\cal P}_{\rm eff}= 0.40^{+0.07}_{-0.24}$. This compares reasonably
well with the averaged ${\cal P}_{\rm ave}=0.27$ .

\end{document}